\documentclass[11pt,letterpaper]{article}

\usepackage[margin=1in]{geometry}
\usepackage{amsmath,amssymb,amsthm,mathtools}
\usepackage[table]{xcolor}
\usepackage{array,multirow,makecell,graphicx,hhline}
\usepackage{tabularray}
\usepackage{pgfplots,pgfplotstable}
\pgfplotsset{compat=1.18}
\usepackage{microtype}
\usepackage[colorlinks=true,citecolor=blue,linkcolor=blue,urlcolor=blue]{hyperref}

\title{Tight Bounds for \texorpdfstring{$k$}{k}-Median++ and Improved Bounds for \texorpdfstring{$k$}{k}-Means++}

\author{
    Haim {Kaplan}%
    \thanks{The Blavatnik School of Computer Science and AI, Tel Aviv University, Tel Aviv, Israel.
    Emails: \texttt{\{haimk, dnaori\}@tau.ac.il}. This research was supported by ISF grant number 1156/23 and the Blavatnik Research Foundation.} 
    \and
    David {Naori}\footnotemark[1]
}

\date{}

\usepackage{amsmath,amssymb,amsthm,mathtools}
\usepackage{ifthen}
\usepackage{dsfont}

\usepackage{float}

\usepackage{tikz}

\usepackage{tikzsymbols}

\newcommand{\R}{\mathbb R}
\newcommand{\E}{\mathbb E}

\newcommand{\OPT}{\operatorname{OPT}}
\newcommand{\cost}{\operatorname{cost}}

\makeatletter
\newcommand{\@giventhatnostar}[2]{#1\;\middle|\;#2}
\newcommand{\@giventhatstar}[3][]{#1(#2\;#1|\;#3#1)}
\newcommand{\given}{\@ifstar\@giventhatstar\@giventhatnostar}
\makeatother

\providecommand{\cost}{\operatorname{cost}}
\providecommand{\OPT}{\operatorname{OPT}}

\newtheorem{theorem}{Theorem}[section]
\newtheorem{lemma}[theorem]{Lemma}

\newtheorem{corollary}[theorem]{Corollary}

\theoremstyle{remark}
\newtheorem{remark}[theorem]{Remark}

\begin{document}

\pagenumbering{roman}

\maketitle
\thispagestyle{empty}

\begin{abstract}
The popular $k$-means++ algorithm, or $D^\ell$-sampling, is a fundamental primitive in both the theory and practice of clustering. In the bi-criteria setting, the algorithm samples $k+\Delta$ centers ($\Delta\ge0$), and its cost is compared with the optimum using $k$ centers. Despite extensive study over the past two decades, its exact approximation ratio and tight bi-criteria guarantees remain unknown.

We give a unified analysis for every distance exponent $\ell>0$, every $k\ge1$, and every $\Delta\ge0$. For metric $k$-median, our bounds are tight for all $k$ and $\Delta$. Without extra centers, the exact approximation ratio is $2(1+H_{k-1})$ (also in Euclidean space), improving the previous upper bound by a factor of $2$ ($H_j$ denotes the $j$th harmonic number). This tight analysis reveals that even one extra center improves the approximation ratio by an additive $\Theta(\log\log k)$, whereas previous bounds show no improvement.

For every $\ell>0$, our analysis gives upper bounds with the same dependence on $k$ and $\Delta$ as our tight bound for metric $k$-median, with leading constants depending only on $\ell$ and whether the metric is Euclidean. Beyond the improved dependence on $\Delta$, our leading constants are smaller than those in previous analyses. In particular, for the widely used Euclidean $k$-means objective, our upper bound without extra centers is $(1+\varphi)(1+H_{k-1})$, where $\varphi\approx1.618$ is the golden ratio, improving upon the previous constant of $5$. With oversampling, a constant fraction of extra centers is known to provide a constant-factor approximation. For example, with $10\%$ extra centers, our guarantee is $5.7$, compared with the best previous guarantee of $25.9$.
\end{abstract}

\clearpage
\pagenumbering{Roman}
\pagestyle{empty}
\tableofcontents

\clearpage
\pagestyle{plain}
\pagenumbering{arabic}
\section{Introduction}
\label{sec:introduction}

In $k$-clustering, we are given a set $X$ of points
in a metric space $(\mathcal M,d)$, an integer $k\ge1$, and a
\emph{distance exponent} $\ell>0$. The goal is to choose a set $C\subseteq\mathcal M$
of $k$ centers minimizing $\cost(C)=\sum_{x\in X}d(x,C)^\ell$, where
$d(x,C)$ is the distance from $x$ to its nearest center in $C$.
The two most prominent cases are $k$-median, with $\ell=1$, and
$k$-means, with $\ell=2$. The Euclidean setting is particularly
important in applications~\cite{Jain2010}.

The fundamental $k$-means++ algorithm of Arthur and
Vassilvitskii~\cite{AV2007} chooses centers by a simple sampling
rule. The first center is chosen uniformly from $X$. Each subsequent
center $x$ is sampled from $X$ with probability proportional to
$D(x)^\ell$, where $D(x)=d(x,C)$ and $C$ is the set of centers
selected so far.
The algorithm stops after $k$ centers have
been chosen. This rule is also known as $D^\ell$-sampling.
While the name $k$-means++ is usually associated with the
$k$-means objective, we use it for all $\ell$.
For $\ell=1$, this algorithm is also called
$k$-median++~\cite{FLL2023}.

$k$-means++ is a prominent example of an algorithm developed in
theoretical computer science that has become widely used in
practice. It and its variants are used in major clustering
libraries and platforms, including scikit-learn, Apache Spark,
and Google BigQuery ML~\cite{ScikitLearnKMeans,SparkClustering,BigQueryKMeans}.
Its influence on the theory of clustering is equally substantial.
It has inspired many variants, including parallel, streaming, and
faster sampling algorithms~\cite{BMVKV2012,AJM2009,BLHK2016}.
Its sampling rule also serves as a building block in other
clustering algorithms~\cite{JKS2014,LS2019,CGPR2020}.
A precise understanding of this rule is therefore important both
for $k$-means++ itself and for the clustering algorithms built on it.
We further explore the connections between $k$-means++
and two fundamental algorithms: the greedy farthest-first algorithm
for $k$-center~\cite{Gonzalez1985} and Meyerson's algorithm
for online facility location~\cite{Meyerson2001}.

A substantial line of work studies the guarantees and limitations
of this sampling
primitive~\cite{ADK2009,BR2013,DasguptaNotes,Wei2016,BJA2016,BERS2020,GOR2023,BNS2024,Rozhon2026}.
With exactly $k$ centers, the best known bounds on the
approximation ratio for distance exponent $\ell\ge1$ are
$c_\ell(1+H_{k-1})$, where $c_\ell=2^{2\ell}$~\cite{AV2007},
except for squared Euclidean distances, for which there is a
better constant $c_2=5$~\cite{MRS2020}.
Here, $H_j$ is the $j$th harmonic number.\footnote{ Wei~\cite{Wei2016}
improves the harmonic bound for $k=2,3$ when $\Delta=0$,
using the same $c_\ell$.
In particular, for Euclidean $k$-means with $k=2$ and
$\Delta=0$, combining Wei's analysis with~\cite{MRS2020} gives
the upper bound $5$, matching the lower bound of
Lemma~\ref{lem:euclidean-two-center-exact}
in Appendix~\ref{app:euclidean-balancing-tight}.}

Bi-criteria guarantees for $k$-means++ and its variants have also
been studied extensively~\cite{ADK2009,AJM2009,BMVKV2012,Wei2016,BLK2017,MRS2020}.
Here the algorithm samples $k+\Delta$ centers, for an
integer $\Delta\ge0$, and its cost is compared with the optimal
cost using $k$ centers.
These bounds quantify how much a few extra centers help
and how many suffice for a desired approximation factor.
In particular, when $\Delta=\Omega(k)$, $k$-means++ gives a
constant-factor approximation~\cite{ADK2009,Wei2016,MRS2020}.
Bi-criteria guarantees also guide the design and analysis of
algorithms that first sample a larger set of centers and then reduce their number to
$k$~\cite{ADK2009,AJM2009,BMVKV2012,BLK2017}.
Tighter constants in the bi-criteria bounds for $k$-means++
therefore improve the approximation guarantees of these algorithms.
Bi-criteria guarantees are also relevant when the number of clusters
is not known in advance and is estimated using heuristics such
as the elbow method~\cite{Wei2016,MRS2020}.

The best known general bi-criteria bounds are due to Makarychev,
Reddy, and Shan~\cite{MRS2020}. For distance exponent
$\ell\ge1$, their analysis gives
two upper bounds on the approximation ratio:
$c_\ell(2+\frac{1}{2e}+\ln\frac{2k}{\Delta})$ for
$1\le\Delta\le2k$ and $c_\ell(1+\frac{k}{e(\Delta-1)})$ for $\Delta>1$,
where, as above, $c_\ell=2^{2\ell}$, except for squared Euclidean distances,
for which $c_2=5$.\footnote{The bounds in~\cite{MRS2020}
are stated for Euclidean $k$-means with $c_2=5$. The same analysis applies to
every $\ell\ge1$ using the bound of~\cite{AV2007},
also used in the bi-criteria analysis of Wei~\cite{Wei2016}.}

Despite the prominence of $k$-means++ in theory and practice
and the many works devoted to its analysis, its performance
as a function of $\Delta$ is still not fully understood.
The best known guarantee captures the asymptotic behavior by
taking the minimum of three bounds: one for exactly $k$ centers
and the two bi-criteria bounds of~\cite{MRS2020}.
The underlying analyses capture different effects in the sampling
process, but their combination does not tightly account for how
these effects interact across the range of $\Delta$.
In particular, the combined bound does not fully capture the
improvement in performance from small amounts of oversampling.
For example, it gives no better guarantee for $k+1$ centers
($\Delta=1$) than for exactly $k$ centers.

The exact performance remains unknown even at $\Delta=0$,
where there is a gap between the leading constants in the
lower and upper bounds: $2^{\ell}$ versus $2^{2\ell}$ in general
metrics, and $2^{\ell/2}$ versus $2^{2\ell}$ in Euclidean
spaces, except for Euclidean $k$-means, where the constants
are $2$ and $5$, respectively.\footnote{The construction of~\cite{AV2007}
gives a lower bound of $2H_k$ for Euclidean $k$-means.
Adapting it to every $\ell\ge1$ gives $2^{\ell/2}H_k$ in
Euclidean space; a modification of the metric gives
$2^\ell H_k$ in general metrics.}
These gaps also leave unclear whether $k$-means++ performs better
for $k$-median than for $k$-means, or
in Euclidean space than in general metrics for a fixed $\ell$.

\subsection{Our Results}
\label{subsec:intro-results}

We give a unified analysis of $k$-means++ for every $\Delta\ge0$
(Theorem~\ref{thm:exact-sampling}).
As in previous analyses, our upper bound on the
approximation ratio has the form $\alpha_\ell B_{k,\Delta}$,
where $\alpha_\ell$ depends only on the distance exponent $\ell$ and on
whether we are in Euclidean space or in a general metric space,
while $B_{k,\Delta}$ depends only on $k$ and $\Delta$.
These upper bounds hold for every $\ell>0$.
We also give lower bounds of the same form: a leading constant
times $B_{k,\Delta}$.
We describe $B_{k,\Delta}$ in the overview of our techniques in
Section~\ref{subsec:intro-overview}.

For metric $k$-median, we obtain $\alpha_1=2$ and prove that
the bound $2B_{k,\Delta}$ is \emph{tight} for every $k$ and $\Delta$ (Theorem~\ref{thm:metric-median-tight}).
With exactly $k$ centers, $B_{k,0}=1+H_{k-1}$, giving the tight
ratio $2(1+H_{k-1})$ and improving the previous leading constant
from $4$ to $2$. For $\Delta=0$, we also give a Euclidean
construction with the same lower bound
(Theorem~\ref{thm:euclidean-median-tight}), showing that the
performance of $k$-median++ is the same in
Euclidean space and in general metrics.

In particular, our analysis captures the benefit of any
number $\Delta>0$ of additional centers for every $k>1$.
For example, it shows that a single extra center gives an
additive $\Theta(\log\log k)$ improvement 
(Corollary~\ref{cor:loglog-bound}).
This improvement comes from the common factor $B_{k,\Delta}$,
so our upper bounds capture it for every distance exponent $\ell$.

For Euclidean $k$-means,~\cite{MRS2020} showed that their
leading constant $5$ is tight for the technique used in their
analysis and previous analyses.
Our analysis breaks this barrier with
$\alpha_2=1+\varphi\approx2.618$, where
$\varphi=(1+\sqrt5)/2$ is the golden ratio.
In particular, with exactly $k$ centers the bound is
$(1+\varphi)(1+H_{k-1})$.
We also show that any bound of the form
$\alpha(1+H_{k-1})$ that holds for every $k$ must have
$\alpha\ge5/2$. Our construction for $k=2$ gives a lower bound
of $5$, matching the upper bound obtained
from~\cite{Wei2016,MRS2020}
(Lemma~\ref{lem:euclidean-two-center-exact}).
This also separates the performance of $k$-means++ for the
$k$-median and $k$-means objectives, even in
Euclidean space.

For $k$-means in general metrics, we obtain an upper bound
with $\alpha_2\approx4.383$, improving the previous leading
constant $16$ (Lemma~\ref{lem:balancing-extensions}).
For the lower bound, the same construction used for metric
$k$-median gives $4B_{k,\Delta}$
(Corollary~\ref{cor:metric-squared-lower}).
This lower bound exceeds our Euclidean upper bound
$(1+\varphi)B_{k,\Delta}$, showing that, for $k$-means,
restricting to Euclidean space gives a better guarantee for
every $k$ and $\Delta$.

For $\ell>2$, we obtain upper bounds with
$\alpha_\ell=3^{\ell-1}+2^\ell$ in general metrics and
$\alpha_\ell=3^{\ell-1}+2^{\ell-1}$ in Euclidean space
(Lemma~\ref{lem:balancing-extensions}).
These constants improve on $2^{2\ell}$, the best
previously known leading constant in both settings.
On the lower-bound side, we obtain
$2^\ell B_{k,\Delta}$ in general metrics and
$2^{\ell/2}B_{k,\Delta}$ in Euclidean space for every $\ell>0$
(Corollary~\ref{cor:euclidean-power-lower}).
For $\Delta=0$, we strengthen the Euclidean lower bound to
$\max\{2,2^{\ell-1}\}B_{k,0}$
(Corollary~\ref{cor:euclidean-power-no-extra}).
For Euclidean $k$-median with $\Delta>0$, the lower bound is
$\sqrt2 B_{k,\Delta}$, leaving a gap to our upper bound of
$2B_{k,\Delta}$ (our bounds are tight in all other $k$-median cases).

For the remaining ranges $0<\ell<1$ and $1<\ell<2$,
we extend our upper bounds by reductions to $k$-median and
$k$-means, respectively (Lemma~\ref{lem:balancing-small-exponents}).
In particular, for $0<\ell<1$, we obtain the upper bound
$2B_{k,\Delta}$ in both settings, which is tight for $\Delta=0$
(Corollary~\ref{cor:euclidean-power-no-extra}).
Table~\ref{tab:bounds-intervals} summarizes all our upper and lower bounds.

\begin{table}[h!]
\centering
\small
\definecolor{boundstight}{RGB}{248,244,227}
{\color{black}
\resizebox{\linewidth}{!}{%
\begin{tblr}{
    colspec={Q[c,m]Q[c,m]Q[c,m]Q[c,m]Q[c,m]},
    colsep=4pt, rowsep=5pt,
    row{1,2}={rowsep=3pt},
    column{2,5}={colsep=6pt},
    column{3}={leftsep=6pt,rightsep=8pt},
    column{4}={leftsep=8pt,rightsep=6pt},
    hline{1,8}={1-5}{0.5pt,black},
    hline{2}={2-5}{0.35pt,black!75},
    hline{3,4,6,7}={1-5}{0.35pt,black!75},
    hline{5}={1-5}{0.55pt,black!75},
    vline{1,6}={0.5pt,black},
    vline{2}={0.35pt,black!75},
    vline{3,5}={2-7}{0.55pt,black!75},
    vline{4}={0.8pt,black!80},
    cell{1}{1}={r=2}{c},
    cell{1}{2,4}={c=2}{c},
    cell{3}{2,3,4}={bg=boundstight},
    cell{5}{2,3}={bg=boundstight},
}
$\ell$ & $\Delta=0$ & & $\Delta>0$ & \\
& Metric & Euclidean & Metric & Euclidean\\
{$\ell=1$\\($k$-median)}
    & $2$ & $\textcolor{orange!70!black}{2}$
    & $2$ & $[\sqrt2,2]$\\
{$\ell=2$\\($k$-means)}
    & $[\textcolor{purple}{4},4.383]$
    & {$\begin{gathered}
            [2,\textcolor{purple}{2.618}]\\[2pt]
            \text{%
              \setlength{\fboxsep}{1.5pt}%
              \colorbox{boundstight}{%
                $\textcolor{orange!70!black}{2.5\ (k=2)}$}%
            }
          \end{gathered}
        $}
    & $[\textcolor{purple}{4},4.383]$
    & $[2,\textcolor{purple}{2.618}]$\\
$0<\ell<1$
    & $2$ & $2$
    & $[2^\ell,2]$ & $[2^{\ell/2},2]$\\
$1<\ell<2$
    & $[2^\ell,\min\{4.383,3^{\ell-1}+2^\ell\}]$
    & $[2,2.618]$
    & $[2^\ell,\min\{4.383,3^{\ell-1}+2^\ell\}]$
    & $[2^{\ell/2},2.618]$\\
$\ell>2$
    & $[2^\ell,3^{\ell-1}+2^\ell]$ & $[2^{\ell-1},3^{\ell-1}+2^{\ell-1}]$
    & $[2^\ell,3^{\ell-1}+2^\ell]$ & $[2^{\ell/2},3^{\ell-1}+2^{\ell-1}]$\\
\end{tblr}%
}
}
\caption{Our bounds on the approximation ratio,
written as $[\text{lower bound},\text{upper bound}]$.
Both endpoints multiply $B_{k,\Delta}$, where
$B_{k,0}=1+H_{k-1}$.
The shaded cells give tight bounds for every $k$.
The highlighted value $2.5\ (k=2)$ is tight for Euclidean
$k$-means with $k=2$ and $\Delta=0$, where the approximation ratio
is exactly $2.5 B_{2,0}=5$
(Lemma~\ref{lem:euclidean-two-center-exact}).
The interval $[2,2.618]$ in the same cell applies to every $k$.
Purple entries mark the separation between metric and Euclidean
$k$-means; orange entries mark the separation between Euclidean
$k$-median and $k$-means at $k=2$ and $\Delta=0$.
The values $2.618$ and $4.383$ are rounded approximations to
$1+\varphi=(3+\sqrt5)/2$ and the solution $a>4$ of
$a=4+a/(a-1)^2$, respectively.}
\label{tab:bounds-intervals}
\end{table}

The function $B_{k,\Delta}$ can be computed efficiently for any
$k$ and $\Delta$ using a simple recurrence
(Section~\ref{subsec:equal-mass}).
We also give a closed-form upper bound that is tight up to an
additive $1/\Delta$
(Theorems~\ref{thm:closed-form} and~\ref{thm:closed-form-lower}).
For every $\Delta\ge1$, 
\[
    B_{k,\Delta}
    \le
    1+W\left(
        \frac{k-1}{e\Delta}e^{1/(2\Delta)}
    \right),
\]
where $W$ is the nonnegative Lambert $W$ function, defined by
$W(x)e^{W(x)}=x$.

In addition to reducing the leading constant, our
analysis improves the dependence on $\Delta$ in the bound
of~\cite{MRS2020} for every $k>1$ and $\Delta>0$.

For example, with $\Delta=\lceil k/10\rceil$ extra
centers, our Euclidean $k$-means guarantee is $5.7$ for every
$k\ge1$. The corresponding previous guarantee uniform over $k$
is at least $5(2+\frac{1}{2e}+\ln20)\approx25.9$.
Indeed, at $k=80$ and $\Delta=8$, this is the minimum of the
harmonic bound and the two bi-criteria bounds of~\cite{MRS2020}.

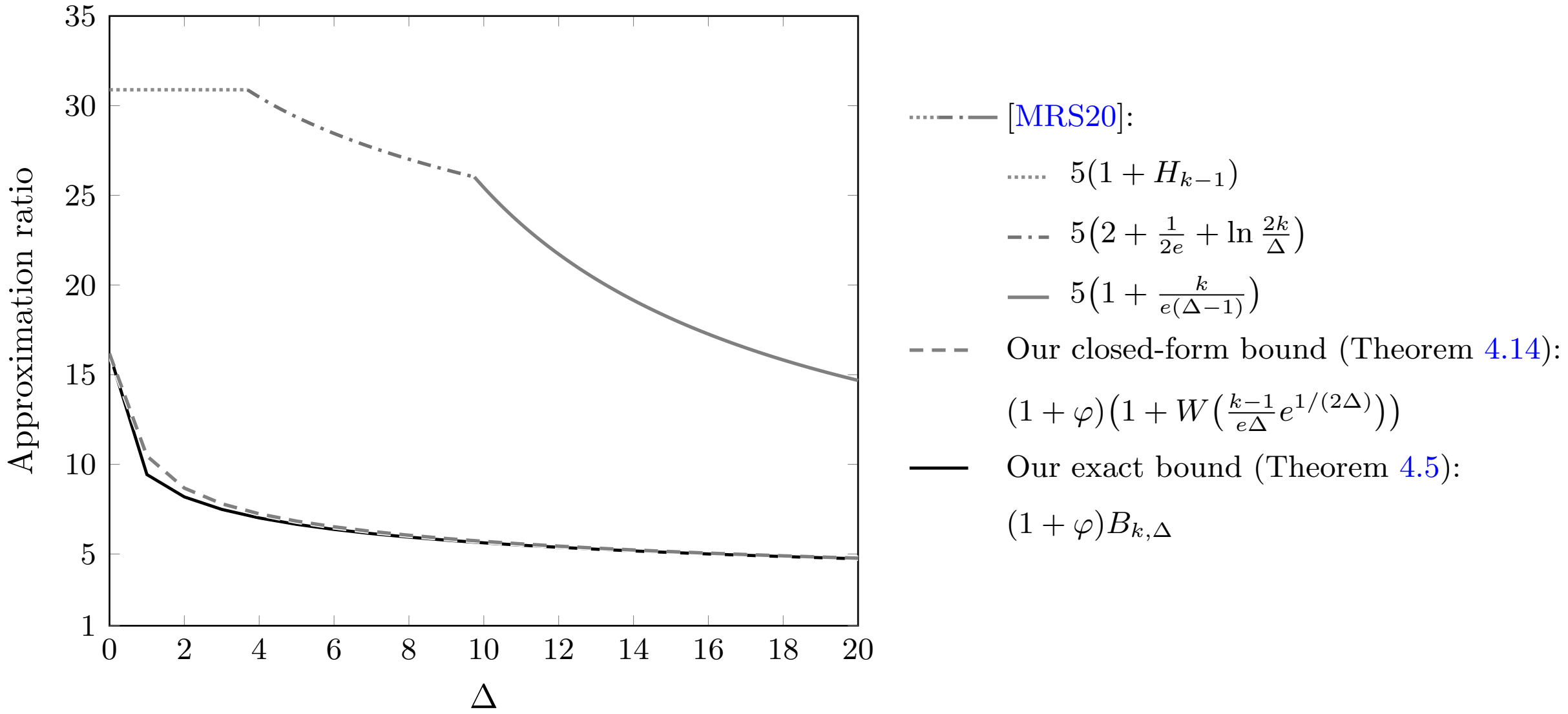
\begin{figure}[!ht]
\centering
{\color{black}\begingroup
\providecommand{\KMeansPlotWidth}{0.56\linewidth}
\providecommand{\KMeansPlotHeight}{3.08in}
\def\KMeansK{100}
\def\KMeansH{5.17737751763962}
\def\KMeansDivisor{1}
\pgfmathdeclarefunction{kmbaseline}{1}{\pgfmathparse{5*(1+\KMeansH)}}
\pgfmathdeclarefunction{kmsmall}{1}{\pgfmathparse{5*(2+1/(2*exp(1))+ln(2*\KMeansK/#1))}}
\pgfmathdeclarefunction{kmlarge}{1}{\pgfmathparse{5*(1+\KMeansK/(exp(1)*(#1-1)))}}
\pgfplotstableread{
delta exact closed
0 16.17258430252 16.17258430252
1 9.4301758084333 10.4652792116015
2 8.1836521034629 8.67432761532149
3 7.48682327126901 7.80187589546174
4 7.0108944541696 7.23999174285128
5 6.65378937306832 6.83217690148779
6 6.3705773065863 6.51564285251985
7 6.13756472631465 6.25913344849784
8 5.94076125613124 6.04491405956094
9 5.77123465535778 5.8619899479273
10 5.62294486554743 5.70309291027281
11 5.49162331860039 5.56317820264912
12 5.37414594908902 5.43860759035425
13 5.26816032875133 5.32667478836458
14 5.17185275655948 5.22531474930583
15 5.08379676490698 5.1329176591196
16 5.00285118208026 5.04820552972434
17 4.92808953108851 4.97014776581823
18 4.85874989986484 4.89790185158618
19 4.79419856708352 4.83077071763459
20 4.73390310130449 4.76817147529444
}\KMeansData
\begin{tikzpicture}[
km mrs baseline/.style={black!45,line width=1pt,densely dotted},
km mrs small/.style={black!55,line width=1pt,dash pattern=on 4pt off 2pt on 1pt off 2pt},
km mrs large/.style={black!50,line width=1pt,solid},
km exact/.style={black,line width=0.9pt},
km closed/.style={black!50,line width=1pt,dash pattern=on 4pt off 2.5pt,
  preaction={draw=white,line width=1.6pt,dash pattern=on 4pt off 2.5pt}},
]
\newcommand{\KMeansLegendLine}[1]{%
  \tikz[baseline=-0.4ex]{\draw[km mrs #1] (0pt,0pt)--(12pt,0pt);}}
\pgfplotsset{km mrs combined/.style={legend image code/.code={%
  \draw[km mrs baseline] (0cm,0cm)--(0.3cm,0cm);
  \draw[km mrs small] (0.3cm,0cm)--(0.6cm,0cm);
  \draw[km mrs large] (0.6cm,0cm)--(0.9cm,0cm);
}}}
\begin{axis}[
  name=kmAxis,width=\KMeansPlotWidth,height=\KMeansPlotHeight,
  xmin=0,xmax=20,ymin=1,ymax=35,
  extra y ticks={1},extra y tick labels={1},
  xlabel={$\Delta$},ylabel={Approximation ratio},
  axis lines=box,axis line style={line width=0.5pt},
  tick align=inside,tick pos=both,major tick length=3pt,scaled ticks=false,
  label style={font=\small},tick label style={font=\footnotesize},title style={font=\small},
  legend style={at={(1,0.5)},anchor=west,xshift=12pt,
    draw=none,fill=none,font=\footnotesize,cells={anchor=west},row sep=4pt,inner sep=2pt},
  legend columns=1,
]
\addplot[km mrs baseline,forget plot,no marks,
  domain=0:3.68724501018314,samples=2,variable=\x]
  ({\x/\KMeansDivisor},{kmbaseline(\x)});
\addplot[km mrs small,forget plot,no marks,
  domain=3.68724501018314:9.74827168877635,samples=100,variable=\x]
  ({\x/\KMeansDivisor},{kmsmall(\x)});
\addplot[km mrs large,forget plot,no marks,
  domain=9.74827168877635:20,samples=100,variable=\x]
  ({\x/\KMeansDivisor},{kmlarge(\x)});
\addplot[km exact,no marks,forget plot] table[x expr=\thisrow{delta}/\KMeansDivisor,y=exact] {\KMeansData};
\addplot[km closed,no marks,forget plot] table[x expr=\thisrow{delta}/\KMeansDivisor,y=closed] {\KMeansData};
\addlegendimage{km mrs combined}
\addlegendentry{\cite{MRS2020}:}
\addlegendimage{empty legend}
\addlegendentry{\KMeansLegendLine{baseline}\; $5(1+H_{k-1})$}
\addlegendimage{empty legend}
\addlegendentry{\KMeansLegendLine{small}\; $5\bigl(2+\frac{1}{2e}+\ln\frac{2k}{\Delta}\bigr)$}
\addlegendimage{empty legend}
\addlegendentry{\KMeansLegendLine{large}\; $5\bigl(1+\frac{k}{e(\Delta-1)}\bigr)$\rule[-4pt]{0pt}{10pt}}
\addlegendimage{km closed}
\addlegendentry{Our closed-form bound (Theorem~\ref{thm:closed-form}):}
\addlegendimage{empty legend}
\addlegendentry{$(1+\varphi)\bigl(1+W\bigl(\frac{k-1}{e\Delta}e^{1/(2\Delta)}\bigr)\bigr)$}
\addlegendimage{km exact}
\addlegendentry{Our exact bound (Theorem~\ref{thm:exact-sampling}):}
\addlegendimage{empty legend}
\addlegendentry{$(1+\varphi)B_{k,\Delta}$}

\end{axis}

\end{tikzpicture}
\endgroup}
\caption[Comparison of bounds for Euclidean $k$-means.]{Upper bounds
on the approximation ratio for Euclidean $k$-means with
$k=100$ and $\Delta\in\{0,\ldots,20\}$.
The figure shows our bound from Theorem~\ref{thm:exact-sampling}
and our closed-form bound from Theorem~\ref{thm:closed-form}.
The top curve is the minimum of the harmonic bound and the two
bi-criteria bounds of~\cite[Theorem~5.1]{MRS2020}.}
\label{fig:intro-kmeans-comparison}
\end{figure}

Figure~\ref{fig:intro-kmeans-comparison} compares the bounds for
Euclidean $k$-means with $k=100$.
Note that the previous best bound does not show an improvement
until $\Delta=4$.

The limit $\ell\to\infty$ connects this family of
objectives to $k$-center, which minimizes the maximum distance
to a center.
On each fixed input, taking the $\ell$-th root of the objective
makes it converge to the $k$-center objective as $\ell$ goes to infinity.
At the same time, $D^\ell$-sampling converges to the greedy
farthest-first algorithm, which gives a
$2$-approximation~\cite{Gonzalez1985,BNS2024}.
One might therefore expect the $\ell$-th root of the approximation
ratio to approach $2$ as well.
We show that it instead approaches $3$, already for $k=2$
and $\Delta=0$ on the real line.
The distinction is that convergence to farthest-first holds on
each fixed input, whereas the inputs in our lower bound depend
on $\ell$.

Specifically, for every integer $\ell\ge1$, we give an instance
with approximation ratio at least $3^\ell/4$
(Lemma~\ref{lem:large-exponent-lower}).
The $\ell$-th root of our upper bound also approaches $3$,
since $\alpha_\ell^{1/\ell}$ approaches $3$ and
$B_{k,\Delta}^{1/\ell}$ approaches $1$ for fixed $k$ and $\Delta$.
The same lower bound also shows that any bound of the form
$\alpha_\ell B_{k,\Delta}$ valid for all $k$ and $\Delta$ must have
$\alpha_\ell=\Omega(3^\ell)$.
Thus the dependence of our leading constants on $\ell$ is
asymptotically tight.

\subsection{Overview of the Analysis}
\label{subsec:intro-overview}

Fix an optimal solution and consider its $k$ clusters.
Previous analyses distinguish clusters from which a center has
been sampled from those from which none has been sampled
~\cite{AV2007,DasguptaNotes,Wei2016,MRS2020}.
They bound the expected cost of a cluster after the first
sample from it, and use these bounds to control the cost of
the unsampled clusters.
Our analysis refines both parts.
It has two main ingredients: \emph{balancing} gives the improved
constant $\alpha_\ell$, and a comparison with an
\emph{equal-mass process} gives the factor $B_{k,\Delta}$.

\paragraph{Balancing.}
Let $\phi(P)$ denote the cost of an optimal cluster $P$ in
the optimal solution.
For Euclidean $k$-means, the bound on its expected cost after
the first sample from it is $5\phi(P)$, and the constant $5$
is tight~\cite{MRS2020}.
We obtain a smaller constant by being more selective about
the sample used to bound a cluster's cost.
Each point's current cost is also its sampling weight,
which we call its \emph{mass}.
For the analysis, we split this mass into a \emph{balanced}
part and an \emph{imbalanced} part.
We call a cluster \emph{settled} once its balanced part has
been sampled.
A sample that settles a cluster is a \emph{hit}; every other
sample is a \emph{miss}.
By treating some of the mass at poor candidate centers as
imbalanced, we improve the bound on the expected cost after
a hit compared with the bound after the first sample from
the cluster, at the expense of treating more sampling mass as misses.

\begin{figure}[!ht]
\centering
{\color{black}\begin{tikzpicture}[x=1cm,y=1cm,yscale=.8,font=\small,
    point/.style={circle,fill=blue!70!black,inner sep=2pt},
    samplingmass/.style={draw=blue!65!black,fill=blue!65!black},
    balanced/.style={draw=green!45!black,fill=green!50!black},
    imbalanced/.style={draw=purple!80!black,fill=purple!55}]
\path[use as bounding box] (0,-9.3) rectangle (15.6,2.4);

\draw[blue!55!black,line width=.65pt] (9.6,.38)
    ellipse [x radius=3.75cm,y radius=1.55cm];
\node[blue!55!black,above=3pt] at (9.6,1.93) {$P$};
\draw[black!60] (2.05,-1.25) -- (13.15,-1.25);
\foreach \x/\label in {2.5/{-1},7.5/{0},12.5/{1}} {
    \draw[black!60] (\x,-1.33) -- (\x,-1.17);
    \node[below=4pt] at (\x,-1.33) {$\label$};
}
\node[circle,fill=red!75!black,inner sep=2pt] at (2.5,.38) {};
\node[red!75!black,above=5pt] at (2.5,.38) {$q$};
\node[below] at (2.5,.05) {Existing center};

\foreach \height/\label in {.95/{z_1},.52/{z_2},-.40/{z_n}} {
    \node[point] at (7.5,\height) {};
    \node[anchor=east] at (7.27,\height) {$\label$};
}
\node[blue!70!black] at (7.5,.05) {$\vdots$};
\node[point] at (12.5,.38) {};
\node[above=5pt] at (12.5,.38) {$y$};
\node at (9.8,.40) {$\phi(P)\approx1$};

\begin{scope}[yshift=-1.5cm]
\foreach \offset/\title/\barstyle in {0/{First sample from $P$}/samplingmass,8.1/{Balancing $P$}/balanced} {
  \begin{scope}[xshift=\offset cm]
    \node at (3.7,-2.20) {\title};
    \begin{scope}[yshift=-.5cm]
    \node[rotate=90,text=black!65] at (.92,-3.85) {Sampling mass};
    \draw[black!45] (1.42,-5.00) -- (6.05,-5.00);
    \foreach \x/\label in {1.8/{z_1},2.6/{z_2},4.2/{z_n}} {
      \draw[\barstyle] (\x-.24,-5.00) rectangle (\x+.24,-4.48);
      \node[above=3pt] at (\x,-4.48) {$1$};
      \node[below=4pt] at (\x,-5.00) {$\label$};
    }
    \node at (3.4,-4.75) {$\cdots$};
    \node[below=4pt] at (5.4,-5.00) {$y$};
    \end{scope}
  \end{scope}
}
\begin{scope}[yshift=-.5cm]
\draw[samplingmass] (5.16,-5.00) rectangle (5.64,-2.92);
\node[above=3pt] at (5.4,-2.92) {$4$};

\draw[balanced] (13.26,-5.00) rectangle (13.74,-4.22);
\draw[imbalanced] (13.26,-4.22) rectangle (13.74,-2.92);
\node[above=3pt] at (13.5,-2.92) {$4$};
\node[anchor=west] at (13.83,-4.61) {$3/2$};
\node[anchor=west] at (13.83,-3.57) {$5/2$};

\node[align=center,font=\footnotesize] at (3.7,-6.585) {Expected cost after first sample:\\[3pt]$\dfrac{n\cdot1+4\cdot n}{n+4}=\dfrac{5n}{n+4}\approx5$};
\node[align=center,font=\footnotesize] at (11.8,-6.585) {Expected cost after a hit:\\[3pt]$\dfrac{n\cdot1+(3/2)\cdot n}{n+3/2}=\dfrac{5n}{2n+3}\approx\dfrac52$};
\end{scope}
\end{scope}
\end{tikzpicture}}
\caption[Balancing on the example giving the factor $5$.]{
Balancing for $k$-means on the line, using the example
of~\cite{MRS2020}. Take $n$ large. The cluster $P$ contains
$n$ points $z_1,\ldots,z_n$ at $0$ and one point $y$ at $1$.
Points at the same location are separated vertically in the drawing.
Assume that the center $q=-1$ has already been sampled.
The optimal center of $P$ is $c^*=1/(n+1)$, so
$\phi(P)=n(c^*)^2+(1-c^*)^2=n/(n+1)\approx1$.
The histograms show the sampling masses: each $z_i$ has mass $1$
and $y$ has mass $4$. If a $z_i$ is sampled, assigning all points
of $P$ to it costs $1$; if $y$ is sampled, this cost is $n$.
Thus the expected cost after the first sample from $P$ is
$\frac{n\cdot1+4\cdot n}{n+4}=\frac{5n}{n+4}\approx5$,
showing that the factor $5$ is tight.
To reduce the expected cost after a hit, we declare $5/2$ units
of $y$'s mass imbalanced (purple), retaining $3/2$ as balanced
mass (green).
Conditioned on a hit, only the balanced masses determine the
sampled point, so the expected cost is
$\frac{n\cdot1+(3/2)\cdot n}{n+3/2}=\frac{5n}{2n+3}\approx5/2$.
Thus this choice makes the imbalanced mass and the expected
cost after a hit approximately equal.
Conditioned on sampling $y$, we declare it a miss with probability
$\frac{5/2}{5/2+3/2}=5/8$ and hit otherwise.}
\label{fig:intro-balancing}
\end{figure}
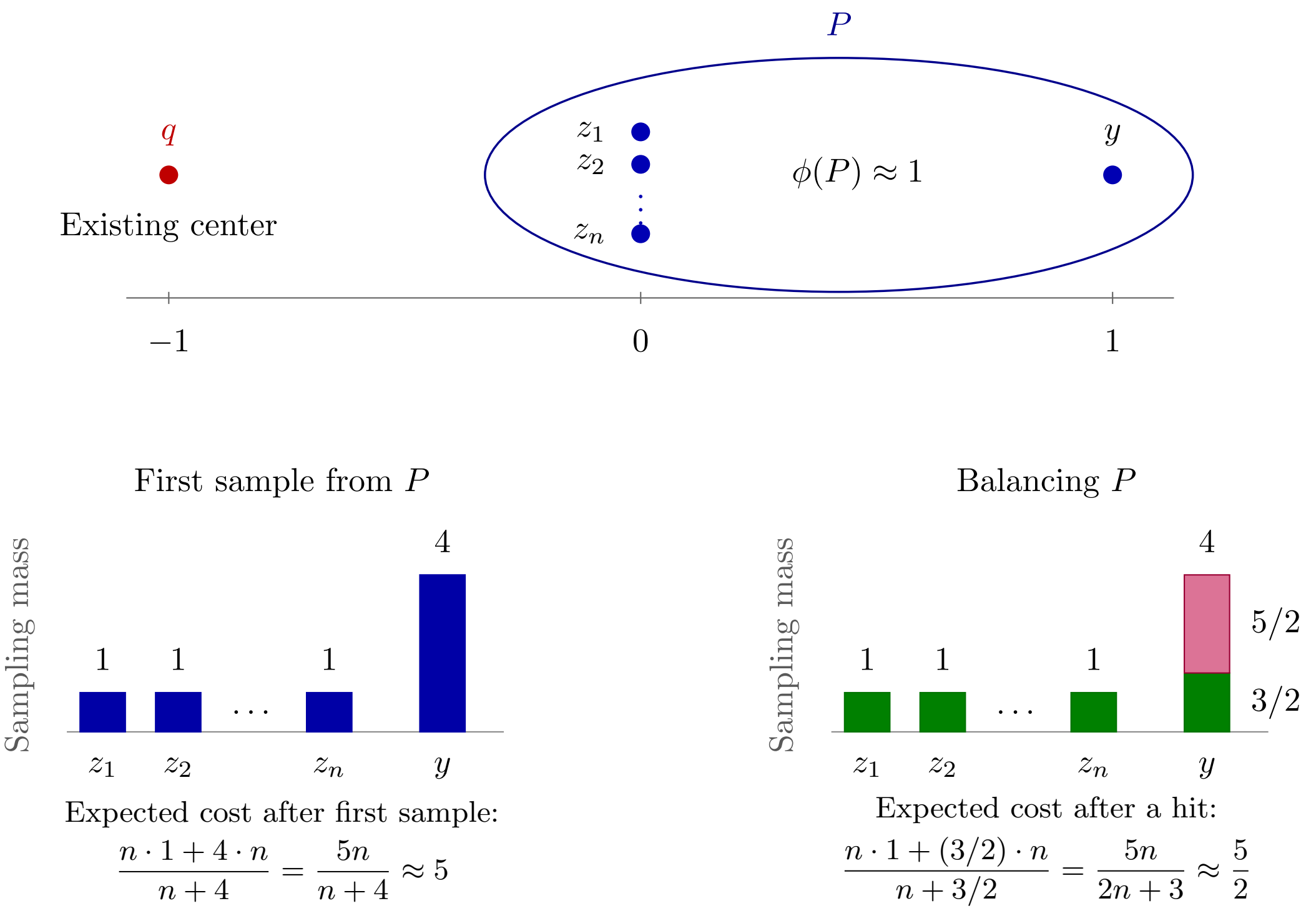

For each unsettled cluster $P$, our constructions declare
at most $\alpha_\ell\phi(P)$ mass imbalanced and guarantee
expected cost at most $\alpha_\ell\phi(P)$ after a hit.
We call $\alpha_\ell$ the \emph{balancing constant}.
Bounding the imbalanced mass in terms of $\phi(P)$ lets us
account for the additional misses.
The total mass of samples treated as misses, which we call
the \emph{miss mass}, consists of the costs of settled clusters
and the imbalanced mass of unsettled clusters.
Together, the two bounds keep the expected miss mass at most
$\alpha_\ell\OPT$ throughout the process.
Figure~\ref{fig:intro-balancing} illustrates
balancing on the example of~\cite{MRS2020}, which shows that
the first-sample bound of $5$ is tight for Euclidean $k$-means.
Sampling the isolated point $y$ gives a much larger cost than
sampling any other point in the cluster.
For a given amount of imbalanced mass, we therefore minimize
the expected cost after a hit by taking this mass from $y$.
Declaring more mass imbalanced further reduces this expected
cost, but increases the mass of misses.
In this example, $y$ is at distance $2$ from the
sampled center $q$, so its sampling mass is $2^2=4$.
Declaring $5/2$ units of this mass imbalanced reduces the expected
cost from approximately $5$ after the first sample from the
cluster $P$ to approximately $5/2$ after a hit, balancing the two
bounds. The caption gives a detailed calculation.

For a fixed budget of imbalanced mass, the best split into
balanced and imbalanced mass in a given state is simple.
Declare mass imbalanced at points whose selection would leave
the largest cluster cost, taking these points in decreasing order
of that cost until the budget is exhausted.
This minimizes the expected cost after a hit.
The difficulty is bounding this expectation over all possible
states (see Remark~\ref{rem:optimal-balancing-discussion} for further discussion). We do not have a direct tight analysis of this rule.
Instead, we analyze simpler balancing constructions
that nevertheless attain the optimal balancing constants for
metric $k$-median and Euclidean $k$-means
(Theorem~\ref{thm:metric-median-tight} and
Lemma~\ref{lem:euclidean-balancing-tight}).
We do not establish optimality of the balancing
constants for the remaining cases with $\ell>1$
(Lemmas~\ref{lem:balancing-extensions}
and~\ref{lem:balancing-small-exponents}).

For metric $k$-median, we adapt the balancing construction
from our joint work with Raz~\cite{KNR23} on Meyerson's
algorithm for online facility location, obtaining $\alpha_1=2$.
This reveals a connection between Meyerson's algorithm and
$k$-means++.
For Euclidean $k$-means, we form each point's balanced part
by capping its mass at a common threshold for its cluster,
chosen to balance these two bounds, giving $\alpha_2=1+\varphi$
(Section~\ref{subsec:balancing}).

An optimal balancing constant need not give a tight
approximation bound. Once a cluster is settled, our analysis
does not capture the further decrease in its cost from later
samples. Exploiting this decrease may improve the approximation
bound, particularly when many extra centers are sampled.
We discuss this further in the Discussion section
(Section~\ref{sec:discussion}).

Balancing is the only part of the upper-bound analysis that
depends on the metric space and distance exponent.
The second part bounds the balanced mass of clusters that
are still unsettled when sampling ends.

\paragraph{The equal-mass comparison.}

A cluster with large balanced mass is costly to leave
unsettled, but is also more likely to be hit.
We analyze this tradeoff by comparing the sampling process
with a simple process without geometry.
It starts with $r$ removable items of weight $1$ and one
permanent item of weight $\Gamma>0$.
Each step samples an item proportionally to its weight.
If a removable item is sampled, we delete it.
If the permanent item is sampled, the state is unchanged.
The removable items represent unsettled clusters, and the
permanent item represents the miss mass.
Let $R_{r,s}$ be the expected number of removable items left
after $s$ steps.
We use $R_{r,s}$ to bound the expected balanced mass of the
clusters that remain unsettled when sampling ends.

Equal weights are motivated by a simple
observation: for a fixed total balanced mass and number of
unsettled clusters, the expected mass of the cluster settled
by a hit is smallest when these masses are equal.
The difficulty is extending this observation from a single
hit to the whole run.
In the actual process, the masses are unequal and can decrease
without a hit.
Such a decrease reduces the remaining cost, but can also
make a later hit less likely.

After the first center is sampled uniformly at random, we
immediately settle the optimal cluster containing it.
We start the comparison at this point, with $r=k-1$ unsettled
clusters and $s=k+\Delta-1$ samples left.
We therefore run the equal-mass process with the same number
$r$ of removable items for $s$ steps.
For $\Delta>0$, we choose the weight of the permanent
item to be $\Gamma=\Delta$.
This choice will become clear in the analysis, and we explain
it in the proof of Lemma~\ref{lem:sampling-comparison}.

At each state, we define $A$ by summing the current
costs of settled clusters and, for each unsettled cluster $P$,
the bound $\alpha_\ell\phi(P)$ on its imbalanced mass.
Thus, $A$ upper-bounds the current miss mass.
The balancing bounds give $\E[A]\le\alpha_\ell\OPT$ throughout
the process. To make the permanent item represent the miss mass,
we scale all weights in the equal-mass process by $A/\Delta$,
so that the permanent item has weight $A$ and each removable
item has weight $A/\Delta$.

Conditioned on the initial state, the expected total
weight of the removable items left after $s$ steps is
$(A/\Delta)R_{r,s}$. We prove that this quantity bounds the
expected final balanced mass of $k$-means++.
Averaging over the first sample gives the bound
$(\alpha_\ell\OPT/\Delta)R_{k-1,k+\Delta-1}$.
Adding the bound $\alpha_\ell\OPT$ on the expected miss mass
gives the approximation ratio $\alpha_\ell B_{k,\Delta}$,
where $B_{k,\Delta}=1+R_{k-1,k+\Delta-1}/\Delta$.
For $\Delta=0$, we use the same framework with a sufficiently
small positive weight $\Gamma$, obtaining $B_{k,0}=1+H_{k-1}$.

We prove the bound on the expected final balanced mass
by charging payments to misses.
If $j$ clusters remain unsettled at the end, there have been
$\Delta+j$ misses in total.
The standard analysis of $k$-means++ ($\Delta=0$) by
Dasgupta~\cite{DasguptaNotes} can be viewed as charging each
miss at the current average cost of
the unsettled clusters.
These payments cover the final cost
of the unsettled clusters in expectation.
For $\Delta>0$, there are more misses than clusters left to
pay for, so we need not charge every miss.
We use the equal-mass process to decide which misses
to charge and at what rate.

At each state with $r$ unsettled clusters and $s$
samples left, we use the equal-mass process as a simplified
continuation of $k$-means++. We run it with $r$ removable items for $s$ steps;
the number of items remaining determines how many payments we assign.
We call its misses the \emph{future misses}.
We combine these with the \emph{past misses} of $k$-means++.
If $J$ removable items remain, these past and future misses
total $\Delta+J$.
We assign $J$ payments, one for each remaining item, and
charge them to $J$ of these misses, chosen uniformly at random.
Each chosen past miss is charged at the current average
balanced mass of the unsettled clusters, as in the standard
analysis.
Each chosen future miss is charged at $A/\Delta$, the weight
of a removable item in the scaled equal-mass process.

The future rate gives the equal-mass bound
at the start, and the past rate gives the actual remaining
balanced mass at the end:
At the initial state, all payments are charged to future
misses, so the expected total payment is
$(A/\Delta)R_{k-1,k+\Delta-1}$.
At the end, all payments are charged to past misses at the
current average balanced mass.
There are as many payments as unsettled clusters, so their
total equals the remaining balanced mass.

We prove by induction on the number of remaining samples
that, from every intermediate state, the expected final balanced
mass is at most the expected total payment defined at that
state (Lemma~\ref{lem:sampling-comparison}).
The equality at the end gives the base case.
For the induction step, we apply the bound to the states
reached after a hit and after a miss.
Our choice of charging probabilities and charging rates
ensures that these bounds, averaged over the possible next
states, are at most the current expected total payment.
Applying the result at the initial state gives the claimed bound.

The equal-mass process also guides the matching lower bound
for metric $k$-median (Section~\ref{sec:lower-bounds}).
The construction has one large cluster and $k-1$ singleton
clusters.
The first center lies in the large cluster with high probability.
Further samples from that cluster remove very little cost,
whereas sampling a singleton removes its entire cost.
Suitable distances make the large cluster behave like the
permanent item and the singletons like the removable items.

\subsection{Additional Related Work}

The literature on $k$-clustering is vast; we cover only
the work most closely related
to $k$-means++ and our analysis.
For broader background, see the surveys~\cite{BLSS2016,JS2018}
and recent advances in constant-factor approximation for
$k$-median, $k$-means, and related
objectives~\cite{BGHLWW2026,BCFGJPSSV2026a,BCFGJPSSV2026b},
together with the references therein.

Ostrovsky, Rabani, Schulman, and Swamy~\cite{ORSS2006} independently proposed
a variant of $k$-means++ with a different rule for choosing the
first two centers.
Brunsch and R{\"o}glin~\cite{BR2013} studied whether independent
repetitions of $k$-means++ can improve the logarithmic dependence on $k$.
They constructed Euclidean instances on which, with high probability,
even the best of polynomially many independent runs (in $k$) has
cost $\Omega(\log k)$ times the optimum.
Bhattacharya, Jaiswal, and Ailon~\cite{BJA2016} showed that this
remains true in two dimensions.

Bi-criteria guarantees for $k$-means++ date
back to Aggarwal, Deshpande, and Kannan~\cite{ADK2009}, who showed
that sampling a sufficiently large constant multiple of $k$
centers gives a constant-factor approximation for Euclidean
$k$-means with constant probability.
Ailon, Jaiswal, and Monteleoni~\cite{AJM2009} used $D^2$-sampling
in batches to obtain a constant-factor approximation with
constant probability using $O(k\log k)$ centers.
They used this construction in a streaming algorithm.
Wei~\cite{Wei2016} subsequently proved the bound
$8(1+\varphi k/\Delta)$ for every $\Delta>0$, giving a constant
approximation ratio for $\Delta=\Omega(k)$.
The bounds of~\cite{MRS2020}, stated in the introduction,
improve Wei's guarantee.
They also prove that the approximation ratio for Euclidean
$k$-means is at least $\frac{1-e^{-\sqrt{k}/2}}{8}\ln(k/\Delta)$
for $1\le\Delta\le k/2$, matching the dependence on $k/\Delta$
in their upper bound up to a constant factor.

Rozho\v{n}~\cite{Rozhon2026} studies how
$k$-means++ performs on a fixed Euclidean input for consecutive values of $k$.
To do this, they consider the setting in which the input
and an integer $K\ge1$ are fixed, $k$ is sampled uniformly from
$\{K,\ldots,2K-1\}$, and $k$-means++ is run for exactly $k$ steps.
They show that, with probability at least $1/4$, $k$-means++ gives
a constant-factor approximation to the optimum for the sampled $k$.

Arthur and Vassilvitskii~\cite{AV2007} noted
that their sampling rule is similar in spirit to earlier algorithms,
including Meyerson's algorithm for online facility
location~\cite{Meyerson2001}.
In Meyerson's algorithm, an arriving point $x$ opens a facility
with probability $\min\{D(x)/f,1\}$, where $f$ is the opening cost
and $D(x)$ is the distance to the nearest open facility.
In random order, conditioned on the history, the next point is
uniform among the remaining points.
We observe that, assuming $D(x)\le f$ for all remaining points
of a cluster $P$, conditioning on a facility opening in $P$ at
the next step makes its location a $D$-sample from those points.
Our joint work with Raz~\cite{KNR23} used balancing to give a
tight analysis of Meyerson's algorithm in random order.
We use this connection to adapt that construction to metric
$k$-median (Section~\ref{subsec:balancing}).

Bamas, Nagarajan, and Svensson~\cite{BNS2024} study
$D^\alpha$-sampling with $\alpha>2$ for the Euclidean $k$-means
objective.
Their work is motivated by experiments showing that larger
sampling exponents can improve on $D^2$-sampling.
The larger exponent gives more weight to distant points, which
can help reach new clusters but also increases sensitivity to outliers.
They show that, for every $\alpha>2$, the approximation ratio can
be arbitrarily large even for $k=2$.
Nevertheless, they give an upper bound that
depends on parameters describing the optimal clusters.
One parameter measures the concentration of distances from each cluster's centroid.
Another is the ratio between the largest and smallest average cost per point.
A third measures the variation in cluster sizes.
For example, for $\alpha>2$, they obtain an
$O(\log^{2/\alpha}k)$ approximation ratio when the first two parameters are bounded by constants.

Bhattacharya, Eube, R{\"o}glin, and Schmidt~\cite{BERS2020} study
noisy $k$-means++, which introduces small noise into the sampling
process.
More concretely, if a point has sampling probability $p$ under
the usual rule given the current centers, an adversary can change
this probability to any value in $[(1-\varepsilon)p,(1+\varepsilon)p]$,
for a fixed
$\varepsilon\in(0,1/2)$.
For Euclidean $k$-means, they prove an $O(\log^2 k)$ approximation
ratio.
Grunau, {\"O}z{\"u}do{\u{g}}ru, and Rozho\v{n}~\cite{GOR2023}
improve this to $O(\log k)$, showing that the logarithmic
dependence on $k$ is preserved despite these errors.

Lattanzi and Sohler~\cite{LS2019} obtain a constant
approximation ratio for Euclidean $k$-means by adding
$O(k\log\log k)$ local-search steps to $k$-means++.
Each step samples a point according to the $k$-means++ rule and
performs the best improving swap with an existing center, if one exists. For the same algorithm, 
Choo, Grunau, Portmann, and Rozho\v{n}~\cite{CGPR2020} show that
$\varepsilon k$ local-search steps suffice for an
$O(\varepsilon^{-3})$-approximation with high probability,
for fixed $\varepsilon\in(0,1]$ and sufficiently large $k$.
Later, Beretta, Cohen-Addad, Lattanzi, and Parotsidis~\cite{BCLP2023}
obtained a smaller approximation constant of $26.64$ with constant
probability for the same algorithm after $O(k\log\log k)$ steps.
They also obtain better guarantees by extending the algorithm to sample several points per step and swap several centers at once,
and by constructing candidate centers from weighted averages of
sampled points.

\section{Preliminaries}
\label{subsec:preliminaries}

In the $k$-clustering problem, we are given a metric space
$(\mathcal M,d)$, a finite (multi)set $X\subseteq\mathcal M$ of input points,
and an integer $k\ge1$.  A solution consists of $k$ points in
$\mathcal M$, called \emph{centers}.  Each input point is assigned to
its nearest center, with ties broken arbitrarily.  The set of input
points assigned to the same center is called a \emph{cluster}.

For a point $x\in\mathcal M$ and a set of centers
$C\subseteq\mathcal M$, let $d(x,C)=\min_{c\in C}d(x,c)$ be the
distance from $x$ to its nearest center in $C$.  For a fixed
$\ell>0$ and a set $Y\subseteq X$, the cost of $Y$ with respect to
$C$ is $\cost(Y,C)=\sum_{x\in Y}d(x,C)^\ell$.  We write
$\cost(C)=\cost(X,C)$ for the total cost.  We refer to $\ell$ as the \emph{distance exponent}.
The goal is to find $k$ centers minimizing $\cost(C)$.
The cases $\ell=1$ and $\ell=2$ are the $k$-median and $k$-means
objectives, respectively.

We analyze the $D^\ell$-sampling algorithm.  The algorithm samples its
first center uniformly from $X$.  Given a current set of centers $C$,
let $D(x)=d(x,C)$ be the distance from $x$ to its nearest center in
$C$.  Each subsequent center is sampled from $X$ with probability
proportional to its current cost $D(x)^\ell$.  More precisely,
conditioned on the current set $C$, a point $x\in X$ is sampled with
probability $\Pr[c=x\mid C]=D(x)^\ell/\cost(C)$.\footnote{If
$\cost(C)=0$, the algorithm terminates as the cost is already
zero.}  The sampled point
is then added to $C$.

For an integer $\Delta\ge0$, we consider the bi-criteria setting in
which the algorithm samples $k+\Delta$ centers, while its cost is
compared with the optimal cost using $k$ centers.  We refer to
$\Delta$ as the number of \emph{additional centers}.

Throughout the analysis, we fix an arbitrary optimal solution.  We
denote its clusters by $P_1,\ldots,P_k$ and call them the
\emph{optimal clusters}.  For each optimal cluster $P_i$, let $c_i^*$
be its center in the optimal solution, which we call the
\emph{optimal center} of $P_i$.  For a point $x\in P_i$, let
$d_x^*=d(x,c_i^*)$ be its distance from the optimal center.  The cost
of $P_i$ in the optimal solution is
$\phi(P_i)=\sum_{x\in P_i}(d_x^*)^\ell$, and the optimal cost is
$\OPT=\sum_{i=1}^k\phi(P_i)$.

Let $C_{\rm final}$ be the set of centers returned by the
algorithm. For fixed $k$, $\Delta$, and $\ell$, the
\emph{approximation ratio} of the algorithm is the smallest
value $\alpha$ for which
$\E[\cost(C_{\rm final})]\le\alpha\OPT$ for every input
instance, where the expectation is over the algorithm's
random choices.

\section{Balancing the Sampling Mass}
\label{subsec:balancing}

The first center is sampled uniformly and will be handled separately.  For the subsequent sampling process, we declare
the optimal cluster containing the first center \emph{settled}, and
all other optimal clusters \emph{unsettled}.  Once a cluster is
settled, it remains settled for the rest of the process.

Given the current set of centers $C$, the quantity $D(x)^\ell$ is
both the current cost of $x$ and its unnormalized weight in the next
sampling step.  We therefore also refer to $D(x)^\ell$ as the
\emph{sampling mass} (or simply \emph{mass}) of $x$.  Similarly, the sampling mass of a set
$Y\subseteq X$ is its current cost
$\cost(Y,C)=\sum_{x\in Y}D(x)^\ell$. 

Following the balancing idea of Kaplan, Naori, and
Raz~\cite{KNR23}, we split, only for the analysis, the sampling mass
of each unsettled optimal cluster into a \emph{balanced} part and an
\emph{imbalanced} part.  A sample from the balanced part of an
unsettled cluster $P$ is called a \emph{balanced hit}, and a balanced
hit settles $P$.  Every other sample is called a \emph{miss}.  Thus,
a sample from a settled cluster, or from the imbalanced part of an
unsettled cluster, is a miss.  For brevity, throughout the analysis
we refer to a balanced hit simply as a \emph{hit}.  Our hit event is
therefore stricter than the usual hit or coverage event in previous
analyses, in which an optimal cluster is hit or covered once any
point from that cluster is sampled
~\cite{AV2007,DasguptaNotes,Wei2016,MRS2020}.

The purpose of the balancing is to improve the bound we get when a hit occurs.
Conditioned on a hit of an unsettled cluster $P$, we can assign all
points of $P$ to the sampled point, which gives a bound on the cost of
$P$.  Thus, we want the sampled point to be a good center for $P$ in
expectation. By declaring some sampling mass imbalanced, we can exclude from the hit event some outcomes in which the sampled point is a bad center for $P$.
The more mass we exclude, however, the more mass must instead be treated
as misses.  This creates a tradeoff between the expected cost
after a hit and the imbalanced mass. We use a common upper bound
on these two quantities. We explain this choice in
Remark~\ref{rem:common-balancing-constant} at the end of
Section~\ref{subsec:hit-miss-bounds}.

Finally, we also require the balanced mass to be monotone.

Accordingly, for each value of the distance exponent $\ell$, we give a balancing
construction together with a \emph{balancing constant} $\alpha_\ell$ satisfying the
following three properties.

\begin{enumerate}
    \renewcommand{\theenumi}{P\arabic{enumi}}
    \renewcommand{\labelenumi}{(\theenumi)}
    \item \label{prop:balancing-hit-cost}\emph{Hit cost.}  Conditioned on a hit of an unsettled optimal cluster $P$,
    its expected cost after the hit is at most
    $\alpha_\ell\phi(P)$.  
    \item \label{prop:balancing-imbalanced-mass} \emph{Imbalanced mass.} The imbalanced mass of every unsettled optimal cluster $P$
    is at most $\alpha_\ell\phi(P)$. 
    \item \label{prop:balancing-monotonicity} \emph{Monotonicity.} The balanced mass of an unsettled optimal cluster can only
    decrease as more centers are added.
\end{enumerate}

The first center settles the optimal cluster it is sampled from without a balanced hit. Conditioned on the event that this cluster is $P$,
the first center is uniform in $P$. The following lemma shows
that the same bound $\alpha_\ell\phi(P)$ holds for the expected
cost of $P$ whenever a balancing construction with constant
$\alpha_\ell$ exists.

\begin{lemma}
\label{lem:balancing-uniform}
Suppose a balancing construction satisfies
Properties~\eqref{prop:balancing-hit-cost}
and~\eqref{prop:balancing-imbalanced-mass} with constant
$\alpha_\ell$.
Then, for every optimal cluster $P$ and a uniformly
random point $Z\in P$, we have
$\E[\cost(P,\{Z\})]\le\alpha_\ell\phi(P)$.
\end{lemma}

The idea of the proof is to consider an input with $P$
and another point placed very far from it. Conditioned on this point being the first
center, the sampling masses in $P$ are almost equal.
As the distance between the point and the cluster grows,
these masses increase, while the imbalanced mass remains at most
$\alpha_\ell\phi(P)$. Thus, the distribution of the second sampled
point, conditioned on a hit of $P$, can be made arbitrarily close
to uniform. The proof appears at the start of
Appendix~\ref{app:balancing-extensions}.

An intermediate state $\mathcal S$ contains all information revealed
by the process so far.  In particular, conditioning on $\mathcal S$ fixes the
current center set $C$, which optimal clusters are settled and
unsettled, and all their current sampling masses.  For any quantity
$Y$ defined during the process, we write $Y_{\rm new}$ for its value
after the next sample and $Y_{\rm final}$ for its value at the end of
the process.

\paragraph{The case $\ell=1$.}
We first describe the balancing for $D$-sampling.  Fix an unsettled
optimal cluster $P$.  We define its balanced and imbalanced parts by
splitting the sampling mass of each point $x\in P$.  Let
\[
    b(x)
    =
    \min\{D(w):w\in P,\ d_w^*\le d_x^*\}.
\]
Note that $b(x)\le D(x)$.  We take $b(x)$ to be the balanced mass of
$x$, and $D(x)-b(x)$ to be its imbalanced mass.  Thus, the balanced
and imbalanced masses of $P$ are
$\sum_{x\in P}b(x)$ and
$\sum_{x\in P}(D(x)-b(x))$, respectively.

If the algorithm samples a point $x\in P$, then $D(x)>0$, and we
declare the sample a hit with probability $b(x)/D(x)$.  Thus, the
sampling mass corresponding to a hit at $x$ is exactly $b(x)$.

For this construction, the hit cost and imbalanced mass bounds in Properties~\eqref{prop:balancing-hit-cost}
and~\eqref{prop:balancing-imbalanced-mass}
hold with $\alpha_1=2$, as proved in~\cite{KNR23}.

\begin{lemma}
\label{lem:D-balancing}
Let $P$ be an unsettled optimal cluster.  Its imbalanced mass is at
most $2\phi(P)$.  Moreover, conditioned on the next sample being a hit
of $P$,
\[
    \E[\cost(P,C_{\rm new})
        \mid \mathcal S,\mathrm{hit\ of\ }P]
    \le
    2\phi(P).
\]
\end{lemma}

The proof is simple. By construction, the balanced mass $b(x)$ is
nonincreasing in the distance $d_x^*$ to the optimal center.
Thus, if $X$ is sampled according to the balanced masses and
$Z$ is uniform in $P$, then
$\E[d_X^*]\le\E[d_Z^*]=\phi(P)/|P|$.
By the triangle inequality, assigning all points of $P$ to $X$
has expected cost at most
$\phi(P)+|P|\E[d_X^*]\le2\phi(P)$, which gives the hit-cost bound.

For the imbalanced-mass bound, fix $x\in P$ and
let $w\in P$ be a point that attains the minimum in the definition
of $b(x)$, so that
$b(x)=D(w)$ and $d_w^*\le d_x^*$.
By the triangle inequality,
$D(x)-b(x)=D(x)-D(w)\le d(x,w)\le d_x^*+d_w^*\le2d_x^*$.
Summing over $x\in P$ gives total imbalanced mass at most
$2\phi(P)$.

Monotonicity also holds, since every distance $D(w)$, and hence
every value $b(x)$, can only decrease as centers are added. Thus Properties~\eqref{prop:balancing-hit-cost}--\eqref{prop:balancing-monotonicity}
hold with $\alpha_1=2$.

\paragraph{The case $\ell=2$.}

We now consider the standard Euclidean $k$-means objective,
where the cost is the sum of squared Euclidean distances: $\sum_{x\in X}D(x)^2$.
Makarychev, Reddy, and Shan~\cite{MRS2020} showed that a point
sampled from an optimal cluster $P$ proportionally to its
$D^2$-mass leaves expected cost at most $5\phi(P)$ and that
the constant $5$ is tight. That is, the constant $5$ cannot be improved while treating every sample from $P$ as a hit. Balancing allows us to improve this constant.
Our construction has balancing constant $\alpha_2=1+\varphi$,
where $\varphi=(1+\sqrt5)/2$ is the golden ratio.

Fix an unsettled optimal cluster $P$, write $n=|P|$, and let $c^*$
be its centroid. For every point $z$, the standard centroid
identity~\cite[Lemma~2.1]{AV2007} gives
\begin{align}
    \cost(P,\{z\})
    &= \phi(P)+n\|z-c^*\|^2.
    \label{eq:euclidean-centroid}
\end{align}
Thus, to bound the expected cost after a hit, it suffices to bound
the expected squared distance of the sampled point from $c^*$.
We do this by capping the largest sampling weights, so that no point
receives too large a share of the balanced mass.

Fix a parameter $\alpha>1$. We allow up to $\alpha\phi(P)$ mass to be
imbalanced and retain the rest as balanced mass. More precisely, we denote the balanced mass by
\begin{align}
    B &= \max\{\cost(P,C)-\alpha\phi(P),0\}.
    \label{eq:euclidean-balanced-mass}
\end{align}

If $B=0$, we set $b(x)=0$ for every $x\in P$, so all the mass
is imbalanced. Otherwise, we choose a \emph{cap} $\tau^2>0$ and set
$b(x)=\min\{D(x)^2,\tau^2\}$, choosing $\tau^2$ so that
$\sum_{x\in P}b(x)=B$.
Such a cap exists because the sum of the capped weights varies
continuously from $0$ to $\cost(P,C)$ as the cap increases from $0$
to $\max_{x\in P}D(x)^2$. We take $b(x)$ to be the balanced mass of
$x$, and $D(x)^2-b(x)$ to be its imbalanced mass. If the algorithm
samples $x\in P$, we declare a hit with probability $b(x)/D(x)^2$.

Increasing $\alpha$ allows us to remove more mass from the hit event.
The following lemma quantifies the resulting improvement in the
expected cost after a hit.

\begin{lemma}
\label{lem:euclidean-balancing}
For every $\alpha>1$, the imbalanced mass of $P$ is at most
$\alpha\phi(P)$, and its total balanced mass is nonincreasing as
centers are added. Conditioned on a hit of $P$,
\begin{align*}
    \E[\cost(P,C_{\rm new})\mid\mathcal S,\mathrm{hit\ of\ }P]
    &\le \left(2+\frac{1}{\alpha-1}\right)\phi(P).
\end{align*}
\end{lemma}

\begin{proof}
The imbalanced mass bound holds by construction. The monotonicity property~\eqref{prop:balancing-monotonicity} follows from
\eqref{eq:euclidean-balanced-mass}, since $\phi(P)$ is fixed and
$\cost(P,C)$ can only decrease as centers are added.

It remains to prove the hit cost bound. If $\phi(P)=0$, all points of $P$ coincide and the cost after a hit
is zero. We therefore assume $\phi(P)>0$. We condition on a hit of
$P$ and let $Z$ be the sampled point. Then $B>0$, and the cap gives, for every $x \in P$, that
$\Pr[Z=x\mid\mathcal S,\mathrm{hit\ of\ }P]=b(x)/B\le\tau^2/B$.
We will show that $\tau^2/B\le\alpha/((\alpha-1)n)$. We first show how this implies the lemma.
Assigning all points of $P$ to the new center $Z$
gives $\cost(P,C_{\rm new})\le\cost(P,\{Z\})$.
The centroid identity~\eqref{eq:euclidean-centroid} and
the probability bound $\Pr[Z=x\mid\mathcal S,\mathrm{hit\ of\ }P] \leq \tau^2/B\le\alpha/((\alpha-1)n)$ then give
\begin{align*}
    \E[\cost(P,C_{\rm new})\mid\mathcal S,\mathrm{hit\ of\ }P] 
    &\le \phi(P)+\sum_{x\in P}
         \Pr[Z=x\mid\mathcal S,\mathrm{hit\ of\ }P]\cdot n (d_x^*)^2  \\
    &\le \phi(P)+\frac{\alpha}{\alpha-1}
         \sum_{x\in P}(d_x^*)^2 
    = \left(2+\frac{1}{\alpha-1}\right)\phi(P),
\end{align*}
which is the claimed hit cost bound.

It remains to show that $\tau^2/B\le\alpha/((\alpha-1)n)$.
We first bound how much the distances $D(x)$ vary within $P$. Let
$\overline D=\sum_{x\in P}D(x)/n$ be the average distance to the
closest current center, and let $D(c^*)$ be this distance for the
centroid. The triangle inequality gives both
$D(x)\le d_x^*+D(c^*)$ and $D(c^*)\le d_x^*+D(x)$.
Hence $|D(x)-D(c^*)|\le d_x^*$. Since the average $\overline D$
minimizes the sum of squared deviations, we obtain
\begin{align}
    \sum_{x\in P}(D(x)-\overline D)^2
    &\le \sum_{x\in P}(D(x)-D(c^*))^2
     \le \phi(P).
    \label{eq:euclidean-distance-variance}
\end{align}

We next relate $\overline D$ to the total balanced mass $B$ and the
threshold $\tau$. To apply Cauchy-Schwarz, we write
$D(x)=\min\{D(x),\tau\}\max\{1,D(x)/\tau\}$.
Cauchy-Schwarz gives
\begin{align}
    n^2\overline D^2
    &\le
    \left(\sum_{x\in P}\min\{D(x)^2,\tau^2\}\right)
    \left(\sum_{x\in P}\max\{1,D(x)^2/\tau^2\}\right).
    \label{eq:euclidean-capped-cauchy}
\end{align}
The first factor is $B$ by definition. For the second factor, a
point with $D(x)\le\tau$ contributes $1$ and has no imbalanced
mass. A point with $D(x)>\tau$ contributes
$D(x)^2/\tau^2=1+(D(x)^2-\tau^2)/\tau^2$, where
$D(x)^2-\tau^2$ is its imbalanced mass.
The total imbalanced mass is exactly $\alpha\phi(P)$ (since $B>0$).
Summing over all points,
the second factor is therefore $n+\alpha\phi(P)/\tau^2$.
Substituting these expressions in~\eqref{eq:euclidean-capped-cauchy}
and dividing by $n$ gives
\begin{align}
    n\overline D^2
    &\le B+\frac{\alpha B\phi(P)}{n\tau^2}.
    \label{eq:euclidean-mean-distance}
\end{align}

We now combine~\eqref{eq:euclidean-distance-variance} and
\eqref{eq:euclidean-mean-distance}. Expanding the squared deviations
and using $\sum_{x\in P}D(x)=n\overline D$ gives
\begin{equation*}
    \sum_{x\in P}(D(x)-\overline D)^2
    = \sum_{x\in P}D(x)^2
       -2\overline D\sum_{x\in P}D(x)+n\overline D^2
    = \sum_{x\in P}D(x)^2-2n\overline D^2+n\overline D^2
    = B+\alpha\phi(P)-n\overline D^2,
\end{equation*}
where the last equality uses
$\sum_{x\in P}D(x)^2=\cost(P,C)=B+\alpha\phi(P)$.
It follows that
\begin{equation*}
    \phi(P)
    \ge \sum_{x\in P}(D(x)-\overline D)^2
    = B+\alpha\phi(P)-n\overline D^2
    \ge B+\alpha\phi(P)
         -\left(B+\frac{\alpha B\phi(P)}{n\tau^2}\right)
    = \alpha\phi(P)\left(1-\frac{B}{n\tau^2}\right).
\end{equation*}
Here the first inequality is~\eqref{eq:euclidean-distance-variance},
and the second follows by substituting the upper bound on
$n\overline D^2$ from~\eqref{eq:euclidean-mean-distance}.
Dividing by $\phi(P)>0$ and rearranging gives
$\tau^2/ B \leq \alpha/ ((\alpha-1)n)$. 
\end{proof}

We now choose $\alpha$ so that the same constant bounds both the
imbalanced mass and the expected cost after a hit. The former bound
increases with $\alpha$, while the latter decreases. Their maximum
is therefore minimized when $\alpha=2+1/(\alpha-1)$, giving
$\alpha=1+\varphi$. We set $\alpha_2=1+\varphi$.
Thus Properties~\eqref{prop:balancing-hit-cost}--\eqref{prop:balancing-monotonicity}
hold with $\alpha_2=1+\varphi$.

\paragraph{Tightness of the balancing constant for Euclidean $k$-means.}
For $\alpha\ge2$, the factor $2+1/(\alpha-1)$ in the hit cost
bound of Lemma~\ref{lem:euclidean-balancing} cannot be improved by a different
choice of balanced mass, as long as the imbalanced mass is at most
$\alpha\phi(P)$. The proof appears in
Appendix~\ref{app:euclidean-balancing-tight}.

\begin{lemma}
\label{lem:euclidean-balancing-tight}
For every $\alpha\ge2$, there is a sequence of Euclidean $2$-means
instances on the line such that, with probability $1-o(1)$, the
state $\mathcal S$ after the first sample has an unsettled optimal
cluster $P$ for which every choice of balanced weights with
imbalanced mass at most $\alpha\phi(P)$ satisfies
\begin{align*}
    \E[\cost(P,C_{\rm new})\mid\mathcal S,\mathrm{hit\ of\ }P]
    &\ge \left(2+\frac{1}{\alpha-1}-o(1)\right)\phi(P).
\end{align*}
Both $o(1)$ terms tend to zero as the number of points in the instance grows.
\end{lemma}

By the centroid identity~\eqref{eq:euclidean-centroid},
a uniformly random point $Z\in P$ satisfies
$\E[\cost(P,\{Z\})]=2\phi(P)$. Thus,
Lemma~\ref{lem:balancing-uniform} implies that any
balancing constant $\gamma$ must satisfy $\gamma\ge2$. Applying
Lemma~\ref{lem:euclidean-balancing-tight} with $\alpha=\gamma$ then
requires $\gamma\ge2+1/(\gamma-1)$, and hence
$\gamma\ge1+\varphi$. Thus $1+\varphi$ is the optimal
balancing constant.

Even without requiring these balancing properties,
a guarantee of the form $\alpha B_{k,0}$ for Euclidean
$k$-means that holds for every $k$ must have $\alpha\ge5/2$.
This follows from the exact approximation ratio $5$ for $k=2$
and $\Delta=0$, since $B_{2,0}=2$.
See Lemma~\ref{lem:euclidean-two-center-exact} in
Appendix~\ref{app:euclidean-balancing-tight}.

\paragraph{Squared metrics and other distance exponents.}
The same capping construction as in
Lemma~\ref{lem:euclidean-balancing} also applies to squared
distances in a general metric and to higher distance exponents.
The hit-cost analysis uses bounds on pairwise distances in
place of the centroid identity, giving the following constants.

\begin{lemma}
\label{lem:balancing-extensions}
For squared distances in a general metric,
Properties~\eqref{prop:balancing-hit-cost}--\eqref{prop:balancing-monotonicity}
hold with $\alpha_2 \approx 4.383$, where $\alpha_2$
is the unique solution greater than $4$ of
$\alpha_2=4+\alpha_2/(\alpha_2-1)^2$.
For all other exponents $\ell>1$, the properties hold
with $\alpha_\ell=3^{\ell-1}+2^\ell$ in general metrics.
In Euclidean space, we obtain the smaller balancing constant
$\alpha_\ell=3^{\ell-1}+2^{\ell-1}$ when $\ell>2$.
\end{lemma}

We prove Lemma~\ref{lem:balancing-extensions} in
Appendix~\ref{app:balancing-powers}.

Appendix~\ref{app:small-exponents} gives further balancing
bounds for $0<\ell<1$ and $1<\ell<2$ by reductions to metric
$k$-median and squared distances.

\begin{remark}
We do not establish optimality of our balancing constants
for general metrics with $\ell\ge2$. See
Remark~\ref{rem:optimal-balancing-discussion} in
Appendix~\ref{app:balancing-powers} for a discussion of the
optimal choice of balanced mass.
\end{remark}

                                                                    \section{Analysis of the Sampling Process}
\label{sec:sampling-analysis}

We now turn to the part of the analysis that is common to all
distance exponents $\ell$. 
From this point on, we use only
Properties~\eqref{prop:balancing-hit-cost}--\eqref{prop:balancing-monotonicity} of the balancing construction, 
with the corresponding balancing constant $\alpha_\ell$.

\medskip

\subsection{Bounds After a Hit or a Miss}
\label{subsec:hit-miss-bounds}

Fix an intermediate state $\mathcal S$.  Let $\mathcal U$ and
$\mathcal H$ be the sets of unsettled and settled optimal clusters,
respectively.  For each $P_i\in\mathcal U$, let $u_i$ be its balanced
mass, and let $U=\sum_{P_i\in\mathcal U}u_i$ be the total balanced
mass.  Thus, $U$ is exactly the sampling mass that produces a hit.
Conditioned on a hit, an unsettled cluster $P_i$ is hit
with probability $u_i/U$. We first describe the behavior of $U$ after one step of the algorithm.

\begin{lemma}
\label{lem:U-hit}
Fix a state $\mathcal S$ with $r$ unsettled optimal clusters and total
balanced mass $U$.  On a miss, $U_{\rm new}\le U$.
On a hit of $P_i\in\mathcal U$, $U_{\rm new}\le U-u_i$.
Conditioned on a hit,
\[
    \E[U_{\rm new}\mid\mathcal S,\mathrm{hit}]
    \le
    \frac{r-1}{r}U.
\]
\end{lemma}

\begin{proof}
Suppose first that the next sample is a miss.  By Property~\eqref{prop:balancing-monotonicity}, the balanced mass of every unsettled cluster can
only decrease, and therefore $U_{\rm new}\le U$.

Now suppose that the next sample is a hit of $P_i$.  The cluster
$P_i$ no longer contributes its balanced mass $u_i$ to $U$, while
the balanced masses of the other unsettled clusters can only
decrease by Property~\eqref{prop:balancing-monotonicity}.  Hence $U_{\rm new}\le U-u_i$.

Let $u_1,\ldots,u_r$ be the balanced masses of the unsettled clusters.
Conditioned on a hit, cluster $i$ is hit with probability $u_i/U$.
As shown above, after such a hit the remaining
balanced mass is at most $U-u_i$.  Therefore,
\begin{align}
    \E[U_{\rm new}\mid\mathcal S,\mathrm{hit}]
    &\le
    \sum_{i=1}^r \frac{u_i}{U}(U-u_i)
    =
    U-\frac{\sum_{i=1}^r u_i^2}{U}
    \le 
    U-\frac{U^2/r}{U}
    =
    \frac{r-1}{r}U,
\end{align}
where the last inequality follows from  $\sum_{i=1}^r u_i^2\ge U^2/r$ by Cauchy-Schwarz.
\end{proof}

Let $M$ be the sampling mass that produces a miss.  A settled cluster
contributes all of its current cost to $M$, whereas an unsettled
cluster contributes only its imbalanced mass.  Hence
\[
    M
    =
    \sum_{P_i\in\mathcal H}\cost(P_i,C)
    +
    \sum_{P_i\in\mathcal U}
    \bigl(\cost(P_i,C)-u_i\bigr).
\]
The total sampling mass is $M+U$.  Therefore, the next sample is a
miss with probability $M/(M+U)$ and a hit with probability
$U/(M+U)$.

For the analysis, it will be useful to replace the actual miss mass 
$M$ by an upper bound with better behavior.  The miss 
mass $M$ is not monotone as centers are added and may either increase 
or decrease. A miss may decrease $M$ by reducing the cost of
settled clusters. A hit can reduce these costs as well, but it also
settles a new cluster. Before the hit, this cluster contributed only
its imbalanced mass to $M$. After the hit, it contributes its entire
remaining cost, which can exceed its previous contribution even
though the cost itself decreases.  We therefore define a \emph{miss proxy} $A$ that upper-bounds 
$M$ and is non-increasing in expectation. We define
\[
    A
    =
    \sum_{P_i\in\mathcal H}\cost(P_i,C)
    +
    \alpha_\ell
    \sum_{P_i\in\mathcal U}\phi(P_i).
\]
By Property~\eqref{prop:balancing-imbalanced-mass}, the imbalanced mass of every unsettled
cluster $P_i$ is at most $\alpha_\ell\phi(P_i)$, and therefore
$M\le A$.

The next lemma describes the behavior of $A$ after one step of the algorithm.

\begin{lemma}
\label{lem:A-evolution}
Fix a state $\mathcal S$.  On a miss,
$A_{\rm new}\le A$.  On a hit of
$P_i\in\mathcal U$,
$\E[A_{\rm new}\mid\mathcal S,\mathrm{hit\ of\ }P_i]\le A$.
\end{lemma}

\begin{proof}
Suppose first that the next sample is a miss.  No new optimal cluster
is settled, so every unsettled cluster continues to contribute the
same term $\alpha_\ell\phi(P_i)$ to $A$.  The cost of every settled
cluster can only decrease when a center is added.  Hence
$A_{\rm new}\le A$.

Now suppose that the next sample is a hit of $P_i$.
Before the hit, $P_i$ contributes
$\alpha_\ell\phi(P_i)$ to $A$.  After the hit, it becomes settled, so
this term is replaced by its actual cost.  By Property~\eqref{prop:balancing-hit-cost},
conditioned on a hit of $P_i$, its expected cost after the sample is
at most $\alpha_\ell\phi(P_i)$.  The costs of previously settled
clusters can only decrease, while the contributions of all other
unsettled clusters to $A$ remain unchanged.  Therefore $\E[A_{\rm new}\mid\mathcal S,\mathrm{hit\ of\ }P_i]
    \le
    A$.
\end{proof}

Lemma~\ref{lem:A-evolution} immediately gives the two properties of the miss
proxy $A$ that we will use later.

\begin{corollary}
\label{cor:A-evolution}
For every state $\mathcal S$,
\begin{equation}
    \E[A_{\rm new}\mid\mathcal S,\mathrm{hit}]
    \le
    A,
    \label{eq:A-hit-update}
\end{equation}
and
\begin{equation}
    \E[A_{\rm new}\mid\mathcal S]
    \le
    A.
    \label{eq:A-supermartingale}
\end{equation}
Thus, $A$ is a supermartingale.
\end{corollary}

The first inequality follows by applying the law of total expectation to the bound
$\E[A_{\rm new}\mid\mathcal S,\mathrm{hit\ of\ }P_i]\le A$
over the cluster that is hit.  Similarly, the second inequality follows by
applying the law of total expectation, conditioning on whether
the next sample is a hit or a miss, using~\eqref{eq:A-hit-update} on
a hit and $A_{\rm new}\le A$ on a miss. Note that only Property~\eqref{prop:balancing-hit-cost} is used for Corollary~\ref{cor:A-evolution}.

We now bound the expected value of the miss proxy throughout the
process.  Conditioned on the first center lying in an optimal cluster
$P_i$, it is uniform in $P_i$, so
Lemma~\ref{lem:balancing-uniform} bounds the expected cost of
$P_i$ by $\alpha_\ell\phi(P_i)$, while every other optimal cluster is
unsettled and contributes $\alpha_\ell\phi(P_j)$ to $A$ by definition.  Hence,
immediately after the first center is sampled,
$\E[A]\le\alpha_\ell\OPT$.  Since $A$ is a supermartingale (Corollary~\ref{cor:A-evolution}), at every
later point in the process,
\begin{equation}
    \E[A]
    \le
    \alpha_\ell\OPT.
    \label{eq:A-bound}
\end{equation}

Let $C_{\rm final}$ be the final set of $k+\Delta$ sampled centers.
At the end of the process, the miss mass $M_{\rm final}$ and the
balanced mass $U_{\rm final}$ partition the total sampling mass.
Therefore, $\cost(C_{\rm final})
    =
    M_{\rm final}+U_{\rm final}$. 
By
Property~\eqref{prop:balancing-imbalanced-mass},
$M_{\rm final}\le A_{\rm final}$, and hence
\begin{equation}
    \cost(C_{\rm final})
    \le
    A_{\rm final}+U_{\rm final}.
    \label{eq:final-cost-AU}
\end{equation}

By~\eqref{eq:A-bound},
$\E[A_{\rm final}]\le\alpha_\ell\OPT$.  Thus, it remains to bound
$\E[U_{\rm final}]$.
In the next section (Section~\ref{subsec:equal-mass}), we bound this quantity in terms of $\E[A]$,
where $A$ is the miss proxy immediately after the first center is sampled.

\begin{remark}
\label{rem:common-balancing-constant}
We can now give some intuition for why we use the same
constant in Properties~\eqref{prop:balancing-hit-cost}
and~\eqref{prop:balancing-imbalanced-mass}.
These constants enter the final guarantee only through the
bound on $\E[A]$ in~\eqref{eq:A-bound}.
If we used different constants $\alpha$ in~\eqref{prop:balancing-hit-cost}
and $\beta$ in~\eqref{prop:balancing-imbalanced-mass}, our analysis
would still require $A$ to satisfy the two conditions established
above: First, $M\le A$. Second, $A$ does not increase in expectation
after a hit. To ensure the first condition, the coefficient of
$\sum_{P_i\in\mathcal U}\phi(P_i)$ in the definition of $A$
must be at least $\beta$. To ensure the second, it must be
at least $\alpha$. We would therefore replace this coefficient
by $\max\{\alpha,\beta\}$, and~\eqref{eq:A-bound} would hold
with this constant in place of $\alpha_\ell$.

Allowing a larger $\beta$ in the imbalanced-mass bound can only
help us obtain a smaller $\alpha$ in the hit-cost bound.
However, once $\beta>\alpha$, the guarantee is determined by
$\beta$, so reducing $\alpha$ further gives no improvement.
Thus, we seek the smallest balancing constant $\alpha_\ell$.
\end{remark}

\subsection{The Equal-Mass Comparison}
\label{subsec:equal-mass}

The bound in Lemma~\ref{lem:U-hit} comes from Cauchy-Schwarz,
and is tight when the unsettled clusters have equal balanced mass.
Thus, for fixed $r$ and $U$, a hit leaves the largest possible
balanced mass, according to this bound, when every unsettled cluster
has balanced mass $\bar u=U/r$.  This motivates comparing the actual
process with an idealized process in which the unsettled clusters always have
equal masses.

\paragraph{The equal-mass process.}
The process starts with $r$ \emph{removable} items, each of weight $1$, and
one \emph{permanent} item of weight $\Gamma>0$.  At each step, an item is
sampled proportionally to its weight.  Sampling a removable item is
a hit and removes that item.  Sampling the permanent item is a miss
and leaves the state unchanged.  Thus, when $j$ removable items
remain, the next step is a hit with probability $j/(j+\Gamma)$
and a miss with probability $\Gamma/(j+\Gamma)$.

Let $R_{r,s}$ be the expected number of
removable items remaining after $s$ steps, starting with $r$ removable
items.  We have $R_{r,0}=r$ and $R_{0,s}=0$. Conditioning on the
first step gives
\begin{equation}
    R_{r,s}
    =
    \frac{\Gamma}{r+\Gamma}R_{r,s-1}
    +
    \frac{r}{r+\Gamma}R_{r-1,s-1}.
    \label{eq:R-recurrence}
\end{equation}

Consider a state $\mathcal S$ of the actual process with $r$
unsettled clusters, $s$ samples remaining, and miss proxy $A$. We
associate with $\mathcal S$ an equal-mass process starting with $r$
removable items and running for $s$ steps. To relate this process to
the actual one, we scale its weights by $A/\Gamma$, so that the
permanent item has weight $A$, matching the miss proxy in state
$\mathcal S$. After scaling, each removable item has weight
$\bar a=A/\Gamma$. This scaling does not change the sampling
probabilities. Hence the expected remaining removable mass after
$s$ steps is $\bar a R_{r,s}$.
Note that the scaling is chosen for the current state $\mathcal S$ and
may differ between states.

For integers $k\ge1$ and $\Delta\ge0$, define
\[
    B_{k,\Delta}
    =
    \begin{cases}
        1+H_{k-1}, & \Delta=0,\\[3pt]
        1+\dfrac{R_{k-1,k+\Delta-1}}{\Delta}, & \Delta>0,
    \end{cases}
\]
where, for $\Delta>0$, the equal-mass process has permanent weight
$\Gamma=\Delta$.

\begin{theorem}
\label{thm:exact-sampling}
For all integers $k\ge1$ and $\Delta\ge0$, after sampling
$k+\Delta$ centers,
\[
    \E[\cost(C_{\rm final})]
    \le
    \alpha_\ell B_{k,\Delta}\OPT.
\]
\end{theorem}

For metric $k$-median, Theorem~\ref{thm:exact-sampling}
gives the bound $2B_{k,\Delta}$, which we show is tight for every $k$ and $\Delta$ by a matching lower bound in Theorem~\ref{thm:metric-median-tight}.
For Euclidean $k$-means, it gives the bound
$(1+\varphi)B_{k,\Delta}$.

The bounds for squared metrics and other distance exponents
follow by combining Theorem~\ref{thm:exact-sampling} with
Lemmas~\ref{lem:balancing-extensions}
and~\ref{lem:balancing-small-exponents}.

We now prove Theorem~\ref{thm:exact-sampling}, first for the case $\Delta>0$, then handle $\Delta=0$
using the same framework.

\subsubsection{Additional Centers $\Delta>0$}
\label{subsec:additional-centers}

We first consider $\Delta>0$. For this case, we choose
$\Gamma=\Delta$. The reason for this choice will become clear
in the analysis. Although equal, we keep separate notation for the
permanent weight $\Gamma$ and the number $\Delta$ of additional
centers. This will be useful for the case $\Delta=0$, discussed in
Section~\ref{subsec:no-extra-centers}, where we choose
$\Gamma\ne\Delta$.

Recall that, as shown above, it remains to bound
$\E[U_{\rm final}]$, the expected total balanced mass of the clusters
that remain unsettled at the end.  Before proving
Theorem~\ref{thm:exact-sampling}, we introduce a payment construction
that will be used to bound this quantity.

\paragraph{Charging misses.}

The idea of the analysis is to charge payments to misses.
Recall that we analyze the process after the first center is sampled.
We call the resulting state the initial state and condition on it
in the comparison argument.

Consider the scaled equal-mass process associated with the initial
state. It starts with $k-1$ removable items and runs for
$k+\Delta-1$ steps. We assign a payment of $\bar a=A/\Gamma$ to each
removable item that remains at its end. The expected total payment is therefore
$\bar a R_{k-1,k+\Delta-1}$, which, as discussed above, exactly covers the
expected remaining removable mass. 
 We will show that
this expected total payment also upper-bounds $\E[U_{\rm final}]$.

To do so, for each state $\mathcal S$ we define an expected
total payment $\Phi_{r,s}(A,U)$ that depends only on $r,s,A$, and $U$. Initially, its
value is $\bar a R_{k-1,k+\Delta-1}$, and at the end its value is
$U_{\rm final}$. We prove by
induction on the number $s$ of remaining samples that
$\E[U_{\rm final}\mid\mathcal S]\leq\Phi_{r,s}(A,U)$.
The induction proceeds backward through the sampling process.
We start with the base case $s=0$, when no samples remain, and
increase $s$ until we reach the initial state, where $s=k+\Delta-1$.
Applying this bound at the initial state gives
$\E[U_{\rm final}]\leq\Phi_{\rm initial}
=\bar a R_{k-1,k+\Delta-1}$.

We now need to describe $\Phi_{r,s}(A,U)$ at an intermediate state $\mathcal S$.
We consider the scaled equal-mass process associated with $\mathcal S$, and assign one payment to each removable item
that remains at its end. We charge these payments to a subset of the
misses: the past misses of the actual process that happened before reaching $\mathcal S$, and the future misses
of this equal-mass process. Each chosen miss receives one payment,
but we use different rates for past and future misses.

A payment charged to a past miss has value $\bar u=U/r$, i.e., the current
average balanced mass of the actual unsettled clusters. A payment
charged to a future miss has value $\bar a=A/\Gamma$, i.e., the mass of a
removable item in the scaled equal-mass process. 
$\Phi$ is the expected total value of these payments. This gives the
desired values at both ends: initially all payments are charged to future misses and have value
$\bar a$, so their expected total value is
$\bar a R_{k-1,k+\Delta-1}$.  At the end, all payments are
charged to past misses, and their total value is
$r\bar u=U_{\rm final}$.

We proceed to define $\Phi$ formally. First, we count the past and
future misses associated with $\mathcal S$, and then we
distribute the payments among these misses.

\paragraph{Counting misses.}
Fix an intermediate state $\mathcal S$ where $r$ clusters are currently unsettled and $s$
samples remain.  After the first center is chosen, there are $k-1$ unsettled clusters and
$k+\Delta-1$ samples remaining.
Since $r$ clusters are unsettled, $k-1-r$ clusters have been hit after the first center was sampled. The number of samples taken since then is $(k+\Delta-1)-s$. Each of these samples is either a hit or a miss, so the number of past misses is $(k+\Delta-1)-s-(k-1-r)=r+\Delta-s$.  We denote the number
of past misses by $m=r+\Delta-s$.

Consider the equal-mass process associated with $\mathcal S$.  Let $J$ be the number of removable
items left at the end. Thus, by definition, $\E[J] = R_{r,s}$.  Since $r-J$ removable items are hit, the
remaining $s$ steps contain $s-(r-J)=\Delta-m+J$ misses. These are the
future misses in the equal-mass process.  Together with the $m$ past
misses that have already occurred in the actual process, there are
$\Delta+J$ misses in total.

\paragraph{Charging the misses.}
We assign one payment to each of the $J$ removable items that remain,
and charge these $J$ payments to $J$ out of the $\Delta+J$ misses, chosen
uniformly at random.  Conditioned on $J$, each past miss is therefore
charged with probability $J/(\Delta+J)$. Taking expectation over
$J$, define $p_{r,s}
    =
    \E[{J}/(\Delta+J)]$.
Thus each past miss is charged with probability $p_{r,s}$, and the expected
number of charged past misses is $m p_{r,s}$.
 Since 
$R_{r,s}=\E[J]$, the expected total number of payments is $R_{r,s}$.
Therefore the expected number of payments charged to future misses is $R_{r,s}-m p_{r,s}$.
Pricing the payments charged to past misses
at $\bar u = U/r$ and those charged to future misses
at $\bar a = A/\Gamma$ gives the expected total payment of
\begin{equation}
    \Phi_{r,s}(A,U)
    =
    m p_{r,s}\bar u
    +
    \bigl(R_{r,s}-m p_{r,s}\bigr)\bar a \quad \text{for } r\geq 1.
    \label{eq:Phi-definition}
\end{equation}
When $r=0$, we have $U=0$, while $\bar u$ is undefined. In this case,
$J=R_{0,s}=0$, so there are no payments to charge. Hence, we define
$\Phi_{0,s}(A,0)=0$.

\paragraph{The initial and final payments.}
Now that we have formally defined $\Phi$, we verify that it has the desired values at both ends of the process. In the initial state, $r=k-1$, $s=k+\Delta-1$, and $m=0$. Therefore
\begin{equation}
    \Phi_{k-1,k+\Delta-1}(A,U)
    =
    \bar a R_{k-1,k+\Delta-1}.
    \label{eq:Phi-initial}
\end{equation}

At the end, $s=0$. If $r>0$, all $r$ removable items remain in the equal-mass process, and all payments are charged to past misses at rate $\bar u$. Hence
\begin{equation}
    \Phi_{r,0}(A,U)
    =
    r\bar u
    =
    U.
    \label{eq:Phi-final}
\end{equation}
For $r=0$, the same identity follows from $\Phi_{0,0}(A,0)=0$.

It remains to show that $\E[U_{\rm final}\mid\mathcal S]\leq\Phi_{r,s}(A,U)$ at every intermediate state. We prove this in the following lemma.

\begin{lemma}
\label{lem:sampling-comparison}
Consider any fixed state $\mathcal S$ of the actual process with $r$ unsettled clusters,
$s$ samples remaining, miss proxy $A$, and total balanced mass $U$.
Then
\[
    \E[U_{\rm final}\mid \mathcal S]
    \le
    \Phi_{r,s}(A,U).
\]
\end{lemma}

Before proving Lemma~\ref{lem:sampling-comparison}, we use it
to prove Theorem~\ref{thm:exact-sampling} for $\Delta>0$.

\begin{proof}[Proof of Theorem~\ref{thm:exact-sampling} for $\Delta>0$]

Taking expectation in~\eqref{eq:final-cost-AU}, we have
\[
    \E[\cost(C_{\rm final})]
    \le
    \E[A_{\rm final}]+\E[U_{\rm final}].
\]
By~\eqref{eq:A-bound},
$\E[A_{\rm final}]\le\alpha_\ell\OPT$. It remains to bound
$\E[U_{\rm final}]$.

Consider a state $\mathcal S$
immediately after the first center is chosen. We have $r=k-1$ and
$s=k+\Delta-1$.  Let $A$ be the miss proxy in this state.
By Lemma~\ref{lem:sampling-comparison} and Equation
\eqref{eq:Phi-initial},
$\E[U_{\rm final}\mid \mathcal S]
    \le
    \frac{A}{\Gamma}R_{k-1,k+\Delta-1}$.
Taking expectation over $\mathcal S$ and using $\Gamma=\Delta$ and
\eqref{eq:A-bound}, which gives
$\E[A]\le\alpha_\ell\OPT$, we get
\[
    \E[U_{\rm final}]
    \le
    \frac{R_{k-1,k+\Delta-1}}{\Delta}\alpha_\ell \OPT.
\]
Combining this with the bound on $\E[A_{\rm final}]$ gives
$\E[\cost(C_{\rm final})]\le\alpha_\ell B_{k,\Delta}\OPT$,
as claimed.
\end{proof}

We now prove Lemma~\ref{lem:sampling-comparison}.

\begin{proof}[Proof of Lemma~\ref{lem:sampling-comparison}]
We prove the lemma by induction on the number $s$ of remaining
samples.  If $s=0$, then no more centers are sampled and
$U_{\rm final}=U$.  By~\eqref{eq:Phi-final},
$\Phi_{r,0}(A,U)=U$, so the claim holds.

Now fix a state $\mathcal S$ with $s\ge1$ and $r$ unsettled clusters,
and consider the next sample.   If $U=0$, then
$U_{\rm final}=0$ by Lemma~\ref{lem:U-hit}, while $\Phi_{r,s}(A,U)$ is non-negative by
definition, so the claim holds.
We therefore assume $U>0$.

We bound the expected final balanced mass separately when the next
sample is a miss and when it is a hit.  We have
\begin{align}
    \E[U_{\rm final}\mid\mathcal S]
    ={}&
    \E[U_{\rm final}\mid\mathcal S,\mathrm{hit}]
    \Pr[\mathrm{hit}\mid\mathcal S]
    +
    \E[U_{\rm final}\mid\mathcal S,\mathrm{miss}]
    \Pr[\mathrm{miss}\mid\mathcal S].
    \label{eq:total-expectation}
\end{align}

\paragraph{The miss case.}
Suppose first that the next sample is a miss, and consider any
resulting state $\mathcal S_{\rm miss}$.  The number of unsettled
clusters remains $r$, the number of remaining samples decreases to
$s-1$, and the number of past misses increases from $m$ to $m+1$.
Let $A_{\rm miss}$ and $U_{\rm miss}$ be the miss proxy and total
balanced mass in $\mathcal S_{\rm miss}$.  By the induction
hypothesis,
\begin{equation}
    \E[U_{\rm final}\mid\mathcal S_{\rm miss}]
    \le
    \Phi_{r,s-1}(A_{\rm miss},U_{\rm miss}).
    \label{eq:miss-induction}
\end{equation}

 Now recall the payment definition~\eqref{eq:Phi-definition}. For fixed
$r$ and $s-1$, the payment $\Phi_{r,s-1}(A',U')$ is nondecreasing in
both $A'$ and $U'$. This is because, by~\eqref{eq:Phi-definition}, the
coefficients of $A'/\Gamma$ and $U'/r$ are non-negative, since they are
the expected numbers of payments charged to future misses and past
misses, respectively. Moreover, after a miss we have $U_{\rm miss}\le U$ and  $A_{\rm miss}\le A$  by Lemmas~\ref{lem:U-hit} and \ref{lem:A-evolution}, respectively.  Therefore,
$\Phi_{r,s-1}(A_{\rm miss},U_{\rm miss})
\le \Phi_{r,s-1}(A,U)$. 
Combining this with~\eqref{eq:miss-induction} gives
$\E[U_{\rm final}\mid\mathcal S_{\rm miss}]
\le \Phi_{r,s-1}(A,U)$, and removing the conditioning on
$\mathcal S_{\rm miss}$ by taking expectation over the possible
states after a miss gives
\begin{equation}
    \E[U_{\rm final}\mid\mathcal S,\mathrm{miss}]
    \le
    \Phi_{r,s-1}(A,U).
    \label{eq:miss-continuation}
\end{equation}

\paragraph{The hit case.} Now suppose that the next sample is a hit. Assume first that $r\ge2$, and consider
any resulting state $\mathcal S_{\rm hit}$ after a hit.  This state has $r-1$
unsettled clusters and $s-1$ samples remaining, while the number $m$
of past misses remains unchanged.  Let $A_{\rm hit}$ and
$U_{\rm hit}$ be the miss proxy and total balanced mass in
$\mathcal S_{\rm hit}$.  By the induction hypothesis,
\begin{align}
\begin{split}
    \E[U_{\rm final}\mid\mathcal S_{\rm hit}]
    & \le  \Phi_{r-1,s-1}(A_{\rm hit}, U_{\rm hit}) \\
    &=
    m p_{r-1,s-1}\frac{U_{\rm hit}}{r-1}
    +
    \bigl(R_{r-1,s-1}-m p_{r-1,s-1}\bigr)
    \frac{A_{\rm hit}}{\Gamma}.
\end{split}
\label{eq:given-hit}
\end{align}

Every possible state $\mathcal S_{\rm hit}$ has the same values of
$r-1$, $s-1$, and $m$.  Hence the coefficients
$m p_{r-1,s-1}$ and
$(R_{r-1,s-1}-m p_{r-1,s-1})$ in~\eqref{eq:given-hit} are the same
for every possible state after the hit.  Removing the conditioning
by taking expectation over $\mathcal S_{\rm hit}$ therefore gives
\begin{align}
    \E[U_{\rm final}\mid\mathcal S,\mathrm{hit}]
    \le{}&
    m p_{r-1,s-1}
    \frac{\E[U_{\rm hit}\mid\mathcal S,\mathrm{hit}]}{r-1}
    +
    \bigl(R_{r-1,s-1}-m p_{r-1,s-1}\bigr)
    \frac{\E[A_{\rm hit}\mid\mathcal S,\mathrm{hit}]}{\Gamma}.
    \label{eq:ufinal_exp_hit}
\end{align}
By
Lemma~\ref{lem:U-hit} and Corollary~\ref{cor:A-evolution},
$\E[U_{\rm hit}\mid\mathcal S,\mathrm{hit}]
\le U\cdot (r-1)/r$ and $\E[A_{\rm hit}\mid\mathcal S,\mathrm{hit}]\le A$, respectively.  Since these two quantities are multiplied by
non-negative coefficients in~\eqref{eq:ufinal_exp_hit}, substituting
these bounds gives
\begin{equation}
    \E[U_{\rm final}\mid\mathcal S,\mathrm{hit}]
    \le
    m p_{r-1,s-1}\bar u
    +
    \bigl(R_{r-1,s-1}-m p_{r-1,s-1}\bigr)\bar a.
    \label{eq:hit-continuation}
\end{equation}

If $r=1$, a hit settles the last unsettled cluster, so
$U_{\rm final}=0$.  Since the right-hand side
of~\eqref{eq:hit-continuation} is non-negative,
\eqref{eq:hit-continuation} holds in this case as well.

\paragraph{Combining the two cases.} We now combine the miss and hit bounds.  In the fixed state
$\mathcal S$, the next sample is a miss with probability
$M/(M+U)$ and a hit with probability $U/(M+U)$.  Substituting the
miss bound~\eqref{eq:miss-continuation} and the hit
bound~\eqref{eq:hit-continuation} into~\eqref{eq:total-expectation}
gives
\begin{align}
    \E[U_{\rm final}\mid\mathcal S]
    \le{}&
    \frac{M}{M+U}\Phi_{r,s-1}(A,U)
    +
    \frac{U}{M+U}
    \left(
        m p_{r-1,s-1}\bar u
        +
        \bigl(R_{r-1,s-1}-m p_{r-1,s-1}\bigr)\bar a
    \right).
    \label{eq:combined-continuation}
\end{align}

Once $\mathcal S$ is fixed, the miss bound~\eqref{eq:miss-continuation}
and the hit bound~\eqref{eq:hit-continuation} are fixed.  Hence the
right-hand side of~\eqref{eq:combined-continuation} is their weighted
average, with the miss probability $M/(M+U)$ as the weight on the miss
bound.  Since $0\le M\le A$, we have $M/(M+U) \in [0,A/(A+U)]$.  Therefore, to prove the inductive claim, it is enough to
show that~\eqref{eq:combined-continuation} is upper bounded by
$\Phi_{r,s}(A,U)$ at the two endpoints: $M=0$, so the next sample is
forced to be a hit, and $M=A$.

\paragraph{The endpoint $M=0$.}
We first consider the forced-hit case $M=0$.  By
\eqref{eq:hit-continuation}, in this case we need to show that
\begin{equation}
\begin{split}
    m p_{r-1,s-1}\bar u
    +
    \bigl(R_{r-1,s-1}-m p_{r-1,s-1}\bigr)\bar a
    &\le
    \Phi_{r,s}(A,U)\\
    &=
    m p_{r,s}\bar u
    +
    \bigl(R_{r,s}-m p_{r,s}\bigr)\bar a.
\end{split}
\label{eq:forced-hit-goal}
\end{equation}
Thus we need to compare the expected numbers of payments charged to
past misses and to future misses in the equal-mass process starting
from $(r-1,s-1)$ with those starting from $(r,s)$.  Intuitively,
forcing the first sample to be a hit should not increase either the
number of payments charged to past misses or the number charged to
future misses. In  Lemma~\ref{lem:forced-hit-payments}, which we prove after
this lemma, we show that $p_{r-1,s-1}\le p_{r,s}$ and
$ R_{r-1,s-1}-m p_{r-1,s-1}
    \le
    R_{r,s}-m p_{r,s}$.
Since $m\ge0$, the first inequality gives
$m p_{r-1,s-1}\le m p_{r,s}$.  Since
$\bar u,\bar a\ge0$, these two inequalities prove
\eqref{eq:forced-hit-goal}.

\paragraph{The endpoint $M=A$.}
We now consider the other endpoint, $M=A$.  In this case, the next
sample is a miss with probability $A/(A+U)$ and a hit with probability
$U/(A+U)$.  By~\eqref{eq:combined-continuation}, it remains to show
that
\begin{align}
&\frac{A}{A+U}\Phi_{r,s-1}(A,U)
+
\frac{U}{A+U}
\left(
m p_{r-1,s-1}\bar u
+
\bigl(R_{r-1,s-1}-m p_{r-1,s-1}\bigr)\bar a
\right) \leq 
\Phi_{r,s}(A,U).
\label{eq:M-equals-A-goal}
\end{align}

To compare the two sides, we first put the current payment and the two
continuation bounds in the same form.  Rewriting the payment
definition~\eqref{eq:Phi-definition}, we have
\begin{align}
\Phi_{r,s}(A,U)
=
\bar a R_{r,s}
+
m p_{r,s}(\bar u-\bar a).
\label{eq:Phi-rewrite}
\end{align}
Thus $\bar a R_{r,s}$ values all expected payments at the equal-mass
value $\bar a$, while the second term changes the value of the
expected $m p_{r,s}$ payments charged to past misses from $\bar a$ to
$\bar u$.

In the miss continuation bound, the number of past misses is $m+1$,
and therefore
\begin{align}
\Phi_{r,s-1}(A,U)
=
\bar a R_{r,s-1}
+
(m+1)p_{r,s-1}(\bar u-\bar a).
\label{eq:miss-bound-rewrite}
\end{align}
Similarly, the hit bound~\eqref{eq:hit-continuation} can be written as
\begin{align}
&m p_{r-1,s-1}\bar u
+
\bigl(R_{r-1,s-1}-m p_{r-1,s-1}\bigr)\bar a =
\bar a R_{r-1,s-1}
+
m p_{r-1,s-1}(\bar u-\bar a).
\label{eq:hit-bound-rewrite}
\end{align}

Substituting~\eqref{eq:miss-bound-rewrite} and
\eqref{eq:hit-bound-rewrite} into the left-hand side of
\eqref{eq:M-equals-A-goal}, and~\eqref{eq:Phi-rewrite} into its
right-hand side, we need to show that
\begin{align}
\begin{split}
&\bar a
\left(
\frac{A}{A+U}R_{r,s-1}
+
\frac{U}{A+U}R_{r-1,s-1}
-
R_{r,s}
\right)\\
&\quad+
\frac{A}{A+U}p_{r,s-1}(\bar u-\bar a)\\
&\quad+
m(\bar u-\bar a)
\left(
\frac{A}{A+U}p_{r,s-1}
+
\frac{U}{A+U}p_{r-1,s-1}
-
p_{r,s}
\right) \leq 0 .
\end{split}
\label{eq:M-equals-A-difference}
\end{align}
The first line is the change in the equal-mass part of the payment.
The second is the contribution of the new past miss created when the
next sample is a miss.  The third is the change in the contribution
of the $m$ past misses that were already present.
We evaluate these contributions for arbitrary $\Gamma>0$
before using our choice $\Gamma=\Delta$.

We begin with the first line. We first examine the change in $R$ due to a miss and compare $R_{r,s-1}$ and $R_{r,s}$. Consider the equal-mass process starting with $r$ removable items. Run it for $s-1$ steps, followed by one additional step. Conditioned on $J$ items remaining after the first $s-1$ steps, the additional step removes one item with probability $J/(J+\Gamma)$.
Taking expectation over $J$, the expected number of items
removed at this step is $q_{r,s-1}=\E[J/(J+\Gamma)]$, the probability that this step is a hit. Hence $R_{r,s-1}-R_{r,s}=q_{r,s-1}$.
Since $\Gamma=\Delta$, we have $q_{r,s-1}=p_{r,s-1}$. Although they are equal here, we keep separate notation
$p_{r,s}$ and $q_{r,s}$ for the case $\Delta=0$ in
Section~\ref{subsec:no-extra-centers}, where $\Gamma\ne\Delta$
and the two probabilities differ.

For the change in $R$ due to a hit, multiplying the recurrence
\eqref{eq:R-recurrence} by $r+\Gamma$ and rearranging gives $
r\bigl(R_{r,s}-R_{r-1,s-1}\bigr)
=
\Gamma\bigl(R_{r,s-1}-R_{r,s}\bigr)
=
\Gamma q_{r,s-1}$, and so $R_{r-1,s-1}-R_{r,s} = -\frac{\Gamma}{r} q_{r,s-1}$. Using these two identities, we rewrite the first line as follows.
\begin{align}
\begin{split}
&\bar a
\left(
\frac{A}{A+U}R_{r,s-1}
+
\frac{U}{A+U}R_{r-1,s-1}
-
R_{r,s}
\right)
\\
&=
\bar a
\left(
\frac{A}{A+U}
\bigl(R_{r,s-1}-R_{r,s}\bigr)
+
\frac{U}{A+U}
\bigl(R_{r-1,s-1}-R_{r,s}\bigr)
\right)
\\
&=
\frac{\bar a q_{r,s-1}}{A+U}
\left(
A-\frac{\Gamma}{r}U
\right)
\\
&=
\frac{\bar a q_{r,s-1}}{A+U}
\bigl(\Gamma\bar a-\Gamma\bar u\bigr)
\\
&=
-\frac{A}{A+U}
q_{r,s-1}(\bar u-\bar a),
\label{eq:equal-mass-change}
\end{split}
\end{align}
where we used $A=\Gamma\bar a$ and $U=r\bar u$. Adding the second line of~\eqref{eq:M-equals-A-difference}
gives
\begin{equation}
    \frac{A}{A+U}
    \bigl(p_{r,s-1}-q_{r,s-1}\bigr)(\bar u-\bar a).
    \label{eq:first-two-lines}
\end{equation}
This explains our choice $\Gamma=\Delta$. The factor
$(\bar u-\bar a)$ can be either positive or negative, while $\Gamma=\Delta$
gives $p_{r,s-1}=q_{r,s-1}$ and makes the first two lines
cancel in either case.

We now show that the third line of
\eqref{eq:M-equals-A-difference}, corresponding to the $m$ existing
past misses, is non-positive.  After a miss, the payment probability of
each such miss changes from $p_{r,s}$ to $p_{r,s-1}$, and after a hit it
changes from $p_{r,s}$ to $p_{r-1,s-1}$.  We first relate these
changes.  Conditioning on the first step of the equal-mass process
gives
$p_{r,s}
=
\frac{\Gamma}{r+\Gamma}p_{r,s-1}
+
\frac{r}{r+\Gamma}p_{r-1,s-1}$. Multiplying by $r+\Gamma$ and rearranging gives
$
p_{r,s-1}-p_{r,s}
=
\frac{r}{\Gamma}
\bigl(p_{r,s}-p_{r-1,s-1}\bigr)$.
Using this identity, the third line of
\eqref{eq:M-equals-A-difference} is
\begin{align}
\begin{split}
&m(\bar u-\bar a)
\left(
\frac{A}{A+U}
\bigl(p_{r,s-1}-p_{r,s}\bigr)
+
\frac{U}{A+U}
\bigl(p_{r-1,s-1}-p_{r,s}\bigr)
\right)
\\
&=
\frac{m(\bar u-\bar a)}{A+U}
\left(
\frac{Ar}{\Gamma}-U
\right)
\bigl(p_{r,s}-p_{r-1,s-1}\bigr)
\\
&=
\frac{rm(\bar u-\bar a)}{A+U}
(\bar a-\bar u)
\bigl(p_{r,s}-p_{r-1,s-1}\bigr)
\\
&=
-\frac{rm}{A+U}
\bigl(p_{r,s}-p_{r-1,s-1}\bigr)
(\bar u-\bar a)^2 \leq 0.
\label{eq:existing-miss-change}
\end{split}
\end{align}
In the third line we used $A=\Gamma\bar a$ and $U=r\bar u$, and the
final inequality follows from
$p_{r,s}\ge p_{r-1,s-1}$ by
Lemma~\ref{lem:forced-hit-payments}.

We have therefore shown that the first two lines of
\eqref{eq:M-equals-A-difference} cancel exactly and that the third is
non-positive, which proves~\eqref{eq:M-equals-A-goal}, and completes the proof of the lemma.
\end{proof}

It remains to prove Lemma~\ref{lem:forced-hit-payments}.
Recall that at an intermediate state with $r$ unsettled clusters,
$s$ samples remaining, and $m=r+\Delta-s$ past misses, the equal-mass
process charges in expectation $m p_{r,s}$ payments to past misses and
$R_{r,s}-m p_{r,s}$ payments to future misses.  The lemma shows that
forcing the next step of the equal-mass process to be a hit cannot
increase either quantity.

\begin{lemma}
\label{lem:forced-hit-payments}
For any weight $\Gamma>0$, let $r,s\ge1$ and
let $m=r+\Delta-s\ge0$. Then $p_{r-1,s-1}\le p_{r,s}$, 
and $R_{r-1,s-1}-m p_{r-1,s-1} \le R_{r,s}-m p_{r,s}$.
\end{lemma}

\begin{proof}

We couple two equal-mass processes: one starting from $(r,s)$,
and one starting from $(r-1,s-1)$.
We think of every step of the equal-mass process as drawing an
independent random variable uniformly from $[0,1]$.  If $j$
removable items are currently present, the step is a hit when this
random variable is at most $j/(j+\Gamma)$, and a miss otherwise.

Run the first step of the $(r,s)$ process.  If it is a hit, it now
has $r-1$ removable items and $s-1$ steps remaining, exactly as the
$(r-1,s-1)$ process.  If it is a miss, it has exactly one more
removable item than the $(r-1,s-1)$ process.  For each of the
remaining $s-1$ steps, use the same uniform random variable in the
two processes. 

If the two processes have the same number of removable items, they
evolve identically from that point on.  Otherwise, the $(r,s)$
process has exactly one extra removable item. Since $j/(j+\Gamma)$
increases with $j$, either both processes make the same move and the
difference remains one, or only the $(r,s)$ process has a hit and the
two processes become equal.  Thus throughout the coupling, the
$(r,s)$ process never has fewer removable items than the
$(r-1,s-1)$ process.

Let $J$ and $J'$ be the numbers of removable items left at the end of
the $(r,s)$ and $(r-1,s-1)$ processes, respectively.  The coupling
gives $J\ge J'$ in every realization.  Since
$j/(\Delta+j)$ is increasing in $j$, taking expectations gives
\begin{equation}
    p_{r,s}
    =
    \E\left[\frac{J}{\Delta+J}\right]
    \ge
    \E\left[\frac{J'}{\Delta+J'}\right]
    =
    p_{r-1,s-1}.
    \label{eq:p-monotonicity}
\end{equation}
This proves the first inequality.  Since $m\ge0$, it also shows
that forcing the first step to be a hit cannot increase the expected
number of payments charged to past misses.

We now compare the payments charged to future misses.
Conditioned on $J$, their expected number in the $(r,s)$ process is
\[
    J-\frac{mJ}{\Delta+J}
    =
    \frac{J}{\Delta+J}(\Delta+J-m).
\]
This expression is increasing in $J$, as
$J/(\Delta+J)$ is non-negative and increasing in $J$, while
$\Delta+J-m$ is also increasing in $J$ and equals the number of
future misses, so it is non-negative for every reachable value of
$J$.  Thus, since $J'\leq J$ in every realization we have
$J'-mJ'/(\Delta+J')\leq J-mJ/(\Delta+J)$. Taking expectations gives
\[
    R_{r-1,s-1}-m p_{r-1,s-1}
    \le
    R_{r,s}-m p_{r,s}.
\]
This proves the second inequality and completes the proof.
\end{proof}

\subsubsection{No Additional Centers $\Delta=0$}
\label{subsec:no-extra-centers}

In Lemma~\ref{lem:sampling-comparison}, we chose
$\Gamma=\Delta$ so that~\eqref{eq:first-two-lines} is zero for
every state. For $\Delta=0$, we cannot use $\Gamma=0$, since
the payment definition~\eqref{eq:Phi-definition} uses
$\bar a=A/\Gamma$. We instead choose a sufficiently
small $\Gamma>0$ for which the analysis still holds. 

\begin{proof}[Proof of Theorem~\ref{thm:exact-sampling} for $\Delta=0$]
Fix an instance. We use the same payment construction with
$\Gamma>0$, whose value will be chosen later, and follow the proof
of Lemma~\ref{lem:sampling-comparison}. The argument applies
unchanged until~\eqref{eq:first-two-lines}. This is the only place
where we use $\Gamma=\Delta$, to obtain
$p_{r,s-1}=q_{r,s-1}$ and cancel the first two lines of~\eqref{eq:M-equals-A-difference}.

Here $\Delta=0$, so every miss is charged and $p_{r,s}=1$.\footnote{For $\Delta=0$, the ratio $J/(\Delta+J)$ is undefined
at $J=0$. We interpret it as $1$ there, consistent with charging
every miss. Thus $p_{r,s}=1$ throughout.}
Thus~\eqref{eq:first-two-lines} becomes
\[
    \frac{A}{A+U}(1-q_{r,s-1})(\bar u-\bar a).
\]
We need to show that this expression is non-positive.
Since $q_{r,s-1}\le1$, it is non-positive if
$\bar u\le\bar a$, i.e., if $U/r\le A/\Gamma$.
We  choose a single $\Gamma>0$ small enough that this holds
at every state with $A>0$ and $r>0$. Such a choice exists because
$A$ takes only finitely many values and $U$ is bounded on the
fixed input. With this choice, the expression is non-positive,
and the rest of the induction proceeds unchanged.

Applying this bound at the initial state $\mathcal S$ gives
$\E[U_{\rm final}\mid\mathcal S]\le\bar a R_{k-1,k-1}$.
It remains to bound this initial payment $ \bar a R_{k-1,k-1}$.
After $k-1$ samples in the equal-mass process starting with $k-1$
removable items, the number of remaining items equals the number
of misses. Before sample $t+1$, 
at least $k-1-t$ items remain,
so its miss probability is at most
$\Gamma/(\Gamma+k-1-t)$. Therefore 
\begin{align*}
    \bar a R_{k-1,k-1}
    &\le
    \frac{A}{\Gamma}
    \sum_{t=0}^{k-2}\frac{\Gamma}{\Gamma+k-1-t}
    =
    \frac{A}{\Gamma}
    \sum_{j=1}^{k-1}\frac{\Gamma}{j+\Gamma}
    =
    A\sum_{j=1}^{k-1}\frac{1}{j+\Gamma}
    \le AH_{k-1}.
\end{align*}
Taking expectation over the initial state and using
\eqref{eq:final-cost-AU} and~\eqref{eq:A-bound} gives
\begin{align*}
    \E[\cost(C_{\rm final})]
    &\le
    \E[A_{\rm final}]+H_{k-1}\E[A]
    \le
    \alpha_\ell B_{k,0}\OPT. \qedhere
\end{align*}
\end{proof}

\subsection{Lower Bounds via the Equal-Mass Process}
\label{sec:lower-bounds}

\paragraph{Tightness for metric $k$-median.}

We now give a simple metric construction that mimics the
equal-mass process. The permanent item is represented by one
large cluster containing many points at small pairwise distances. The
removable items are represented by singleton clusters, each at the same
distance from the large cluster's center.  The $D$-sampling algorithm samples its first center from
the large cluster with high probability. Each later sample from it
removes only a small amount of its mass. Sampling a singleton
removes its entire mass. By choosing the distances within the large
cluster, we can control its mass relative to that of the singletons
to match the ratio of the weights in the equal-mass process.

We use this construction to show that our bounds are tight for
metric $k$-median.
Recall that $\alpha_1=2$, so Theorem~\ref{thm:exact-sampling}
gives an upper bound of $2B_{k,\Delta}$ on the approximation
ratio.

\begin{theorem}
\label{thm:metric-median-tight}
For fixed integers $k\ge1$ and $\Delta\ge0$, there is an
infinite sequence of metric $k$-median instances with an increasing
number of input points, on which the approximation ratio
of $D$-sampling with $k+\Delta$ centers approaches $2B_{k,\Delta}$.
\end{theorem}

\begin{proof}
We construct an instance with one large cluster $P=\{x_1,\ldots,x_n\}$ and
$k-1$ singleton clusters $y_1,\ldots,y_{k-1}$. The metric space also contains
a point $o$. Let $0<\varepsilon<1/2$ be a small distance, to be chosen
later. The distances from $o$ are as follows. For each $x_i\in P$,
we set $d(x_i,o)=\varepsilon$, and for each singleton $y_i$, we
set $d(y_i,o)=1-\varepsilon$. For distinct input points, we define
the distances so that the triangle inequality through $o$ holds
with equality. That is, we set $d(x_i,x_j)=2\varepsilon$,
$d(y_i,x_j)=1$, and $d(y_i,y_j)=2(1-\varepsilon)$.

To upper bound $\OPT$, we place the $k$ centers at
$o,y_1,\ldots,y_{k-1}$. Each point in $P$ contributes
$\varepsilon$ to the cost, and each singleton contributes $0$,
so $\OPT\le n\varepsilon$.

We now consider the cost of $D$-sampling. The first center belongs
to $P$ with probability $n/(n+k-1)$. Condition on this event.
Every unsampled point in $P$ has cost $2\varepsilon$, and every
unsampled singleton has cost $1$. Sampling a singleton does not
reduce the cost of any other point (since $2\varepsilon < 1 < 2(1-\varepsilon)$ by the choice $\varepsilon < 1/2$). Sampling another point in $P$
also leaves the costs of all other points unchanged. Let $q$ be the number of points sampled from $P$ so far,
including the first center. The cost of $P$ is then $2\varepsilon(n-q)$. 

First consider $\Delta>0$, for which the equal-mass bound uses
permanent weight $\Gamma=\Delta$. We choose $\varepsilon$ so that
$P$ retains mass at least $\Delta$ throughout. Since at most
$k+\Delta$ points of $P$ are sampled (so $q\le k+\Delta$), we set
$\varepsilon=\Delta/(2(n-k-\Delta))$, taking $n$ sufficiently large
that $\varepsilon<1/2$.
This ensures that
$2\varepsilon(n-q)\ge 2\varepsilon(n-k-\Delta)=\Delta$.

We use a coupling, as in the proof of
Lemma~\ref{lem:forced-hit-payments}, to compare the remaining
$k+\Delta-1$ samples with the equal-mass process with $k-1$
removable items and permanent weight $\Delta$.
When $j$ singletons remain and $q$ points from $P$ have been sampled, the probability of sampling a singleton
is $j/(j+2\varepsilon(n-q))\le j/(j+\Delta)$. At each step, we
use the same uniform random number for the hit decision in both
processes. Whenever
their numbers of remaining items agree, a hit in the actual
process also gives a hit in the equal-mass process. Since each
step removes at most one item, the actual process never has fewer
remaining singletons than the equal-mass process has removable items.

It follows that the expected final cost, conditioned on the first
center belonging to $P$, is at least
$\Delta+R_{k-1,k+\Delta-1}=\Delta B_{k,\Delta}$.
The first term accounts for $P$, and the second for the remaining
singletons. Using $\OPT\le n\varepsilon=n\Delta/(2(n-k-\Delta))$
and taking expectation over the first sample gives 
\begin{align*}
    \frac{\E[\cost(C_{\rm final})]}{\OPT}
    &\ge
    \frac{n}{n+k-1}\cdot
    \frac{\Delta B_{k,\Delta}}{n\varepsilon}
    =
    \frac{2(n-k-\Delta)}{n+k-1}B_{k,\Delta}.
\end{align*}
This lower bound approaches $2B_{k,\Delta}$ as $n$ approaches
infinity, proving tightness in this case.

For $\Delta=0$, take $n>k$ and use the same construction with
$\Gamma=1/n$ and $\varepsilon=\Gamma/(2(n-k))$. We bound the cost directly.
Again, condition on the first center
belonging to $P$. There are $k-1$ singletons and $k-1$ samples
remaining. Thus any further sample from $P$ leaves at least one
singleton unsampled, contributing $1$ to the final cost. We lower
bound the expected total cost of the singletons by the probability
that at least one of them remains unsampled.

For all singletons to be sampled, every remaining sample must be
a singleton. Along this sequence of samples, the cost of $P$
stays at $2\varepsilon(n-1)\ge\Gamma$, while the number of
unsampled singletons decreases from $k-1$ to $0$. The probability
that all singletons are sampled is therefore
\begin{align*}
    \prod_{j=1}^{k-1}\frac{j}{j+2\varepsilon(n-1)}
    &\le \frac{1}{\prod_{j=1}^{k-1}(1+\Gamma/j)}
    \le \frac{1}{1+\Gamma H_{k-1}}.
\end{align*}
The last inequality follows by expanding the product in the
denominator and keeping $1$ and the terms $\Gamma/j$, since all
other terms are non-negative. Hence the probability that at least
one singleton remains unsampled is at least
$\Gamma H_{k-1}/(1+\Gamma H_{k-1})$. In addition, $P$ always
contributes at least $2\varepsilon(n-k)=\Gamma$ to the final cost.
Taking expectation over the first sample and using
$\OPT\le n\varepsilon =n\Gamma/(2(n-k))$, we obtain
\begin{align*}
    \frac{\E[\cost(C_{\rm final})]}{\OPT}
    &\ge
    \frac{n}{n+k-1}\cdot
    \frac{\Gamma+\Gamma H_{k-1}/(1+\Gamma H_{k-1})}{n\Gamma/(2(n-k))}
    =
    \frac{2(n-k)}{n+k-1}
    \left(1+\frac{H_{k-1}}{1+H_{k-1}/n}\right),
\end{align*}
where the last equality uses $\Gamma=1/n$. This lower bound
approaches $2(1+H_{k-1}) = 2B_{k,0}$ as $n$ approaches infinity.
\end{proof}

\paragraph{Euclidean $k$-median without additional centers.}

For $\Delta=0$, the bound $2B_{k,0}$ is also tight for
Euclidean $k$-median.

\begin{theorem}
\label{thm:euclidean-median-tight}
For every fixed integer $k\ge1$, there is an infinite sequence of
Euclidean $k$-median instances in $\R^k$ with an increasing number
of input points, on which the approximation ratio of
$D$-sampling with exactly $k$ centers approaches
$2B_{k,0}=2(1+H_{k-1})$.
\end{theorem}

We give the construction and proof in
Appendix~\ref{app:euclidean-median-tight}.

\paragraph{General distance exponents.}
By adapting the distances in the construction of
Theorem~\ref{thm:metric-median-tight}, we obtain lower bounds
for every positive distance exponent in both general metrics
and Euclidean space.

\begin{corollary}[Construction of Theorem~\ref{thm:metric-median-tight}]
\label{cor:metric-squared-lower}
\label{cor:euclidean-power-lower}
For every $\ell>0$ and all integers $k\ge1$ and $\Delta\ge0$,
the approximation ratio of $D^\ell$-sampling
with $k+\Delta$ centers is at least $2^\ell B_{k,\Delta}$ in
general metrics and at least $2^{\ell/2}B_{k,\Delta}$ in
Euclidean space.
\end{corollary}

We prove both bounds in Appendix~\ref{app:euclidean-power-lower}. Without additional centers, adapting the Euclidean construction
of Theorem~\ref{thm:euclidean-median-tight} gives stronger bounds.

\begin{corollary}
\label{cor:euclidean-power-no-extra}
\label{cor:euclidean-means-lower}
For every $\ell>0$ and every integer $k\ge1$, the approximation
ratio of $D^\ell$-sampling with exactly $k$ centers is at least
$\max\{2,2^\ell\}B_{k,0}$ in general metrics and at least
$\max\{2,2^{\ell-1}\}B_{k,0}$ in Euclidean space.
\end{corollary}

We prove this in Appendix~\ref{app:euclidean-median-tight}.

\subsection{A Closed-Form Bound for $\Delta>0$}
\label{subsec:closed-form}

We return to $\Delta>0$ and the choice $\Gamma=\Delta$.
Recall that $B_{k,\Delta}=1+R_{k-1,k+\Delta-1}/\Delta$ is the
factor multiplying $\alpha_\ell$ in Theorem~\ref{thm:exact-sampling}.
To obtain a closed-form upper bound on $B_{k,\Delta}$, we bound
the expected number of remaining items in the equal-mass process.

We start with an intuitive discussion to get a rough idea of how
many removable items should remain.  Consider the equal-mass process
starting with $r$ removable items and running for $r+\Delta$ steps.
Suppose that the process currently has $j$ removable items.  The next
step is a hit with probability $j/(j+\Delta)$.  Hence the number of
steps up to and including the next hit has a geometric distribution
with success probability $p=j/(j+\Delta)$ and expectation $1/p$.
The expected number of misses before that hit is therefore
$1/p-1=\Delta/j$.
Thus, the expected number of misses needed to decrease from $r$
removable items to $j$ is $\Delta(H_r-H_j)$.

To get some intuition, suppose that the number of misses needed to
reach $j$ were exactly this expectation.  If the process ended with
$j$ removable items after $r+\Delta$ steps, then $r-j$ steps would be
hits and the remaining $\Delta+j$ steps would be misses.  We would
therefore have
\begin{align}
    \Delta+j
    =
    \Delta(H_r-H_j)
    \le
    \Delta\ln\frac{r}{j}.\label{eq:rough_estimate}
\end{align}
As we will see below, an inequality of this form gives an upper bound
on the number of removable items that remain.

The difficulty in turning the calculation above into a formal
bound is that our process ends with a random number of items $J$.
An averaging argument based on~\eqref{eq:rough_estimate} would
require an upper bound on $\E[\ln(r/J)]$ in terms of $\E[J]$.
However, $\ln(r/j)$ is convex in $j$, so Jensen's inequality gives
a lower bound instead.
We therefore take a different approach. The following lemma
shows that the expected number of remaining items satisfies
the desired inequality up to an additive $1/2$.

\begin{lemma}
\label{lem:R-implicit-bound}
Let $r\ge1$ and $\Delta\ge1$, and let
$R=R_{r,r+\Delta}$.  Then
\begin{equation}
    \Delta+R
    \le
    \Delta\ln\frac{r}{R}+\frac12.
    \label{eq:R-implicit-bound}
\end{equation}
\end{lemma}

\begin{proof}
Let $b>0$ be the unique solution of
$\Delta+b=\Delta\ln(r/b)+1/2$.\footnote{Such a solution exists and is unique: the left-hand side is
strictly increasing in $b$, whereas the right-hand side is strictly
decreasing, tends to infinity as $b\to0$, and is smaller than the
left-hand side at $b=r$.} 
It is enough to prove $R\le b$.  Indeed, in this case
\[
    \Delta+R
    \le
    \Delta+b
    =
    \Delta\ln\frac{r}{b}+\frac12
    \le
    \Delta\ln\frac{r}{R}+\frac12,
\]
which gives~\eqref{eq:R-implicit-bound}.

Let $J$ be the number of removable items left at the end of the
process, so that $R=\E[J]$.  We prove $R\le b$ by a potential-function
argument.  We construct a potential that starts at $1$, does not
increase in expectation, and whose final value is at least $J/b$.
These properties give
$R/b=\E[J]/b\le\E[\Psi_{\rm final}]\le1$, and therefore $R\le b$.
We now construct this potential.

Let $q=\Delta/(\Delta+b)$.  When the process has $j$ removable items
and $m$ misses have occurred so far, define
\begin{equation}
    \Psi_{j,m}
    =
    \left(\frac{j}{r}\right)^q
    \left(1+\frac{1}{\Delta+b}\right)^m.
    \label{eq:Psi-definition}
\end{equation}
A hit decreases the first factor, while a miss increases the second.
We first show that the potential does not increase in expectation.

Suppose that $j>0$.  A miss multiplies $\Psi$ by
$1+1/(\Delta+b)$, whereas a hit multiplies it by
$(1-1/j)^q$.  Since $0<q<1$, the reversed Bernoulli inequality gives
$(1-1/j)^q\le1-q/j$.  Therefore
\begin{align}
    \E[\Psi_{\rm new}\mid j,m]
    &\le
    \Psi_{j,m}
    \left[
        \frac{\Delta}{j+\Delta}
        \left(1+\frac{1}{\Delta+b}\right)
        +
        \frac{j}{j+\Delta}
        \left(1-\frac{q}{j}\right)
    \right]
    =
    \Psi_{j,m},
    \label{eq:Psi-supermartingale}
\end{align}
where the equality uses $q=\Delta/(\Delta+b)$.
If $j=0$, then $\Psi_{j,m}=0$ and remains zero.  Thus $\Psi$ does not
increase in expectation.  Since the process starts from
$\Psi_{r,0}=1$, its expected final value is at most $1$.

It remains to lower bound the final value of the potential.  If $j$
removable items remain at the end, then $r-j$ steps were hits and
$\Delta+j$ steps were misses.  The final potential is therefore
$\Psi_{j,\Delta+j}$, and we need to show that $\Psi_{j,\Delta+j}\ge\frac{j}{b}$.

To prove this inequality, we first lower bound the factor in the
potential contributed by the $\Delta+j$ misses.  Using
$\ln(1+1/x)\ge 1/(x+1/2)$ with $x=\Delta+b$, 
we have
\begin{align*}
    \left(1+\frac{1}{\Delta+b}\right)^{\Delta+j}
    &=
    e^{(\Delta+j)\ln\left(1+\frac{1}{\Delta+b}\right)}
\ge
    e^{(\Delta+j)/(\Delta+b+1/2)}
\ge
    e\frac{\Delta+j}{\Delta+b+\frac12},
\end{align*}
where the last inequality uses $e^y\ge ey$ for $y\ge0$.  Moreover,
\[
    \frac{1}{\Delta+b+\frac12}
    =
    \frac{1}{\Delta+b}
    \frac{1}{1+\frac{1}{2(\Delta+b)}}
    \ge
    \frac{1}{\Delta+b}
    e^{-\frac{1}{2(\Delta+b)}},
\]
where we used $1+x\le e^x$.  Hence
\begin{equation}
    \left(1+\frac{1}{\Delta+b}\right)^{\Delta+j}
    \ge
    e^{1-\frac{1}{2(\Delta+b)}}
    \frac{\Delta+j}{\Delta+b}.
    \label{eq:miss-factor}
\end{equation}

We now rewrite the exponential term $e^{1-\frac{1}{2(\Delta+b)}}$ in~\eqref{eq:miss-factor} using
the definition of $b$.  Since
$\Delta+b=\Delta\ln(r/b)+1/2$ and
$q=\Delta/(\Delta+b)$, we have
\[
    q\ln\frac{r}{b}
    =
    \frac{\Delta+b-\frac12}{\Delta+b}
    =
    1-\frac{1}{2(\Delta+b)}.
\]
Exponentiating gives
\begin{equation}
    e^{1-\frac{1}{2(\Delta+b)}}
    =
    \left(\frac{r}{b}\right)^q.
    \label{eq:other-factor}
\end{equation}

We can now lower bound the final potential.  By definition,
\[
    \Psi_{j,\Delta+j}
    =
    \left(\frac{j}{r}\right)^q
    \left(1+\frac{1}{\Delta+b}\right)^{\Delta+j}.
\]
Using~\eqref{eq:miss-factor} and~\eqref{eq:other-factor}, we get
\begin{align*}
    \Psi_{j,\Delta+j}
    &\ge
    \left(\frac{j}{r}\right)^q
    \left(\frac{r}{b}\right)^q
    \frac{\Delta+j}{\Delta+b}=
    \left(\frac{j}{b}\right)^q
    \left(
        q+(1-q)\frac{j}{b}
    \right),
\end{align*}
where in the last equality we used
$q=\Delta/(\Delta+b)$. By weighted AM-GM,\footnote{The weighted AM-GM inequality says that for non-negative $x_1, x_2$ and weights $w_1, w_2 > 0$ such that $w_1 + w_2 = 1$, we have $w_1 x_1 + w_2 x_2 \ge x_1^{w_1} x_2^{w_2}$. Here, we took $x_1 = 1$, $x_2 = j/b$, $w_1 = q$, and $w_2 = 1-q$.}
$q+(1-q)j/b\ge(j/b)^{1-q}$.  Hence
\[
    \Psi_{j,\Delta+j}
    \ge
    \left(\frac{j}{b}\right)^q
    \left(\frac{j}{b}\right)^{1-q}
    =
    \frac{j}{b}.
\]
This proves the required lower bound on the final potential and
completes the proof.
\end{proof}

We can now solve the inequality in Lemma~\ref{lem:R-implicit-bound}
to get a closed-form upper bound.  For $x\ge0$, let $W(x)$ denote
the Lambert $W$ function, that is, $W(x)$ is the unique nonnegative value
satisfying $W(x)e^{W(x)}=x$.

\begin{corollary}
\label{cor:R-lambert}
For every $r\ge1$ and $\Delta\ge1$,
\[
    \frac{R_{r,r+\Delta}}{\Delta}
    \le
    W\left(
        \frac{r}{e\Delta} 
        e^{1/(2\Delta)}
    \right).
\]
\end{corollary}

\begin{proof}
Let $R=R_{r,r+\Delta}$.  By
Lemma~\ref{lem:R-implicit-bound}, $\Delta+R \le \Delta\ln({r}/{R})+1/2$.
Isolating $\ln(r/R)$ and exponentiating gives
\[
    \frac{r}{R}
    \ge
    e^{1+\frac{R}{\Delta}-\frac{1}{2\Delta}},
\]
or equivalently,
\[
    (R/\Delta)
    e^{R/\Delta}
    \le
    \frac{r}{\Delta}
    e^{-1+\frac{1}{2\Delta}}.
\]
Applying $W$ to both sides gives the claimed bound.
\end{proof}

Setting $r=k-1$ in Corollary~\ref{cor:R-lambert} and using the
definition of $B_{k,\Delta}$ gives the following bound.
\begin{theorem}
\label{thm:closed-form}
For all integers $k\ge1$ and $\Delta\ge1$,
\[
    B_{k,\Delta}
    \le
    1+
    W\left(
        \frac{k-1}{e\Delta}
        e^{1/(2\Delta)}
    \right).
\]
\end{theorem}

We now prove a lower bound on $B_{k,\Delta}$, showing that our
closed-form bound exceeds it by at most $1/\Delta$.
To this end, we return to the
harmonic-number calculation at the start of Section~\ref{subsec:closed-form}.
Jensen's inequality now gives the direction we need.

\begin{theorem}
\label{thm:closed-form-lower}
For all integers $k\ge1$ and $\Delta\ge1$,
\[
    B_{k,\Delta}\ge
    1+W\left(\frac{k-1}{e\Delta}e^{1/(2\Delta)}\right)
    -\frac1\Delta.
\]
\end{theorem}

\begin{proof}
Let $r=k-1$. If we allow the equal-mass process to run until all
$r$ removable items have been removed, the expected number of
steps is $r+\Delta H_r$. Indeed, the calculation preceding
\eqref{eq:rough_estimate} gives $\Delta H_r$ misses in expectation,
and removing the items takes exactly $r$ hits.
We upper bound this expectation by first counting $r+\Delta$ steps
and then the additional steps needed to remove any remaining items.

Let $J$ be the number of removable items left after the first
$r+\Delta$ steps, and write $R=\E[J]=R_{r,r+\Delta}$.
Conditioned on $J=j$, removing these $j$ items takes
$j+\Delta H_j$ additional steps in expectation, by the same
calculation. Counting the first $r+\Delta$ steps and these
additional steps can only overestimate the number of steps needed
to remove all $r$ items. The count may be larger if all items are
removed before step $r+\Delta$, since we still count all
$r+\Delta$ steps. Taking expectations therefore gives
$r+\Delta H_r\le r+\Delta+R+\Delta\E[H_J]$.
Rearranging and dividing by $\Delta$, we obtain
\begin{equation}
    H_r-\E[H_J]\le1+\frac{R}{\Delta}.
    \label{eq:R-harmonic-lower}
\end{equation}

We now use~\eqref{eq:R-harmonic-lower} to lower bound $R$.
Define $f(x)=\ln((r+1)/(x+1))$ for $x\ge0$.
For every integer $0\le j\le r$, the sum
$H_r-H_j=\sum_{i=j+1}^{r}1/i$ is at least
$\int_{j+1}^{r+1}dx/x=f(j)$. Taking expectations gives
$H_r-\E[H_J]\ge\E[f(J)]$.
The function $f$ is convex, so Jensen's inequality gives
$f(R)=f(\E[J])\le\E[f(J)]$.
Combining these two bounds with~\eqref{eq:R-harmonic-lower} gives
\[
    \ln\frac{r+1}{R+1}
    =f(R)
    \le\E[f(J)]
    \le H_r-\E[H_J]
    \le1+\frac{R}{\Delta}.
\]

To express this bound in terms of $W$, we rewrite it as
$\frac{R+1}{\Delta}e^{(R+1)/\Delta}\ge\frac{r+1}{e\Delta}e^{1/\Delta}$.
Applying $W$ to both sides and subtracting $1/\Delta$ gives
\[
    \frac{R}{\Delta}
    \ge W\left(\frac{r+1}{e\Delta}e^{1/\Delta}\right)-\frac1\Delta
    \ge W\left(\frac{r}{e\Delta}e^{1/(2\Delta)}\right)-\frac1\Delta,
\]
where the last inequality uses that $W$ is increasing and
$(r+1)e^{1/\Delta}\ge r e^{1/(2\Delta)}$.
Substituting $r=k-1$ and $B_{k,\Delta}=1+R/\Delta$ completes
the proof.
\end{proof}

\begin{corollary}
\label{cor:loglog-bound}
The additive improvement
in $B_{k,\Delta}$ from one extra center is $B_{k,0}-B_{k,1}=\Theta(\log\log k)$.
\end{corollary}

\begin{proof}
For every $x\ge e$, we have the elementary bounds
$\ln x-\ln\ln x\le W(x)\le\ln x-\ln\ln x+1$.
Set $x=(k-1)/\sqrt e$.
With $\Delta=1$, Theorems~\ref{thm:closed-form} and~\ref{thm:closed-form-lower}
give $W(x)\le B_{k,1}\le1+W(x)$.
Together with $B_{k,0}=1+H_{k-1}$, this gives
\[
    H_{k-1}-W(x)
    \le B_{k,0}-B_{k,1}
    \le 1+H_{k-1}-W(x).
\]
For $k\ge6$, we have $x\ge e$, so the elementary bounds give
$W(x)=\ln k-\ln\ln k+O(1)$.
Since $H_{k-1}=\ln k+O(1)$, both sides of the inequality are
$\ln\ln k+O(1)$, which proves the claim.
\end{proof}

We now show that our bound for squared Euclidean distances improves
both bi-criteria bounds of Makarychev, Reddy, and
Shan~\cite[Theorem~5.1]{MRS2020} throughout their respective ranges
of $\Delta$. Recall that their bounds on the approximation
ratio are $5\left(1+\frac{k}{e(\Delta-1)}\right)$ for $\Delta>1$
and $5\left(2+\frac{1}{2e}+\ln\frac{2k}{\Delta}\right)$ for
$1\le\Delta\le2k$. For this comparison, we derive a slightly
weaker but simpler expression for our bound.

 Using $W(x)\le\ln(1+x)$ and
$e^{1/(2\Delta)}\le\Delta/(\Delta-\frac12)$, where the latter follows
from $1-x\le e^{-x}$ with $x=1/(2\Delta)$,
Theorems~\ref{thm:closed-form} and~\ref{thm:exact-sampling} give
\begin{equation}
    \E[\cost(C_{\rm final})]
    \le
    \alpha_\ell
    \left(
        1+
        \ln\left(
            1+\frac{k-1}{e(\Delta-\frac12)}
        \right)
    \right)
    \OPT.
    \label{eq:simple-closed-form}
\end{equation}

For squared Euclidean distances, our leading constant is
$\alpha_2=(1+\varphi)$, compared with $5$ in the bounds by~\cite{MRS2020}.
The factor multiplying $\alpha_\ell$ in~\eqref{eq:simple-closed-form}
is strictly smaller than the factor multiplying $5$ in each of their bounds, over its stated range of $\Delta$.
First, for $\Delta>1$, using
$\ln(1+x)\leq x$ gives
$1+\ln\left(1+\frac{k-1}{e(\Delta-\frac12)}\right)
<1+\frac{k}{e(\Delta-1)}$,
which improves their large-$\Delta$ bound.

For their second bound that applies for $1\le\Delta\le2k$, we have
$(k-1)/(\Delta-\frac12)<2k/\Delta$.  Hence
$\ln\left(1+\frac{k-1}{e(\Delta-\frac12)}\right)
<\ln\left(1+\frac{2k}{e\Delta}\right)$.
Since $2k/\Delta\ge1$,
$1+\frac{2k}{e\Delta}
\le(1+\frac1e)\frac{2k}{\Delta}$.
Therefore
\[
    1+
    \ln\left(
        1+\frac{k-1}{e(\Delta-\frac12)}
    \right)
    <
    1+\ln\left(1+\frac1e\right)
    +\ln\frac{2k}{\Delta}
    <
    2+\frac{1}{2e}
    +\ln\frac{2k}{\Delta},
\]
which improves their small-$\Delta$ bound.

\section{Discussion}
\label{sec:discussion}

We give a tight, unified analysis of metric $k$-median++ for every
$k$ and every number $\Delta$ of extra centers. Through balancing,
we reduce the analysis to a hit-and-miss abstraction of the
sampling process, which we then relate to the simple equal-mass
process. The equal-mass process also guides our improved
lower-bound constructions. By adapting the balancing
construction, the same analysis extends to every distance
exponent $\ell>0$, giving improved upper bounds.

Our balancing constants are optimal for $k$-median and
Euclidean $k$-means. Determining the optimal balancing constants
for the remaining cases with $\ell>1$, including squared distances
in general metrics, remains open. Such an analysis may be of independent
interest, for example in analyses of online facility
location~\cite{KNR23,HJ2027}.

Balancing suffices for a tight analysis of metric
$k$-median. For Euclidean $k$-means, however, a gap remains even
though our balancing constant is optimal. One possible source
of loss is how we account for samples from settled clusters.
These samples are classified as misses and help pay for the
costs of unsettled clusters, but we do not quantify how much
they further reduce the settled cluster's own cost. These
improvements can occur even when $\Delta=0$, and may become more
substantial when $\Delta$ is large, as repeated samples from the
same clusters become more frequent. Can a richer
analysis exploit this additional benefit to improve the
approximation bound? Such an analysis may also reveal how the
leading coefficient depends on both $k$ and $\Delta$, beyond the
dependence captured by $B_{k,\Delta}$.

We highlight two particularly interesting cases
in this regard. For Euclidean $k$-means without extra centers,
the case $k=2$ has exact approximation ratio $5$
(Lemma~\ref{lem:euclidean-two-center-exact}). For $k\ge3$, our
lower and upper bounds are $2B_{k,0}$ and
$(1+\varphi)B_{k,0}$, respectively. Can a richer analysis improve
the upper bound? On the other hand, can we construct a lower
bound of $(2+\varepsilon)B_{k,0}$ for some fixed $\varepsilon>0$
and every $k\ge2$?

For Euclidean $k$-median, the exact approximation ratio without
extra centers is $2B_{k,0}$, but with extra centers, our Euclidean
lower and upper bounds are $\sqrt2 B_{k,\Delta}$ and
$2B_{k,\Delta}$, respectively. Does restricting to Euclidean
space improve the approximation ratio when $\Delta=1$?

\appendix

\section[Additional Balancing Proofs]{Additional Balancing Proofs}
\label{app:balancing-extensions}

\begin{proof}[Proof of Lemma~\ref{lem:balancing-uniform}]
Fix a cluster $P$ of diameter $\delta$.
We construct a new input $P\cup\{q\}$ with $k=2$ by
placing $q$ arbitrarily far from $P$, and let $R=d(q,P)$.
For sufficiently large $R$, the optimal clusters are $P$ and
$\{q\}$. The first sample is $q$ with probability $1/(|P|+1)$.
Condition on this event, and let $\mathcal S$ be the resulting
state, with $C=\{q\}$.

The triangle inequality gives $R\le D(x)\le R+\delta$ for
every $x\in P$. By
Property~\eqref{prop:balancing-imbalanced-mass}, balancing removes
at most $\alpha_\ell\phi(P)$ mass, which is fixed as $R$ grows.
Let $b(x)$ be the balanced mass of $x$. We therefore have
$R^\ell-\alpha_\ell\phi(P)\le b(x)\le(R+\delta)^\ell$.
We choose $R$ large enough so that $R^\ell>\alpha_\ell\phi(P)$. 
Conditioned on a hit of $P$ in state $\mathcal S$, the probability
$p_R(x)$ of sampling $x\in P$ satisfies
\begin{align*}
    \frac{R^\ell-\alpha_\ell\phi(P)}{|P|(R+\delta)^\ell}
    \le p_R(x)=\frac{b(x)}{\sum_{y\in P}b(y)}
    \le \frac{(R+\delta)^\ell}{|P|(R^\ell-\alpha_\ell\phi(P))}.
\end{align*}
Both bounds approach $1/|P|$ as $R\to\infty$, so
$p_R(x)\to1/|P|$ for every $x\in P$.

For $R>\delta$, every point of $P$ is closer to any sampled
point in $P$ than to $q$. Thus, conditioned on a hit of $P$,
Property~\eqref{prop:balancing-hit-cost} gives
\begin{align*}
    \E[\cost(P,C_{\rm new})\mid\mathcal S,\mathrm{hit\ of\ }P]
    &=\sum_{x\in P}p_R(x)\cost(P,\{x\}) \le\alpha_\ell\phi(P).
\end{align*}
Taking $R\to\infty$ gives
$\frac{1}{|P|}\sum_{x\in P}\cost(P,\{x\})\le\alpha_\ell\phi(P)$.
The left-hand side is $\E[\cost(P,\{Z\})]$ for a uniformly
random point $Z\in P$.
\end{proof}

\subsection{Balancing for Squared Metrics and Higher Powers}
\label{app:balancing-powers}

We use the same capping construction as in
Section~\ref{subsec:balancing}, with $D(x)^\ell$ in place of
$D(x)^2$. Fix an unsettled optimal cluster $P$, let $c^*$ be its
optimal center, and write $n=|P|$. For a parameter $\alpha>1$,
we declare up to $\alpha\phi(P)$ sampling mass imbalanced and
retain total balanced mass
$B=\max\{\cost(P,C)-\alpha\phi(P),0\}$.
If $B=0$, we set $b(x)=0$ for every $x\in P$.
Otherwise, we choose a cap $\tau^\ell>0$ so that the balanced
weights $b(x)=\min\{D(x)^\ell,\tau^\ell\}$ sum to $B$.
The same continuity argument as in the squared Euclidean case
shows that such a cap exists. If the algorithm samples $x\in P$,
we declare a hit with probability $b(x)/D(x)^\ell$.

This construction makes the imbalanced mass at most
$\alpha\phi(P)$. Also, $B$ can only decrease as centers are added,
because $\phi(P)$ is fixed and $\cost(P,C)$ can only decrease.
Thus Properties~\eqref{prop:balancing-imbalanced-mass}
and~\eqref{prop:balancing-monotonicity} hold for every choice of
$\alpha$. It remains to choose $\alpha$ so that the
expected cost after a hit is at most $\alpha\phi(P)$.

Let $Z$ be the sampled center. We bound the expected
cost after adding $Z$ by the expected cost of assigning all
of $P$ to $Z$.
For squared Euclidean distances, the centroid identity
expresses this assignment cost as $\phi(P)+n(d_Z^*)^2$, reducing
the problem to a bound on $\E[(d_Z^*)^2]$.
For the cases considered here, we express the same
assignment cost using a uniform point $Y\in P$, sampled
independently of $Z$:
\begin{align}
    \E[\cost(P,\{Z\})]
    &=\sum_{x\in P}\E[d(x,Z)^\ell]
      =n\E[d(Y,Z)^\ell].
    \label{eq:power-uniform-demand}
\end{align}
Thus we need to bound the expected cost between a
uniform point $Y$ and the sampled center $Z$.
As an intermediate step,
Lemma~\ref{lem:pair-sampling-cost} below proves the pairwise bound
$\E[d(Y,Z)^\ell]\le\beta_\ell\E[(d_Y^*)^\ell]$
when $Y$ and $Z$ are independent samples from the \emph{same}
distribution, where $\beta_\ell=2^\ell$ in a general
metric for $\ell>1$, and $\beta_\ell=2^{\ell-1}$ in Euclidean
space for $\ell\ge2$.

If $Z$ is uniform, we can apply the pairwise bound
directly, since $Y$ and $Z$ are independent uniform points of
$P$. As $\E[(d_Y^*)^\ell]=\phi(P)/n$, this gives
$n\E[d(Y,Z)^\ell]\le\beta_\ell\phi(P)$.
Conditioned on a hit, however, $Z$ has the balanced
sampling distribution, so the pairwise bound in
Lemma~\ref{lem:pair-sampling-cost} does not apply directly
to the full cost.
We split the cost into a part that can be handled by this bound
and a remainder that we bound using the triangle inequality.
We show below that together they give the constants in
Lemma~\ref{lem:balancing-extensions}.

\paragraph{The cost between two sampled points.}
We first prove a pairwise bound on $\E[d(Y,Z)^\ell]$
for two independent samples $Y,Z$ from a common distribution. We will apply it to the uniform
distribution and to the balanced sampling distribution.

\begin{lemma}[Pairwise bound]
\label{lem:pair-sampling-cost}
Let $Y$ and $Z$ be independent points drawn from the same
distribution on $P$. For $\ell>1$,
\begin{align}
    \E[d(Y,Z)^\ell]
    &\le \beta_\ell\,\E[(d_Y^*)^\ell],
    \label{eq:pair-sampling-cost}
\end{align}
where $\beta_\ell=2^\ell$ in a general metric, and
$\beta_\ell=2^{\ell-1}$ in Euclidean space for $\ell\ge2$.
\end{lemma}

\begin{proof}
For a general metric, we use the argument
of~\cite[Lemma~5.1]{AV2007}, which proves the uniform-sampling
case. The same argument applies to any common distribution:
we bound the distance from $Y$ to $Z$ by
the path through $c^*$, whose length is $d_Y^*+d_Z^*$.
Convexity gives $((a+b)/2)^\ell\le(a^\ell+b^\ell)/2$ for
$a,b\ge0$, or equivalently
$(a+b)^\ell\le2^{\ell-1}(a^\ell+b^\ell)$. Hence
\begin{align*}
    \E[d(Y,Z)^\ell]
    &\le 2^{\ell-1}
        \bigl(\E[(d_Y^*)^\ell]+\E[(d_Z^*)^\ell]\bigr) =2^\ell\,\E[(d_Y^*)^\ell].
\end{align*}
The equality holds because $Y$ and $Z$ have the same
distribution.

For Euclidean distances with $\ell\ge2$,
Lemma~\ref{lem:euclidean-pair-sampling} below gives
$\E[d(Y,Z)^\ell]\le2^{\ell-1}\E[(d_Y^*)^\ell]$.
\end{proof}

As in Lemma~\ref{lem:euclidean-balancing}, we use a
cap bound on $n\tau^\ell/B$. For squared distances we reuse
that lemma's bound; for the other exponents we use the following
lemma. We prove it after completing the proof of
Lemma~\ref{lem:balancing-extensions}.

\begin{lemma}
\label{lem:power-cap}
For $\ell>1$, if $B>0$ and the imbalanced mass of $P$,
$\cost(P,C)-B$, is at least $3^{\ell-1}\phi(P)>0$, then
$n\tau^\ell/B\le1+2^{\ell-1}$.
\end{lemma}

For $\phi(P)>0$ and $\alpha\ge3^{\ell-1}$, the balancing
construction meets the lemma's imbalanced-mass condition whenever
$B>0$: it removes exactly $\alpha\phi(P)$ mass.
If it removes less than $\alpha\phi(P)$, all sampling mass
is imbalanced, so $B=0$ and no hit is possible.

\begin{proof}[Proof of Lemma~\ref{lem:balancing-extensions}]
We have already verified the imbalanced-mass bound and
monotonicity. If $\phi(P)=0$, all points of $P$ coincide,
so the cost after a hit is zero. Assume henceforth that
$\phi(P)>0$.

Condition on the state $\mathcal S$ and a hit of $P$, and let $Z$
be the sampled point. Then $B>0$ and
$\Pr[Z=z\mid\mathcal S,\mathrm{hit\ of\ }P]=b(z)/B$.
All expectations in the rest of the hit-cost argument are under
this conditioning. As before, assigning all points of $P$ to $Z$
gives $\cost(P,C_{\rm new})\le\cost(P,\{Z\})$.

In~\eqref{eq:power-uniform-demand}, $Y$ is uniform
so that every point's cost is counted equally. But $Z$ now
has the balanced probabilities $b(z)/B$, which need not be
uniform.
We therefore cannot apply
Lemma~\ref{lem:pair-sampling-cost} directly to the full cost:
it requires $Y$ and $Z$ to have the same distribution.

To apply the lemma, choose $Y$ independently of $Z$
with the same balanced probabilities $b(x)/B$. For a fixed
center $z$, the expectation over $Y$ is the weighted sum
$\E_Y[d(Y,z)^\ell]=\sum_{x\in P}(b(x)/B)d(x,z)^\ell$.
We scale this weighted sum up to account for as much
of the full cost as possible. Multiplying by $B/\tau^\ell$ gives
$(B/\tau^\ell)\E_Y[d(Y,z)^\ell]
=\sum_{x\in P}(b(x)/\tau^\ell)d(x,z)^\ell$.
Thus a point at the cap contributes its full cost,
and every other point contributes the fraction $b(x)/\tau^\ell$.
We call this its \emph{balanced fraction} and call
$1-b(x)/\tau^\ell$ its \emph{remaining fraction}
(these are fractions of the cost in this proof, rather than
probabilities of declaring a balanced hit).
Summing over all points gives
\begin{align}
    \cost(P,\{Z\})
    &=\sum_{x\in P}\frac{b(x)}{\tau^\ell}d(x,Z)^\ell
      +\sum_{x\in P}\left(1-\frac{b(x)}{\tau^\ell}\right)d(x,Z)^\ell.
    \label{eq:power-cost-split}
\end{align}
The first sum is exactly
$(B/\tau^\ell)\cdot\E_Y[d(Y,Z)^\ell]$ for each fixed $Z$,
so its expectation is the part we can bound with
Lemma~\ref{lem:pair-sampling-cost}.
The second sum contains the remaining fractions.
Both sums are nonnegative, and together they include every
point's full cost.

In the optimal solution, assigning $x$ to $c^*$ costs
$(d_x^*)^\ell$. We take the same fraction $b(x)/\tau^\ell$
of this assignment cost and write
$\phi_b=\sum_{x\in P}(b(x)/\tau^\ell)(d_x^*)^\ell$
for the sum of these contributions.
The remaining fraction of each optimal assignment
cost is $(1-b(x)/\tau^\ell)(d_x^*)^\ell$.
These contributions sum to $\phi(P)-\phi_b = \sum_{x\in P} (1-b(x)/\tau^\ell)(d_x^*)^\ell$.

We first bound the first sum
in~\eqref{eq:power-cost-split}, which contains the balanced
fractions of the assignment costs to $Z$.
Taking the expectation over $Z$ and applying
Lemma~\ref{lem:pair-sampling-cost} gives
\begin{align}
\begin{split}
    \E_Z\left[\sum_{x\in P}\frac{b(x)}{\tau^\ell}d(x,Z)^\ell\right]
    &=\frac{B}{\tau^\ell}\E_Z\left[\E_Y[d(Y,Z)^\ell]\right]\\
    &=\frac{B}{\tau^\ell}\E_{Y,Z}[d(Y,Z)^\ell]\\
    &\le\beta_\ell\frac{B}{\tau^\ell}\E_Y[(d_Y^*)^\ell]
      =\beta_\ell\frac{B}{\tau^\ell}
         \sum_{x\in P}\frac{b(x)}{B}(d_x^*)^\ell\\
    &=\beta_\ell\sum_{x\in P}\frac{b(x)}{\tau^\ell}(d_x^*)^\ell
      =\beta_\ell\phi_b.
\end{split}
\label{eq:power-first-part}
\end{align}
Thus the balanced fractions of the assignment costs
to $Z$ have expected sum at most $\beta_\ell\phi_b$, where
$\phi_b$ is the sum of the same fractions of the assignment
costs to $c^*$.

For the second sum in~\eqref{eq:power-cost-split},
the triangle inequality gives
\begin{align*}
    \E\left[\sum_{x\in P}\left(1-\frac{b(x)}{\tau^\ell}\right)d(x,Z)^\ell\right]
    &\le\sum_{x\in P}\left(1-\frac{b(x)}{\tau^\ell}\right)
         \E\left[(d_x^*+d_Z^*)^\ell\right].
\end{align*}
Using the symmetric bound from the proof of
Lemma~\ref{lem:pair-sampling-cost},
$(d_x^*+d_Z^*)^\ell\le2^{\ell-1}((d_x^*)^\ell+(d_Z^*)^\ell)$,
would give the same coefficient $2^{\ell-1}$ to both terms.
We allow different coefficients, since the first term is summed
over the remaining fractions, whereas the second is averaged
over the balanced sample $Z$.
For $t>1$, the same convexity argument, now with weights
$1/t$ and $1-1/t$, gives
\begin{align}
    d(x,z)^\ell
    &\le(d_x^*+d_z^*)^\ell
     \le t^{\ell-1}(d_x^*)^\ell
        +\left(\frac{t}{t-1}\right)^{\ell-1}(d_z^*)^\ell.
    \label{eq:power-weighted-triangle}
\end{align}
Here $d_x^*+d_z^*$ is the weighted average of
$t d_x^*$ and $t d_z^*/(t-1)$.
The choice $t=2$ gives the coefficient $2^{\ell-1}$ on each
term. Increasing $t$ increases the coefficient of $(d_x^*)^\ell$
and decreases that of $(d_z^*)^\ell$. We choose $t$ below.

Applying~\eqref{eq:power-weighted-triangle} to the
remaining fractions and taking the expectation over $Z$ gives
\begin{align}
    \E\left[\sum_{x\in P}\left(1-\frac{b(x)}{\tau^\ell}\right)d(x,Z)^\ell\right]
    &\le t^{\ell-1}\sum_{x\in P}\left(1-\frac{b(x)}{\tau^\ell}\right)(d_x^*)^\ell +\left(\frac{t}{t-1}\right)^{\ell-1}
        \sum_{x\in P}\left(1-\frac{b(x)}{\tau^\ell}\right)\E[(d_Z^*)^\ell] \notag \\
    &=t^{\ell-1}(\phi(P)-\phi_b)
      +\left(\frac{t}{t-1}\right)^{\ell-1}
        \left(n-\frac{B}{\tau^\ell}\right)\E[(d_Z^*)^\ell].
\label{eq:power-remaining-part}
\end{align}
The equality uses the definition of $\phi_b$ and
$\sum_x b(x)=B$.

Combining~\eqref{eq:power-first-part}
and~\eqref{eq:power-remaining-part} in the cost split
\eqref{eq:power-cost-split}, we obtain
\begin{align}
\begin{split}
    \E[\cost(P,C_{\rm new})]
    &\le\beta_\ell\phi_b+t^{\ell-1}(\phi(P)-\phi_b)+\left(\frac{t}{t-1}\right)^{\ell-1}
        \left(n-\frac{B}{\tau^\ell}\right)\E[(d_Z^*)^\ell].
\end{split}
\label{eq:power-hit-before-substitution}
\end{align}
We now express the remaining expectation $\E[(d_Z^*)^\ell]$ in terms of
$\phi_b$. By the balanced sampling probabilities,
\begin{align*}
    \E[(d_Z^*)^\ell]
    &=\sum_{x\in P}\frac{b(x)}{B}(d_x^*)^\ell
      =\frac{\tau^\ell}{B}
         \sum_{x\in P}\frac{b(x)}{\tau^\ell}(d_x^*)^\ell
      =\frac{\tau^\ell}{B}\phi_b.
\end{align*}
Using this in~\eqref{eq:power-hit-before-substitution}, we obtain
\begin{align}
\begin{split}
    \E[\cost(P,C_{\rm new})]
    &\le t^{\ell-1}(\phi(P)-\phi_b)
       +\left[\beta_\ell+
          \left(\frac{t}{t-1}\right)^{\ell-1}
          \left(n-\frac{B}{\tau^\ell}\right)
          \frac{\tau^\ell}{B}\right]\phi_b\\
    &=t^{\ell-1}(\phi(P)-\phi_b)
       +\left[\beta_\ell+
          \left(\frac{t}{t-1}\right)^{\ell-1}
          \left(\frac{n\tau^\ell}{B}-1\right)\right]\phi_b.
\end{split}
    \label{eq:power-hit-cost}
\end{align}
The quantities $\phi_b$ and $\phi(P)-\phi_b$ are the
two parts of the optimal assignment cost: each point contributes
$(b(x)/\tau^\ell)(d_x^*)^\ell$ to the first and
$(1-b(x)/\tau^\ell)(d_x^*)^\ell$ to the second.
Both parts are nonnegative and sum to $\phi(P)$.
Consequently, if both coefficients in~\eqref{eq:power-hit-cost}
are at most $\alpha$, the whole bound is at most
$\alpha(\phi(P)-\phi_b)+\alpha\phi_b=\alpha\phi(P)$.

It remains to choose $\alpha$ and $t$ using the
cap bounds: Lemma~\ref{lem:euclidean-balancing} for squared
distances and Lemma~\ref{lem:power-cap} for all other
exponents $\ell>1$. 

\paragraph{Squared distances in a general metric.}
We use the cap bound $n\tau^2/B\le\alpha/(\alpha-1)$
already proved in Lemma~\ref{lem:euclidean-balancing}.
Every step in the proof of that bound also applies
in a general metric; in particular, it does not use the
centroid identity.
Here $\beta_2=4$. Taking $t=\alpha$ in
\eqref{eq:power-hit-cost} gives
\begin{align*}
    \E[\cost(P,C_{\rm new})]
    &\le \alpha(\phi(P)-\phi_b)
        +\left(4+\frac{\alpha}{(\alpha-1)^2}\right)\phi_b.
\end{align*}
Choosing $\alpha=\alpha_2$, where
$\alpha_2=4+\alpha_2/(\alpha_2-1)^2$, makes both coefficients
equal to $\alpha_2$. The left-hand side of this defining equation
increases with $\alpha_2$, while the right-hand side decreases
for $\alpha_2>1$, so there is a unique solution greater than $4$.
Equivalently, $\alpha_2^3-6\alpha_2^2+8\alpha_2-4=0$,
whose solution is
$\alpha_2=2+\sqrt[3]{2+2\sqrt{33}/9}
             +\sqrt[3]{2-2\sqrt{33}/9}\approx4.383$.
This proves the hit-cost bound for squared metrics.

\paragraph{General powers.}
For the other exponents $\ell>1$, we choose
$\alpha=\alpha_\ell=3^{\ell-1}+\beta_\ell$.
Since $\alpha_\ell\ge3^{\ell-1}$, the cap bound in
Lemma~\ref{lem:power-cap} gives $n\tau^\ell/B-1\le2^{\ell-1}$.
We take $t=3$ in~\eqref{eq:power-hit-cost}.
The extra term in the coefficient of $\phi_b$ is then at most
$(3/2)^{\ell-1}2^{\ell-1}=3^{\ell-1}$.
Hence
\begin{align*}
    \E[\cost(P,C_{\rm new})]
    &\le3^{\ell-1}(\phi(P)-\phi_b)
        +(\beta_\ell+3^{\ell-1})\phi_b\\
    &=3^{\ell-1}\phi(P)+\beta_\ell\phi_b
      \le(3^{\ell-1}+\beta_\ell)\phi(P).
\end{align*}
This is the required bound $\alpha_\ell\phi(P)$.
For $\ell>2$, we substitute $\beta_\ell=2^{\ell-1}$
in Euclidean space and $\beta_\ell=2^\ell$ in a general metric.
For $1<\ell<2$, the general-metric choice $\beta_\ell=2^\ell$
applies in both settings.
\end{proof}

\begin{proof}[Proof of Lemma~\ref{lem:power-cap}]
Suppose for a contradiction that
$B<n\tau^\ell/(1+2^{\ell-1})$.
Call $F=\{y\in P:D(y)>\tau\}$ the far points and
$N=P\setminus F$ the near points.
All imbalanced mass comes from $F$, so $F$ is nonempty.
Each far point retains $\tau^\ell$, and hence
$|F|\tau^\ell\le B<n\tau^\ell/(1+2^{\ell-1})$.
Rearranging gives
$(1+2^{\ell-1})|F|<n=|F|+|N|$, and hence
$2^{\ell-1}|F|<|N|$. In particular, $N$ is nonempty and
$|F|/|N|<1/2^{\ell-1}$.

Next consider the average sampling mass of the near
points. Each near point retains all its mass $D(x)^\ell$,
which is at most $\tau^\ell$. Adding the far points, each
with retained mass $\tau^\ell$, can only increase this average.
The average retained mass over all points is $B/n$, so
\begin{align}
    \frac{1}{|N|}\sum_{x\in N}D(x)^\ell
    &\le\frac{B}{n}
      <\frac{\tau^\ell}{1+2^{\ell-1}}
      <\frac{\tau^\ell}{2^{\ell-1}}.
    \label{eq:power-cap-near-mass}
\end{align}

The first strict inequality uses our assumption
$B<n\tau^\ell/(1+2^{\ell-1})$.

Fix a far point $y\in F$. For every near point $x\in N$,
the path from $y$ through $c^*$ and $x$ to a nearest current
center of $x$ gives $D(y)\le d_y^*+d_x^*+D(x)$.
To bound $(d_y^*+d_x^*+D(x))^\ell$, we apply
the convexity inequality from~\eqref{eq:power-weighted-triangle}
twice:
\begin{align*}
    D(y)^\ell
    &\le\bigl(d_y^*+d_x^*+D(x)\bigr)^\ell\\
    &\le2^{\ell-1}D(x)^\ell
       +2^{\ell-1}(d_y^*+d_x^*)^\ell\\
    &\le2^{\ell-1}D(x)^\ell
       +2^{\ell-1}\left[
          (3/2)^{\ell-1}(d_y^*)^\ell
          +3^{\ell-1}(d_x^*)^\ell\right]\\
    &=3^{\ell-1}(d_y^*)^\ell
       +6^{\ell-1}(d_x^*)^\ell+2^{\ell-1}D(x)^\ell.
\end{align*}
The first use of convexity takes $t=2$ to separate
$D(x)$ from $d_y^*+d_x^*$; the second takes $t=3/2$ to split
$d_y^*+d_x^*$.
Averaging over the near points gives
\begin{align*}
    D(y)^\ell
    &\le3^{\ell-1}(d_y^*)^\ell
       +\frac{6^{\ell-1}}{|N|}\sum_{x\in N}(d_x^*)^\ell
       +\frac{2^{\ell-1}}{|N|}\sum_{x\in N}D(x)^\ell\\
    &<3^{\ell-1}(d_y^*)^\ell
       +\frac{6^{\ell-1}}{|N|}\sum_{x\in N}(d_x^*)^\ell
       +\tau^\ell.
\end{align*}
The last inequality uses~\eqref{eq:power-cap-near-mass}.
Consequently, the imbalanced mass at $y$ is
$D(y)^\ell-\tau^\ell
<3^{\ell-1}(d_y^*)^\ell
 +(6^{\ell-1}/|N|)\sum_{x\in N}(d_x^*)^\ell$.
Near points contribute no imbalanced mass, so summing
over the far points bounds the total imbalanced mass:
\begin{align*}
    \sum_{y\in F}\bigl(D(y)^\ell-\tau^\ell\bigr)
    &<3^{\ell-1}\sum_{y\in F}(d_y^*)^\ell
       +6^{\ell-1}\frac{|F|}{|N|}\sum_{x\in N}(d_x^*)^\ell\\
    &\le3^{\ell-1}\sum_{y\in F}(d_y^*)^\ell
       +3^{\ell-1}\sum_{x\in N}(d_x^*)^\ell
       =3^{\ell-1}\phi(P).
\end{align*}
The second inequality uses
$6^{\ell-1}|F|/|N|\le3^{\ell-1}$, since
$|F|/|N|<1/2^{\ell-1}$.
This contradicts the assumption that the cap removes at least
$3^{\ell-1}\phi(P)$.
\end{proof}

\paragraph{The Euclidean pair bound.}
We obtain the Euclidean bound from the weighted-pair
inequality in~\cite[Theorem~5.3(c)]{Ajiev2013}.
We state the case we need for Euclidean vectors and then
apply it to our sampling distribution.

\begin{lemma}
\label{lem:euclidean-pair-sampling}
Let $Y$ and $Z$ be independent samples from the same
distribution on the Euclidean cluster $P$. For $\ell\ge2$,
\begin{align*}
    \E[\|Y-Z\|^\ell]
    &\le2^{\ell-1}\E[(d_Y^*)^\ell].
\end{align*}
\end{lemma}

\begin{proof}
The Euclidean case of~\cite[Theorem~5.3(c)]{Ajiev2013}
states that, for any vectors $x_1,\ldots,x_n\in\mathbb R^d$,
nonnegative weights $\alpha_1,\ldots,\alpha_n$ with
$\sum_i\alpha_i=1$, and real exponent $\sigma\ge2$,
\begin{align}
    \left(\sum_{i,j=1}^n\alpha_i\alpha_j
                  \|x_i-x_j\|^\sigma\right)^{1/\sigma}
    &\le2^{1/\sigma'}
        \left(\sum_{i=1}^n\alpha_i\|x_i\|^\sigma\right)^{1/\sigma}.
    \label{eq:ajiev-pair-bound}
\end{align}
Here $\|\cdot\|$ is the usual Euclidean norm, and
$\sigma'=\sigma/(\sigma-1)$ is the conjugate exponent.\footnote{Ajiev states the exponent
condition as $\sigma\ge\max\{p_{\min}',p_{\max}\}$.
The parameters $p_{\min}$ and $p_{\max}$ are the smallest
and largest exponents used to define the norm. For the
Euclidean norm both equal $2$. The prime again denotes the conjugate
exponent, so $p_{\min}'=2/(2-1)=2$.
Thus the condition is  $\sigma\ge2$.}

Write $P=\{z_1,\ldots,z_n\}$ and apply
\eqref{eq:ajiev-pair-bound} with $\sigma=\ell$,
$x_i=z_i-c^*$, and
$\alpha_i=\Pr[Y=z_i]=\Pr[Z=z_i]$.
Substituting and raising both sides to the power $\ell$ gives
\begin{align*}
    \sum_{i,j}\alpha_i\alpha_j\|z_i-z_j\|^\ell
    &=\sum_{i,j}\alpha_i\alpha_j
        \|(z_i-c^*)-(z_j-c^*)\|^\ell\\
    &\le2^{\ell-1}\sum_i\alpha_i\|z_i-c^*\|^\ell\\
    &=2^{\ell-1}\sum_{x\in P}\Pr[Y=x](d_x^*)^\ell
      =2^{\ell-1}\E[(d_Y^*)^\ell].
\end{align*}
Here $(2^{1/\sigma'})^\ell=2^{\ell-1}$, since $\sigma=\ell$. 
By the independence of $Y$ and $Z$, the left-hand side is
$\E[\|Y-Z\|^\ell]$, which proves the bound.
\end{proof}

\begin{remark}[Optimal balancing]
\label{rem:optimal-balancing-discussion}
For a fixed removal budget, the balancing that minimizes
the expected cost after a hit is simple to describe. For a parameter
$\alpha$, remove $\min\{\alpha\phi(P),\cost(P,C)\}$ units of sampling
mass in decreasing order of $\cost(P,C\cup\{x\})$, the cluster cost
resulting from selecting $x$, splitting the last point's mass if
necessary. For any fixed remaining mass $B>0$, this rule minimizes
the conditional expected cost by retaining the cheapest outcomes.
Its total balanced mass is
$\max\{\cost(P,C)-\alpha\phi(P),0\}$, so it also satisfies
monotonicity.

The difficulty is determining the smallest $\alpha$ for
which the expected cost after a hit is at most $\alpha\phi(P)$ in
every state with positive balanced mass. We do not have a direct
tight analysis of this rule in general. For Euclidean $k$-means,
the simpler capping analysis in Lemma~\ref{lem:euclidean-balancing}
already attains the balancing constant $\alpha=1+\varphi$, which is optimal by
Lemma~\ref{lem:euclidean-balancing-tight}.
We therefore retain capping, which also gives the extensions in
Lemma~\ref{lem:balancing-extensions}.

For squared distances in general metrics, an analysis
proposed by ChatGPT claims a balancing constant of approximately
$4.162$, together with a matching lower bound within our
balancing framework. The proposed argument relies on a
computer-generated certificate intended to establish a polynomial
inequality through exact-arithmetic checks. We have not independently
verified this certificate, and we do not include the claimed result
or its proof in this paper.
\end{remark}

\subsection{Distance Exponents Below Two}
\label{app:small-exponents}
\label{subsec:small-exponents}

\begin{lemma}
\label{lem:balancing-small-exponents}
For $0<\ell\le1$,
Properties~\eqref{prop:balancing-hit-cost}--\eqref{prop:balancing-monotonicity}
hold with $\alpha_\ell=2$ in both general metrics and
Euclidean space. For $1<\ell<2$, they hold with
$\alpha_\ell=\min\{a,3^{\ell-1}+2^\ell\}$ in general metrics
and $\alpha_\ell=1+\varphi$ in Euclidean space, where
$a\approx4.383$ is the constant for squared distances in
general metrics from Lemma~\ref{lem:balancing-extensions}.
\end{lemma}

\begin{proof}
For $0<\ell\le1$, define $\widetilde d(x,y)=d(x,y)^\ell$.
The triangle inequality and
$(u+v)^\ell\le u^\ell+v^\ell$ show that $\widetilde d$ is
a metric. We view the instance as metric $k$-median under
$\widetilde d$.
Every sampling mass and every assignment cost are unchanged,
including $\phi(P)$ for each optimal cluster $P$.
Thus the balancing construction of Lemma~\ref{lem:D-balancing}
gives the required properties with $\alpha_\ell=2$.

For $1<\ell<2$, define $\widetilde d(x,y)=d(x,y)^{\ell/2}$.
Since $0<\ell/2<1$, the same argument shows that
$\widetilde d$ is a metric. We now use squared distances
under this metric: $\widetilde d(x,y)^2=d(x,y)^\ell$.
Thus every sampling mass and assignment cost equals its
original value, and the sampling probabilities and each
$\phi(P)$ are unchanged.
Applying the squared-metric balancing construction of
Lemma~\ref{lem:balancing-extensions} under $\widetilde d$
therefore gives the constant $a$.
That lemma also gives the constant $3^{\ell-1}+2^\ell$
directly under the original metric $d$ with exponent $\ell$.
We therefore get $\min\{a,3^{\ell-1}+2^\ell\}$.

For the Euclidean case, let $C^*$ be the fixed optimal set of centers.
By Schoenberg's theorem~\cite[Corollary~1]{Schoenberg1938},
there is an embedding $f$ of $X\cup C^*$ into Euclidean space
such that
$\|f(x)-f(y)\|^2=\|x-y\|^\ell$ for all $x,y\in X\cup C^*$.
Thus squared distances in the image give the same sampling
masses and assignment costs as the original instance.

For each optimal cluster $P$ with center $c^*$, the squared-distance
cost of assigning $f(P)$ to its centroid is at most $\phi(P)$, since
$f(c^*)$ already gives cost $\phi(P)$.
The proof of Lemma~\ref{lem:euclidean-balancing} applies to any
cluster using its centroid. Apply that construction to $f(P)$
with $\alpha=1+\varphi$, and assign its balanced weights to the
corresponding points of $P$.
All the required properties therefore transfer from $f(P)$ to
$P$ with $\alpha_\ell=1+\varphi$.
\end{proof}

\begin{remark}[AI assistance]
\label{rem:balancing-ai-assistance}
The initial proofs of the extensions in
Lemmas~\ref{lem:balancing-extensions} and~\ref{lem:balancing-small-exponents}
were produced by ChatGPT and refined through repeated exchanges
with the authors. In particular, ChatGPT suggested using Ajiev's
inequality for the Euclidean pairwise bound and Schoenberg's
embedding theorem for the Euclidean reduction for $1<\ell<2$.
\end{remark}

\section{Tightness of Balancing for Squared Euclidean Distances}
\label{app:euclidean-balancing-tight}

Our example adapts the example of~\cite[Appendix~E.1]{MRS2020}.
It is a Euclidean $2$-means instance on the line with two clusters.
One contains many points at $-t$, where $t\ge1$. The
other, denoted by $P$, contains many points at $0$ and a single
point at $1$. The first center lies at $-t$ with probability
approaching $1$. A hit at $1$ then leaves each point at $0$ with
cost $1$. To keep the expected cost after a hit of $P$ small, we
would like the balancing construction to reduce the balanced mass
assigned to $1$, but the imbalanced mass bound limits how much we can reduce it.

\begin{proof}[Proof of Lemma~\ref{lem:euclidean-balancing-tight}]
Fix $\alpha\ge2$ and set $t=\alpha-1\ge1$. For
$n\ge2$, we construct a Euclidean $2$-means instance with $n^2$ points at
$-t$, $n$ points at $0$, and one point at $1$. Let $P$ consist of
the points at $0$ and $1$. Its centroid is $1/(n+1)$, so $\phi(P)=n/(n+1)^2 + (1- 1/(n+1))^2=n/(n+1)$.

An optimal clustering on the line can be chosen to split the three
locations at one of the two gaps. Separating $-t$ from $P$ costs
$\phi(P)$. The other split groups the points at $-t$ and $0$,
and costs $n^2t^2/(n+1)\ge\phi(P)$. Thus $P$ is an optimal cluster.
The first center is at $-t$ with probability
$n^2/(n^2+n+1)$, which approaches $1$. Condition on this event.
Then $C=\{-t\}$ and $P$ is unsettled.

Consider any choice of balanced weights with imbalanced mass at
most $\alpha\phi(P)$. The point at $1$ has sampling mass
$(t+1)^2$. Even if all the allowed imbalanced mass is removed
from this point, its balanced mass is at least
$(t+1)^2-\alpha\phi(P)\ge(t+1)^2-\alpha
=\alpha(\alpha-1)>0$, where we used $t=\alpha-1$.
Thus the total balanced mass is positive. Condition on a hit of
$P$, and let $Z$ be the sampled point. Since the total balanced
mass is at most $\cost(P,C)=nt^2+(t+1)^2$, we have
\begin{align*}
    \Pr[Z=1]
    &\ge\frac{(t+1)^2-\alpha}{nt^2+(t+1)^2}.
\end{align*} 

If $Z=0$, the cost of $P$ becomes $1$. If $Z=1$, each point at
$0$ has cost $1$, since $t\ge1$, and the cost of $P$ becomes $n$.
Therefore
\begin{align*}
    \frac{\E[\cost(P,C\cup\{Z\})]}{\phi(P)}
    &=\frac{1\cdot\Pr[Z=0]+n\cdot\Pr[Z=1]}{\phi(P)}
     =\frac{1+(n-1)\Pr[Z=1]}{\phi(P)}\\
    &\ge1+(n-1)\frac{(t+1)^2-\alpha}{nt^2+(t+1)^2},
\end{align*}
where we used $\phi(P)<1$ and the lower bound on $\Pr[Z=1]$.
As $n$ grows, the right-hand side approaches
\begin{align*}
    1+\frac{(t+1)^2-\alpha}{t^2}
    &=1+\frac{\alpha^2-\alpha}{(\alpha-1)^2}
    =1+\frac{\alpha}{\alpha-1}
    =2+\frac{1}{\alpha-1},
\end{align*}
where we substituted $t=\alpha-1$. This proves the lemma.
\end{proof}

\paragraph{The exact ratio for two centers.}
\begin{lemma}
\label{lem:euclidean-two-center-exact}
For Euclidean $k$-means with $k=2$ and $\Delta=0$, the
approximation ratio of $D^2$-sampling is $5$.
Consequently, any bound of the form $\alpha(1+H_{k-1})$ that
holds for every $k\ge1$ must have $\alpha\ge5/2$.
\end{lemma}

\begin{proof}
The upper bound follows from~\cite[Lemma~4]{Wei2016} using
the first-hit bound $5$ of~\cite[Lemma~4.1]{MRS2020}.

For the lower bound, set $t=1$ in the construction used to prove
Lemma~\ref{lem:euclidean-balancing-tight}. This gives $n^2$
points at $-1$, $n$ points at $0$, and one point at $1$, with
$\OPT=n/(n+1)$.
The first center is at $-1$ with probability $n^2/(n^2+n+1)$.
Conditioned on this event, the second center is at $0$ with
probability $n/(n+4)$ and at $1$ with probability $4/(n+4)$.
These outcomes give final costs $1$ and $n$, respectively, as
illustrated in Figure~\ref{fig:intro-balancing}. Thus the
conditional expected cost is $5n/(n+4)$, and
\[
    \frac{\E[\cost(C_{\rm final})]}{\OPT}
    \ge \frac{n^2}{n^2+n+1}\cdot\frac{5(n+1)}{n+4},
\]
which approaches $5$ as $n$ grows.
Finally, $1+H_1=2$, so a bound of the stated form requires
$2\alpha\ge5$.
\end{proof}

\section{Lower Bounds for General Distance Exponents}
\label{app:euclidean-power-lower}

\begin{proof}[Proof of Corollary~\ref{cor:metric-squared-lower}]
We first prove the metric bound. Use the construction and the
choices of $\varepsilon$ from the proof of
Theorem~\ref{thm:metric-median-tight}, and let
$\rho=(2\varepsilon)^{1/\ell}/2$.
Replace the distance parameter $\varepsilon$ in the
construction by $\rho$. Since
$0<\rho<1/2$, the modified construction still defines a metric.

Condition on the first center lying in $P$, an event whose
probability is unchanged. Each unsampled point of $P$ has cost
$(2\rho)^\ell=2\varepsilon$, and each unsampled singleton has cost $1$.
Sampling a point makes its cost zero and leaves all other costs
unchanged. Thus the costs and sampling probabilities are
exactly those of $D$-sampling in the original
construction, and the same lower bounds on the algorithm's
expected final cost apply.

What changes is our upper bound on $\OPT$.
Placing centers at $o,y_1,\ldots,y_{k-1}$ gives
$\OPT\le n\rho^\ell=\frac{n\varepsilon}{2^{\ell-1}}$.
Replacing the original upper bound $n\varepsilon$ on $\OPT$
by this bound multiplies the approximation-ratio lower bounds
by $2^{\ell-1}$.
Taking $n$ to infinity gives $2^\ell B_{k,\Delta}$ for both
$\Delta=0$ and $\Delta>0$.

For the Euclidean bound, use the same $n$ and $\rho$.
In $\R^{n+k-1}$, place the points of $P$ at
$x_i=\sqrt2\rho e_i$ for $i=1,\ldots,n$ and the singletons at
$y_j=\sqrt{1-2\rho^2}\,e_{n+j}$ for $j=1,\ldots,k-1$,
where the $e_i$ are the standard unit vectors.
Distinct points of $P$ have distance $2\rho$, each point of
$P$ has distance $1$ from every singleton, and distinct
singletons have distance $\sqrt{2-4\rho^2}>1$.
Thus, conditioned on the first center lying in $P$, the costs
and sampling probabilities are those in the metric construction,
and the same lower bounds on the algorithm's expected final
cost apply.

Placing centers at the origin
and at the singletons gives
$\OPT\le n(\sqrt2\rho)^\ell=2^{\ell/2}n\rho^\ell$.
This is $2^{\ell/2}$ times the upper bound $n\rho^\ell$ used
for the metric construction above. This divides the approximation ratio lower bound above by $2^{\ell/2}$. Thus, we get $\frac{2^\ell}{2^{\ell/2}}B_{k,\Delta}=2^{\ell/2}B_{k,\Delta}$. \end{proof}

\paragraph{Exponential dependence on the distance exponent.}
We now show that the base $3$ in our upper bounds is necessary,
even for two centers on the real line.

\begin{lemma}
\label{lem:large-exponent-lower}
For every integer $\ell\ge1$, the approximation ratio of $D^\ell$-sampling
with $k=2$ and $\Delta=0$ is at least $3^\ell/4$, even for inputs
on the real line.
\end{lemma}

\begin{proof}
Fix an integer $\ell\ge1$. Place $n$ input points at $0$,
$2^\ell$ points at $3/2$, and one point at $3$.
The centers $0$ and $2$ have cost $2^\ell(1/2)^\ell+1=2$,
so $\OPT\le2$.

The first center is $0$ with probability
$n/(n+2^\ell+1)$. Conditioned on this event, the locations
$3/2$ and $3$ each have sampling mass $3^\ell$, so the second
center is $3$ with probability $1/2$. The final cost is then
$2^\ell(3/2)^\ell=3^\ell$. Considering only this event gives
\[
    \frac{\E[\cost(C_{\rm final})]}{\OPT}
    \ge
    \frac{n}{n+2^\ell+1}\cdot\frac12\cdot\frac{3^\ell}{2}
\]
which approaches $3^\ell/4$ as $n$ approaches infinity.
\end{proof}

Since $B_{2,0}=2$, Lemma~\ref{lem:large-exponent-lower} implies
that any bound of the form $\alpha_\ell B_{k,\Delta}$ valid for
all $k,\Delta$ requires $\alpha_\ell = \Omega(3^\ell)$ for every 
$\ell\ge1$. Our coefficients in Lemma~\ref{lem:balancing-extensions}
are $O(3^\ell)$, so their dependence on the distance exponent is
tight up to a constant factor in both general metrics and
Euclidean space.

Together with our upper bounds, Lemma~\ref{lem:large-exponent-lower}
shows that, for $k=2$ and $\Delta=0$, the $\ell$-th root of the
approximation ratio approaches $3$, even on the real line.
This differs from the factor $2$ of the classical farthest-first
algorithm~\cite{Gonzalez1985} because the convergence to $k$-center
is for a fixed input. Here, the input changes with $\ell$: the
$2^\ell$ points at $3/2$ keep its sampling mass equal to that of
the point at $3$. Thus $D^\ell$-sampling still chooses the nearer
location with probability $1/2$ after choosing $0$, however
large $\ell$ is.

\section{Tightness for Euclidean $k$-Median without Extra Centers}
\label{app:euclidean-median-tight}

Recall the metric construction in the proof of
Theorem~\ref{thm:metric-median-tight}. It consists of one large
cluster and $k-1$ singletons. The large cluster contains many points
at distance $\varepsilon$ from its optimal center $o$ and pairwise
distance $2\varepsilon$. This gives the factor $2$, since choosing
any input point of this cluster as a center nearly doubles its cost
compared with choosing $o$ as a center. In Euclidean space, two
points at distance $\varepsilon$ from $o$ can be $2\varepsilon$
apart only if one is the reflection of the other through $o$.
Thus we cannot place more than two distinct points this way.

To recover the factor $2$, we place the large cluster $P$ at two
such locations, with $n$ points at one and one at the other.
Choosing the first location as a center costs $2\varepsilon$,
for serving the single point. A uniform sample from $P$ chooses
this location with probability $n/(n+1)$. With probability
$1/(n+1)$, it chooses the other location and incurs cost
$2n\varepsilon$, for serving the $n$ points. The two outcomes
contribute equally to the expected cost, which is
$4n\varepsilon/(n+1)$, which approaches twice the cost of choosing the first
location. As for the $k-1$ singletons, we place them so that each has cost $1$
until sampled, as in the metric construction of
Theorem~\ref{thm:metric-median-tight}.

\begin{proof}[Proof of Theorem~\ref{thm:euclidean-median-tight}]
Let $n\ge2$ and let
$0<\varepsilon<1/(2n)$ be a small distance, to be chosen later.
In $\R^k$, we place $n$ input points at $x=-\varepsilon e_1$ and
one at $z=\varepsilon e_1$, where $e_1,\ldots,e_k$ are the standard
unit vectors. These $n+1$ points form the large cluster $P$.
We place the singletons at
$y_i=\sqrt{1-\varepsilon^2}\,e_{i+1}$ for $i=1,\ldots,k-1$.
Each singleton is at distance $1$ from both $x$ and $z$, and
distinct singletons are at distance $\sqrt{2(1-\varepsilon^2)}>1$
from one another. Thus, once a point of $P$ is sampled, every
unsampled singleton has cost $1$, and sampling a singleton leaves
the costs of all other points unchanged, just as in the metric
construction of Theorem~\ref{thm:metric-median-tight}.

To upper bound $\OPT$, we place centers at $x,y_1,\ldots,y_{k-1}$.
Only $z$ contributes to the cost, so $\OPT\le2\varepsilon$.

The first center $c_1$ lies in $P$ with probability $(n+1)/(n+k)$.
Condition on $c_1\in P$, and write $w=\cost(P,\{c_1\})$ for the
cost of $P$ after this sample. It is $2\varepsilon$ if $c_1=x$ and
$2n\varepsilon$ if $c_1=z$. Hence $w\le2n\varepsilon<1$, and
\begin{align}
    \E[w\mid c_1\in P]
    &=2\varepsilon\cdot\frac{n}{n+1}
    +2n\varepsilon\cdot\frac{1}{n+1}
    =\frac{4n\varepsilon}{n+1}.
    \label{eq:euclidean-initial-cluster-cost}
\end{align}

We next condition on the specific first center $c_1\in P$, so
that $w$ is fixed. Let $F$ be the event that $P$ is sampled again.
Until then, its cost remains $w$. As in the $\Delta=0$ case of
Theorem~\ref{thm:metric-median-tight}, the probability that all
$k-1$ remaining samples are singletons is
\begin{equation}
    \Pr[\overline F\mid c_1]
    =\prod_{j=1}^{k-1}\frac{j}{j+w}
    \le\frac{1}{1+wH_{k-1}}.
    \label{eq:euclidean-all-singletons}
\end{equation}

On $\overline F$, all singletons are sampled and the final cost
is $w$. On $F$, the second sample from $P$ makes both $x$ and $z$
centers, so $P$ has zero cost. All remaining samples are then
singletons, leaving exactly one unsampled and giving final cost
$1$. Hence
\begin{align*}
    \E[\cost(C_{\rm final})\mid c_1]
    &=w\Pr[\overline F\mid c_1]
    +\Pr[F\mid c_1]\\
    &=1-(1-w)\Pr[\overline F\mid c_1]\\
    &\ge1-\frac{1-w}{1+wH_{k-1}}
    =\frac{w(1+H_{k-1})}{1+wH_{k-1}}
    \ge\frac{w(1+H_{k-1})}{1+2n\varepsilon H_{k-1}}.
\end{align*}
The first inequality uses the bound on
$\Pr[\overline F\mid c_1]$ and $w<1$. The last inequality uses
$w\le2n\varepsilon$ in the denominator.
Averaging over the first center and
using~\eqref{eq:euclidean-initial-cluster-cost} gives
\begin{align*}
    \E[\cost(C_{\rm final})]
    &\ge\Pr[c_1\in P]\cdot
    \E[\cost(C_{\rm final})\mid c_1\in P]\\
    &\ge\Pr[c_1\in P]\cdot\E[w\mid c_1\in P]
    \cdot\frac{1+H_{k-1}}{1+2n\varepsilon H_{k-1}}\\
    &=\frac{n+1}{n+k}\cdot\frac{4n\varepsilon}{n+1}
    \cdot\frac{1+H_{k-1}}{1+2n\varepsilon H_{k-1}}
    =\frac{4n\varepsilon}{n+k}
    \cdot\frac{1+H_{k-1}}{1+2n\varepsilon H_{k-1}}.
\end{align*}
Finally, choose $\varepsilon=1/(2n^2)$. Using
$\OPT\le2\varepsilon$, we obtain
\begin{align*}
    \frac{\E[\cost(C_{\rm final})]}{\OPT}
    &\ge\frac{2n}{n+k}\cdot\frac{1+H_{k-1}}{1+H_{k-1}/n},
\end{align*}
which approaches $2B_{k,0}$ as $n$ approaches infinity.
\end{proof}

\begin{proof}[Proof of Corollary~\ref{cor:euclidean-power-no-extra}]
We first prove the Euclidean lower bound $2B_{k,0}$ for every
$\ell>0$.
Use the construction from the proof of
Theorem~\ref{thm:euclidean-median-tight}, replacing its distance parameter
$\varepsilon=1/(2n^2)$ by
$\rho=1/(2n^{2/\ell})$.
Then $d(x,z)^\ell=(2\rho)^\ell=1/n^2=2\varepsilon$, so
serving one location from the other has the same cost $2\varepsilon$ as in
the original construction. Each singleton remains at
distance $1$ from both locations, and distinct singletons have
distance greater than $1$.
The probability that the first center lies in $P$ is unchanged.
Conditioned on this event, all costs and sampling probabilities
are those in the original proof, so the same lower bound on
the algorithm's expected final cost applies.
Placing centers at $x,y_1,\ldots,y_{k-1}$ gives
$\OPT\le(2\rho)^\ell=1/n^2=2\varepsilon$, as before, yielding the same
approximation-ratio lower bound $2B_{k,0}$.

We now prove the lower bound $2^{\ell-1}B_{k,0}$ using the
same construction with one change: we place $n$ points at
each of $x$ and $z$, keeping all locations unchanged.
The first center lies in $P$ with
probability $2n/(2n+k-1)$. Condition on this event. The cost
of $P$ is $n(2\rho)^\ell=1/n$, regardless of which location
is sampled, and stays unchanged until $P$ is sampled again.
Applying~\eqref{eq:euclidean-all-singletons} from the proof of
Theorem~\ref{thm:euclidean-median-tight} with $w=1/n$ shows
that the probability of sampling $P$ again is at least
$H_{k-1}/(n+H_{k-1})$.
If this happens, the final cost is $1$, since exactly one
singleton remains unsampled. Otherwise, the final cost is
the cost of $P$, namely $n(2\rho)^\ell=1/n$.
Thus the conditional expected final cost is at least
$\frac1n+(1-\frac1n)\frac{H_{k-1}}{n+H_{k-1}}
=\frac{B_{k,0}}{n+H_{k-1}}$,
where we used $B_{k,0}=1+H_{k-1}$.
Placing centers at the midpoint of $x$ and $z$ and at the
singletons gives $\OPT\le2n\rho^\ell=\frac{1}{2^{\ell-1}n}$.
Taking expectation over the first sample and dividing by this
upper bound gives
\[
    \frac{\E[\cost(C_{\rm final})]}{\OPT}
    \ge \frac{2n}{2n+k-1}\cdot
        \frac{2^{\ell-1}B_{k,0}}{1+H_{k-1}/n},
\]
which approaches $2^{\ell-1}B_{k,0}$ as $n$ approaches infinity.

It remains to prove the metric lower bound. Every Euclidean
instance is also a metric instance, so combining the lower
bound $2B_{k,0}$ proved above with the
bound $2^\ell B_{k,0}$ from
Corollary~\ref{cor:metric-squared-lower} gives the metric
lower bound $\max\{2,2^\ell\}B_{k,0}$.
\end{proof}

\par\medskip
\phantomsection\label{sec:ai-methodology}
\noindent\textbf{AI methodology.}
We used ChatGPT, with several models from the GPT-5 family and GPT-6 Astra,
throughout the research process to explore ideas, develop proofs,
and improve the exposition.
The balancing framework and the comparison with an
equal-mass process originated with the authors. The analysis of the sampling process
(Section~\ref{sec:sampling-analysis}) evolved
through several improvements and reformulations in repeated exchanges
with ChatGPT.
ChatGPT produced the initial proofs of the balancing extensions in
Sections~\ref{app:balancing-powers} and~\ref{app:small-exponents}
(see Remark~\ref{rem:balancing-ai-assistance}).
All proofs in the paper were refined through repeated exchanges
between the authors and ChatGPT. The authors verified the correctness
of the results and proofs presented in this paper and take responsibility
for all its content.

\newcommand{\etalchar}[1]{$^{#1}$}


\begin{thebibliography}{BCAF{\etalchar{+}}26b}

\bibitem[ADK09]{ADK2009}
Ankit Aggarwal, Amit Deshpande, and Ravi Kannan.
\newblock Adaptive sampling for {$k$-Means} clustering.
\newblock In {\em Approximation, Randomization, and Combinatorial Optimization.
  Algorithms and Techniques (APPROX/RANDOM)}, volume 5687 of {\em Lecture Notes
  in Computer Science (LNCS)}, pages 15--28, 2009.

\bibitem[Aji13]{Ajiev2013}
Sergey Ajiev.
\newblock Quantitative {Hahn--Banach} theorems and isometric extensions for
  wavelet and other {Banach} spaces.
\newblock {\em Axioms}, 2(2):224--270, 2013.

\bibitem[AJM09]{AJM2009}
Nir Ailon, Ragesh Jaiswal, and Claire Monteleoni.
\newblock Streaming {$k$-means} approximation.
\newblock In {\em Advances in Neural Information Processing Systems (NeurIPS)},
  volume~22, pages 10--18, 2009.

\bibitem[{Apa}]{SparkClustering}
{Apache Software Foundation}.
\newblock Clustering: {Apache Spark MLlib} documentation.
\newblock \url{https://spark.apache.org/docs/latest/ml-clustering.html}.
\newblock Accessed: 2026-10-23.

\bibitem[AV07]{AV2007}
David Arthur and Sergei Vassilvitskii.
\newblock {$k$-means++}: the advantages of careful seeding.
\newblock In {\em Proceedings of the Eighteenth Annual ACM-SIAM Symposium on
  Discrete Algorithms (SODA)}, pages 1027--1035, 2007.

\bibitem[BCAF{\etalchar{+}}26a]{BCFGJPSSV2026a}
Rares-Darius Buhai, Vincent Cohen-Addad, Henry Fleischmann, Anupam Gupta,
  Antoni Jub{\'e}s~Monforte, Madhusudhan~Reddy Pittu, Miltiadis Stouras, Ola
  Svensson, and Lukas Vogl.
\newblock A better-than-2 approximation for {$k$}-median {Part I}: The power of
  general bids.
\newblock arXiv:2610.18144, 2026.

\bibitem[BCAF{\etalchar{+}}26b]{BCFGJPSSV2026b}
Rares-Darius Buhai, Vincent Cohen-Addad, Henry Fleischmann, Anupam Gupta,
  Antoni Jub{\'e}s~Monforte, Madhusudhan~Reddy Pittu, Miltiadis Stouras, Ola
  Svensson, and Lukas Vogl.
\newblock A better-than-2 approximation for {$k$}-median {Part II}: Stable
  instances.
\newblock arXiv:2610.17172, 2026.

\bibitem[BCALP23]{BCLP2023}
Lorenzo Beretta, Vincent Cohen-Addad, Silvio Lattanzi, and Nikos Parotsidis.
\newblock Multi-swap {$k$}-means++.
\newblock In {\em Advances in Neural Information Processing Systems (NeurIPS)},
  volume~36, pages 26069--26091, 2023.

\bibitem[BERS20]{BERS2020}
Anup Bhattacharya, Jan Eube, Heiko R{\"o}glin, and Melanie Schmidt.
\newblock Noisy, greedy and not so greedy {$k$-Means++}.
\newblock In {\em 28th Annual European Symposium on Algorithms (ESA)}, volume
  173 of {\em Leibniz International Proceedings in Informatics (LIPIcs)}, pages
  18:1--18:21, 2020.

\bibitem[BGH{\etalchar{+}}26]{BGHLWW2026}
Jaros{\l}aw Byrka, Yuhao Guo, Yang Hu, Shi Li, Chengzhang Wan, and Zaixuan
  Wang.
\newblock {$k$}-clustering via iterative randomized rounding.
\newblock arXiv:2604.06046, 2026.
\newblock To appear in FOCS 2026.

\bibitem[BJA16]{BJA2016}
Anup Bhattacharya, Ragesh Jaiswal, and Nir Ailon.
\newblock Tight lower bound instances for {$k$-means++} in two dimensions.
\newblock {\em Theoretical Computer Science (TCS)}, 634:55--66, 2016.

\bibitem[BLHK16]{BLHK2016}
Olivier Bachem, Mario Lucic, S.~Hamed Hassani, and Andreas Krause.
\newblock Fast and provably good seedings for {$k$-Means}.
\newblock In {\em Advances in Neural Information Processing Systems (NeurIPS)},
  volume~29, pages 55--63, 2016.

\bibitem[BLK17]{BLK2017}
Olivier Bachem, Mario Lucic, and Andreas Krause.
\newblock Distributed and provably good seedings for {$k$-Means} in constant
  rounds.
\newblock In {\em Proceedings of the 34th International Conference on Machine
  Learning (ICML)}, volume~70 of {\em Proceedings of Machine Learning Research
  (PMLR)}, pages 292--300, 2017.

\bibitem[BLSS16]{BLSS2016}
Johannes Bl{\"o}mer, Christiane Lammersen, Melanie Schmidt, and Christian
  Sohler.
\newblock Theoretical analysis of the {$k$}-means algorithm---a survey.
\newblock In Lasse Kliemann and Peter Sanders, editors, {\em Algorithm
  Engineering: Selected Results and Surveys}, volume 9220 of {\em Lecture Notes
  in Computer Science}, pages 81--116. Springer, 2016.

\bibitem[BMV{\etalchar{+}}12]{BMVKV2012}
Bahman Bahmani, Benjamin Moseley, Andrea Vattani, Ravi Kumar, and Sergei
  Vassilvitskii.
\newblock Scalable {$k$-means++}.
\newblock {\em Proceedings of the VLDB Endowment (PVLDB)}, 5(7):622--633, 2012.

\bibitem[BNS24]{BNS2024}
Etienne Bamas, Sai~Ganesh Nagarajan, and Ola Svensson.
\newblock Analyzing {$D^\alpha$} seeding for {$k$}-means.
\newblock In {\em Proceedings of the 41st International Conference on Machine
  Learning (ICML)}, volume 235 of {\em Proceedings of Machine Learning Research
  (PMLR)}, pages 2673--2699, 2024.

\bibitem[BR13]{BR2013}
Tobias Brunsch and Heiko R{\"o}glin.
\newblock A bad instance for {$k$-means++}.
\newblock {\em Theoretical Computer Science (TCS)}, 505:19--26, 2013.

\bibitem[CGPR20]{CGPR2020}
Davin Choo, Christoph Grunau, Julian Portmann, and V\'aclav Rozho\v{n}.
\newblock {$k$-means++}: few more steps yield constant approximation.
\newblock In {\em Proceedings of the 37th International Conference on Machine
  Learning (ICML)}, volume 119 of {\em Proceedings of Machine Learning Research
  (PMLR)}, pages 1909--1917, 2020.

\bibitem[Das13]{DasguptaNotes}
Sanjoy Dasgupta.
\newblock Lecture 3 --- algorithms for {$k$-means} clustering.
\newblock University of California, San Diego,
  \url{https://cseweb.ucsd.edu/~dasgupta/291-geom/kmeans.pdf}, 2013.
\newblock Accessed: 2025-11-27.

\bibitem[FLL23]{FLL2023}
Chenglin Fan, Ping Li, and Xiaoyun Li.
\newblock {$k$}-median clustering via metric embedding: Towards better
  initialization with differential privacy.
\newblock In {\em Advances in Neural Information Processing Systems (NeurIPS)},
  volume~36, pages 73817--73838, 2023.

\bibitem[Gon85]{Gonzalez1985}
Teofilo~F. Gonzalez.
\newblock Clustering to minimize the maximum intercluster distance.
\newblock {\em Theoretical Computer Science}, 38:293--306, 1985.

\bibitem[{Goo}]{BigQueryKMeans}
{Google Cloud}.
\newblock The {CREATE MODEL} statement for {K-means} models: {BigQuery}
  documentation.
\newblock
  \url{https://docs.cloud.google.com/bigquery/docs/reference/standard-sql/bigqueryml-syntax-create-kmeans}.
\newblock Accessed: 2026-11-01.

\bibitem[G{\"O}R23]{GOR2023}
Christoph Grunau, Ahmet~Alper {\"O}z{\"u}do{\u{g}}ru, and V{\'a}clav
  Rozho{\v{n}}.
\newblock Noisy {$k$-Means++} revisited.
\newblock In {\em 31st Annual European Symposium on Algorithms (ESA)}, volume
  274 of {\em Leibniz International Proceedings in Informatics (LIPIcs)}, pages
  55:1--55:7, 2023.

\bibitem[HJ26]{HJ2027}
Yichen Huang and Shaofeng H.-C. Jiang.
\newblock The power of arrival times in random-order online facility location.
\newblock arXiv:2607.10564, 2026.
\newblock To appear in SODA 2027.

\bibitem[Jai10]{Jain2010}
Anil~K. Jain.
\newblock Data clustering: 50 years beyond {K}-means.
\newblock {\em Pattern Recognition Letters}, 31(8):651--666, 2010.

\bibitem[JKS14]{JKS2014}
Ragesh Jaiswal, Amit Kumar, and Sandeep Sen.
\newblock A simple {$D^2$}-sampling based {PTAS} for {$k$}-means and other
  clustering problems.
\newblock {\em Algorithmica}, 70(1):22--46, 2014.

\bibitem[JS18]{JS2018}
Ragesh Jaiswal and Sandeep Sen.
\newblock Approximate clustering.
\newblock In Teofilo~F. Gonzalez, editor, {\em Handbook of Approximation
  Algorithms and Metaheuristics, Volume 2: Contemporary and Emerging
  Applications}, chapter~9. Chapman and Hall/CRC, 2 edition, 2018.

\bibitem[KNR23]{KNR23}
Haim Kaplan, David Naori, and Danny Raz.
\newblock Almost tight bounds for online facility location in the random-order
  model.
\newblock In {\em Proceedings of the 2023 Annual ACM-SIAM Symposium on Discrete
  Algorithms (SODA)}, pages 1523--1544, 2023.

\bibitem[LS19]{LS2019}
Silvio Lattanzi and Christian Sohler.
\newblock A better {$k$-means++} algorithm via local search.
\newblock In {\em Proceedings of the 36th International Conference on Machine
  Learning (ICML)}, volume~97 of {\em Proceedings of Machine Learning Research
  (PMLR)}, pages 3662--3671, 2019.

\bibitem[Mey01]{Meyerson2001}
Adam Meyerson.
\newblock Online facility location.
\newblock In {\em Proceedings of the 42nd IEEE Symposium on Foundations of
  Computer Science (FOCS)}, pages 426--431, 2001.

\bibitem[MRS20]{MRS2020}
Konstantin Makarychev, Aravind Reddy, and Liren Shan.
\newblock Improved guarantees for {$k$-means++} and {$k$-means++} parallel.
\newblock In {\em Advances in Neural Information Processing Systems (NeurIPS)},
  volume~33, pages 16142--16152, 2020.

\bibitem[ORSS06]{ORSS2006}
Rafail Ostrovsky, Yuval Rabani, Leonard~J. Schulman, and Chaitanya Swamy.
\newblock The effectiveness of {Lloyd}-type methods for the {$k$}-means
  problem.
\newblock In {\em Proceedings of the 47th Annual IEEE Symposium on Foundations
  of Computer Science (FOCS)}, pages 165--176, 2006.

\bibitem[Roz26]{Rozhon2026}
V{\'a}clav Rozho{\v{n}}.
\newblock Randomizing the number of centers in {$k$-means++}.
\newblock arXiv:2607.26202, 2026.

\bibitem[Sch38]{Schoenberg1938}
I.~J. Schoenberg.
\newblock Metric spaces and positive definite functions.
\newblock {\em Transactions of the American Mathematical Society},
  44(3):522--536, 1938.

\bibitem[{sci}]{ScikitLearnKMeans}
{scikit-learn developers}.
\newblock {kmeans\_plusplus}: {scikit-learn} {API} reference.
\newblock
  \url{https://scikit-learn.org/stable/modules/generated/sklearn.cluster.kmeans_plusplus.html}.
\newblock Accessed: 2026-10-23.

\bibitem[Wei16]{Wei2016}
Dennis Wei.
\newblock A constant-factor bi-criteria approximation guarantee for
  {$k$-means++}.
\newblock In {\em Advances in Neural Information Processing Systems (NeurIPS)},
  volume~29, pages 604--612, 2016.

\end{thebibliography}
\end{document}